\documentclass[journal]{IEEEtran}
\usepackage{amsfonts}
\usepackage{amsmath}
\usepackage{algorithm}
\usepackage{algpseudocode}
\usepackage{array}
\usepackage{booktabs}
\usepackage{graphicx}
\newcolumntype{P}[1]{>{\centering\arraybackslash}p{#1}}
\usepackage[caption=false,font=small]{subfig}
\usepackage{url}
\usepackage{verbatim}
\usepackage{xcolor}
\usepackage[sort&compress,numbers]{natbib}

\usepackage[normalem]{ulem}

\newcommand{\noline}{}
\newcommand{\new}{\textcolor{black}}
\usepackage{enumitem}
\setlist[itemize]{leftmargin=*, noitemsep}
\setlist[enumerate]{leftmargin=*, noitemsep} 

\begin{document}

\title{Cyber Deception for Mission Surveillance via Hypergame-Theoretic Deep Reinforcement Learning}

\author{Zelin Wan, Jin-Hee Cho,~\IEEEmembership{Senior Member, IEEE}, Mu Zhu, Ahmed H. Anwar, and Charles Kamhoua,~\IEEEmembership{Senior Member, IEEE}, Munindar P. Singh,~\IEEEmembership{IEEE Fellow}
\IEEEcompsocitemizethanks{\IEEEcompsocthanksitem This research was partly sponsored by the Army Research Laboratory and was accomplished under Cooperative Agreement Number W911NF-19-2-0150 and W911NF-23-2-0012. In addition, this research is also partly supported by the Army Research Office under Grant Contract Numbers W911NF-20-2-0140 and W911NF-17-1-0370. The views and conclusions contained in this document are those of the authors and should not be interpreted as representing the official policies, either expressed or implied, of the Army Research Laboratory or the U.S. Government. The U.S. Government is authorized to reproduce and distribute reprints for Government purposes, notwithstanding any copyright notation herein ({\em Corresponding author: Zelin Wan}). Zelin Wan and Jin-Hee Cho are with the Department of Computer Science, Virginia Tech, Falls Church, VA, USA. Email: \{zelin, jicho\}@vt.edu. Mu Zhu is with the Computer Network Information Center, Chinese Academy of Sciences, Beijing, China. Email: zhumu@cnic.cn. Munindar P. Singh is with the Department of Computer Science, North Carolina State University, Raleigh, NC 27695. Email: mpsingh@ncsu.edu. Ahmed H. Anwar and Charles A. Kamhoua are with the US Army Research Laboratory, Adelphi, MD, USA. Email: a.h.anwar@knights.ucf.edu; charles.a.kamhoua.civ@mail.mil.}}



\maketitle

\begin{abstract}
Unmanned Aerial Vehicles (UAVs) are valuable for mission-critical systems like surveillance, rescue, or delivery. Not surprisingly, such systems attract cyberattacks, including Denial-of-Service (DoS) attacks to overwhelm the resources of mission drones (MDs). How can we defend UAV mission systems against DoS attacks? We adopt cyber deception as a defense strategy, in which honey drones (HDs) are proposed to bait and divert attacks. The attack and deceptive defense hinge upon radio signal strength: The attacker selects victim MDs based on their signals, and HDs attract the attacker from afar by emitting stronger signals, despite this reducing battery life. We formulate an optimization problem for the attacker and defender to identify their respective strategies for maximizing mission performance while minimizing energy consumption. To address this problem, we propose a novel approach, called {\em HT-DRL}. HT-DRL identifies optimal solutions without a long learning convergence time by taking the solutions of hypergame theory into the neural network of deep reinforcement learning. This achieves a systematic way to intelligently deceive attackers. We analyze the performance of diverse defense mechanisms under different attack strategies. Further, the HT-DRL-based HD approach outperforms existing non-HD counterparts up to two times better in mission performance while incurring low energy consumption.
\end{abstract}

\begin{IEEEkeywords}
Cyber deception, deep reinforcement learning, game theory, unmanned aerial vehicle, mission effectiveness
\end{IEEEkeywords}

\section{Introduction} \label{sec:introduction}
Unmanned Aerial Vehicles (UAVs) have been widely adopted in mission systems to improve energy efficiency,
foster autonomous control,
and enhance communication capabilities \cite{kurunathan2023machine}.
However, ensuring the security of UAVs is an ongoing challenge.
Denial-of-Service (DoS) attacks are especially harmful because they can result in data invalidation, data leakage, and physical damage through drone crashes~\cite{hooper2016securing}.
To solve this problem, we take a cyber deception-based defense approach, called {\em defensive deception} (DD). The DD strategy can confuse and mislead attackers into choosing sub-optimal attack strategies~\cite{zhu2021survey}. We aim to design a surveillance mission system using honey drones (HDs) to combat DoS attacks~\cite{daubert2018honeydrone}. HDs, equipped with lightweight virtual machines (VMs) running vulnerable software, can lure potential cyberattacks at higher signal strengths and collect attack intelligence to update system settings in realtime.

Deep reinforcement learning (DRL) has been applied to cybersecurity problems~\cite{li2022defensive, li2022optimal, huang2022reinforcement}.
However, DRL algorithms are challenged by long convergence times to solutions under non-stationary settings.  On the other hand, game theory has been substantially used in cybersecurity to model strategic attack-defense decision-making processes. The synergistic effect of combining game theory and deep learning has been recognized as a promising direction~\cite{zhu2021survey}. However, such hybrid approaches have been significantly less studied.

We propose a novel hybrid approach, {\em hypergame theory-guided DRL} (HT-DRL), integrating DRL with game theory to identify the optimal settings of the proposed honey drone (HD)-based system. By taking the merit of DRL for autonomy and hypergame theory for strategic decision-making, HT-DRL can empower UAVs to make proactive, intelligent decisions in uncertain and adversarial environments.  Hypergame theory~\cite{vane2000hypergame} can better handle an agent's perceived uncertainty than conventional game theory by estimating its utility, {\em hypergame expected utility} (HEU). Hypergame theory helps reduce the number of strategies an agent takes based on the subgames each agent perceives towards the opponent's moves.  A game-theoretic solution is more efficient than its DRL counterparts due to no training time needed in game theory. However, in terms of solution optimality, DRL can perform better than game theory. Therefore, taking a hybrid approach by combining these two is a natural direction to tackle this problem. 

\new{In our UAV DoS setting, the attacker does not know which nodes are honey drones versus mission drones and must instead infer targets from noisy radio-signal patterns and limited mission information. The defender, in turn, only partially observes how the attacker thresholds signal levels and adapts to the current honey–drone configuration. These perceptions evolve over time and may be inaccurate. Conventional game-theoretic defenses for UAVs and cyber–physical systems, including our prior honey–drone study~[11], typically assume a single, fully specified game with a fixed payoff structure and optimize strategies within that static model. DRL-based defenses can adapt in high-dimensional mission states but often treat the opponent as part of a stationary environment, which leads to instability when the attacker is itself adaptive.}

\new{Hypergame theory is therefore directly relevant, as it provides a principled framework for representing and updating the differing perceived games held by the attacker and defender. Building on this insight, our HT-DRL framework formulates a dynamic hypergame between an adaptive attacker and a honey–drone defender, derives HEUs for both players, and uses the defender’s HEU to shape the actor–critic policy logits through a filter layer. This integration enables early exploration and subsequent policy adaptation to be guided by the defender’s evolving perception of the opponent, rather than assuming a fixed or stationary adversarial model.}

In sum, our \textbf{key contributions} are as follows:
\begin{itemize}
\item We propose a surveillance mission system using honey drones to effectively thwart cyber threats, protect valuable resources, and maintain mission integrity, which has not been explored in the literature.
\item We provide a novel way of integrating game theory with DRL, called hypergame theory-guided DRL (HT-DRL), to enable a shorter convergence time for better solutions. Prior studies~\cite{zheng2022stackelberg, cao2022game} using game theory and DRL have not studied this hybrid approach.
\item We model an attack-defense game where the attacker and defender take intelligent strategies via HT-DRL to evaluate defense strategies under various attack scenarios.
\item We validate the superiority of HT-DRL defense strategy through performance analyses under various attack scenarios. Via extensive experiments, we compare the performance of HT-DRL using HDs over non-HD counterparts. Our results demonstrate the outperformance of our approach in mission performance and energy consumption. Due to space constraints, we demonstrate the additional comparative performance results and analyses in the supplement document.  
\end{itemize}
In our preliminary work~\cite{wan2023deception}, we investigated the effectiveness of using honey drones in a UAV-based mission system for a surveillance mission. And the prior considered attack-defense interactions based on game theory.

\new{Our prior work~\cite{wan2023deception} investigated a honey--drone (HD) defense mechanism that adjusts the signal strengths of mission and honey drones using either a game-theoretic strategy selector or a DRL-based selector. That study focused on determining the appropriate HD signal level and comparing the resulting performance against existing baselines (e.g., HD-based IDS and ContainerDrone) in terms of mission completion, energy consumption, and connectivity. The underlying adversarial interaction assumed a standard game model in which the attacker and defender share a common and accurate view of the environment. Moreover, the DRL agent (using a conventional A2C policy) and the game-theoretic agent each optimized a hand-crafted expected-utility function; the two approaches were evaluated independently rather than being integrated into a unified framework.}

\new{In contrast, the present work introduces a fundamentally different modeling and algorithmic perspective. We formulate a \emph{dynamic hypergame} in which the attacker and defender may hold distinct and evolving perceptions of the game state and payoff structure. Building on this formulation, we derive hypergame-expected utilities (HEUs) for both players and leverage the defender’s HEU to construct a filter layer that shapes the DRL policy logits. This coupling of hypergame analysis with policy learning yields the proposed HT-DRL framework, which explicitly addresses cold-start and non-stationarity challenges posed by adaptive DoS attackers. Rather than offering another empirical comparison between DRL and game-theoretic defenses, this work integrates the two viewpoints into a principled, perception-aware learning architecture that is robust to adversarial adaptation.}

In this work, we introduced {\em Hypergame Theory (HT)}~\cite{vane2000hypergame} to address each player's different view about the same game, which reflects real-world scenarios more appropriately. Hence, players' action choices are based on the Hypergame Expected Utility (HEU), representing their utility depending on the level of perceived uncertainty.  Further, the merit of using HT is to reduce the size of an action space by utilizing the concept of a subgame to avoid a full game with all possible actions. \new{Our prior work~\cite{wan2023deception} introduces the basic UAV surveillance system and simulation environment. In contrast, the present work advances the field by developing a hypergame-theoretic DRL framework that supports scalable HD deployment, incorporates refined energy and threat models, and conducts substantially expanded and more rigorous experimental evaluations.} Further, we substantially extended our experimental results and analyses based on four key metrics (see Section~\ref{subsec:metrics}) to investigate the effectiveness of HD-based deception defenses.

Our proposed HT-DRL framework adapts to varying levels of prior knowledge. When no prior knowledge is available, HT-DRL gracefully degrades to operate as a standard DRL model, ensuring applicability even in scenarios with limited historical data. The framework's flexibility allows it to leverage any available domain knowledge while remaining functional without it. With no prior knowledge, it operates as a full DRL model, while with complete prior knowledge, it fully leverages HT-based decision-making. This flexibility ensures its applicability across diverse scenarios.

\section{System Model} \label{sec:system-model}
\subsection{Network Model} \label{subsec:network-model}
This study examines a drone fleet deployed for surveillance in a target area. The network consists of a Charging Station (CS) for drone recharging, a Ground Control Station (GCS) responsible for mission assignments, and UAVs for mission execution. We consider three types of UAVs: Regional Leader Drone (RLD), Mission Drones (MDs), and Honey Drones (HDs). Considering the limited transmission range of a GCS, we propose a network architecture where an RLD is connected to the GCS through a satellite network, and MDs and HDs establish connections with the RLD via WiFi, forming a ``flying ad hoc network'' (FANET)~\cite{kim2016multi}. The FANET facilitates multi-hop communications between drones coordinated by the RLD. MDs send real-time data to the RLD. We assume the GCS possesses high computational power and a firewall to filter out malicious streams, making it a trusted entity not vulnerable to DoS attacks (see our threat model in Section~\ref{subsec:threat-model}).

\begin{figure*}[t]
\centering
\subfloat[Example multi-hop communication network of a drone fleet.]{\includegraphics[width=0.7\textwidth]{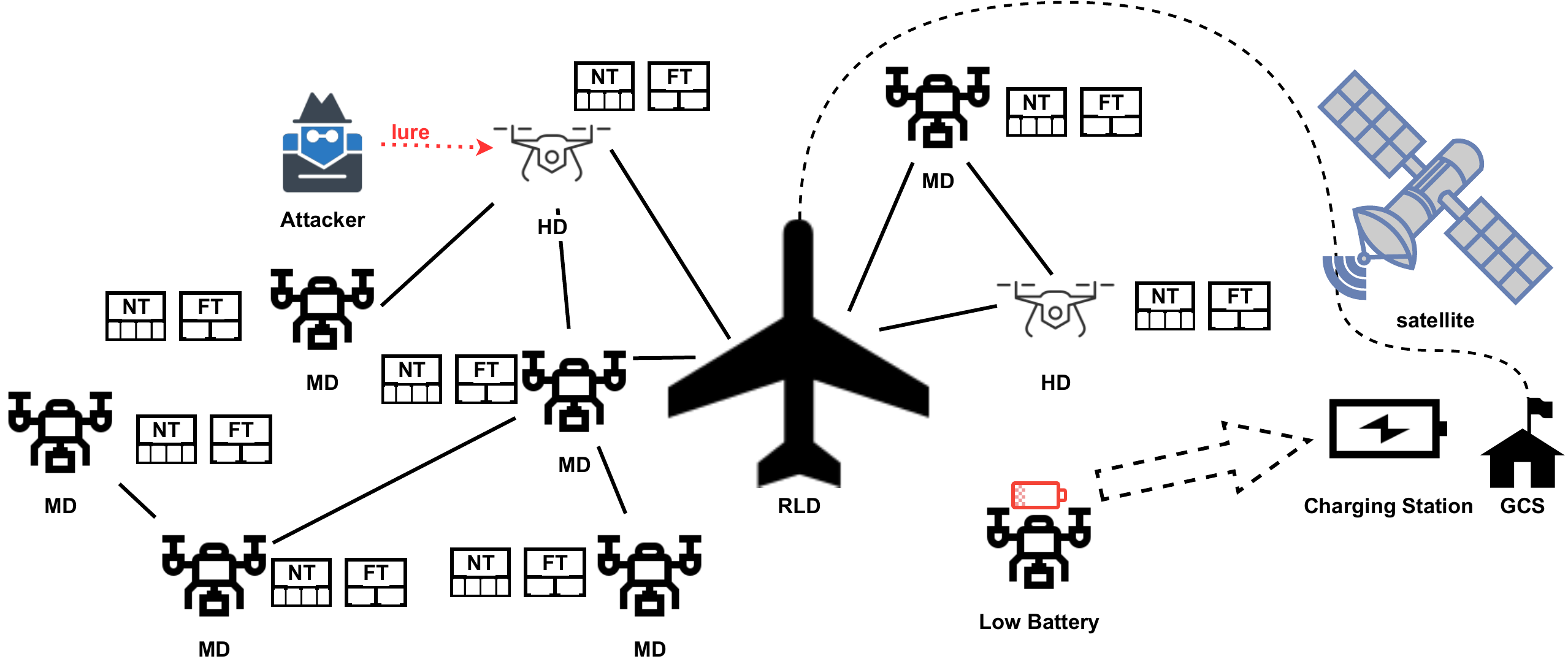}\label{fig: network-model}} 
\hfil
\subfloat[Example neighbor table (NT) and fleet table (FT).]{\includegraphics[width=0.25\textwidth]{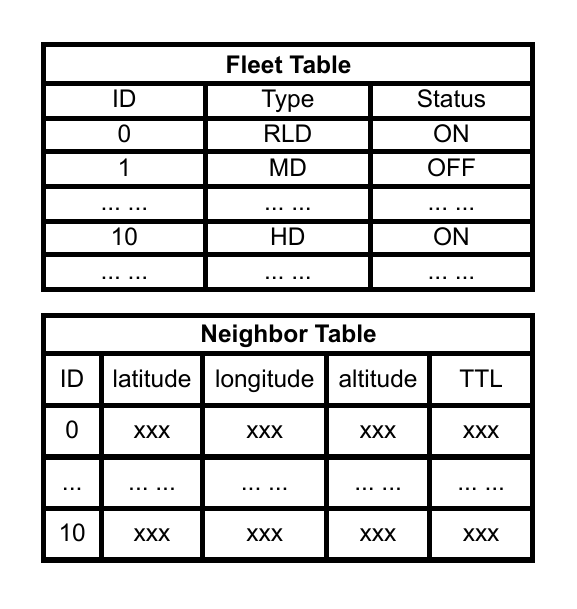}\label{fig: two-table-UAV}}
\caption{\textbf{(1) Example drone fleet with multi-hop communications} consisting of MDs, HDs, RLD, and GCS. \textbf{(2) Example of the Neighbor Table (NT) and Fleet Table (FT)} where NT provides information about the neighboring drones of a given drone, while FT tracks the membership of the mission team, including drones executing missions, leaving the team to recharge batteries, or compromised by the attacker. Drones with a battery level lower than the energy threshold $T_e$ will leave the mission team and move to the charging station.}
\end{figure*}
As Fig.~\ref{fig: two-table-UAV} shows, each drone maintains two tables: The Neighbor Table (NT) and the Fleet Table (FT). The NT contains location information (i.e., latitude, longitude, and altitude) and the Time-to-Live (TTL) values of neighboring drones~\cite{kim2016multi}. The FT includes a drone's ID, type, and membership to monitor the mission team's status.  To ensure continuous network connectivity, each drone (i.e., MD or HD) broadcasts a periodic hello message containing its ID and location. Upon receiving a hello message from drone $j$, drone $i$ searches for drone $j$'s status information in its NT. If drone $j$ is not found in the NT, drone $i$ initiates connection establishment and updates its NT accordingly. If drone $j$ is present in the NT, drone $i$ updates the location information and TTL values based on drone $j$'s hello message~\cite{kim2016multi}. The connection process begins with a TCP handshake where drones exchange JSON records to determine a communication channel. Drone $i$ sends its record to drone $j$, and drone $j$ responds with acceptance or rejection based on whether drone $i$'s ID is found in the FT (for authentication). Once both drones accept the connection, they add each other to their respective NTs. Then, data collection and transmission occur via UDP. We consider the target region composed of multiple discrete cells~\cite{khiati2019air}, allowing MD and HD to move between them.

To mitigate packet loss in unstable wireless environments, our network model integrates TTI bundling~\cite{ahmed2021hybrid}. The receiver can perform soft combining by transmitting repeated copies of a packet in short, consecutive intervals, significantly improving decoding reliability under adverse channel conditions.

To facilitate scalability in larger or more complex missions, our network architecture was designed to be inherently modular and could be extended by integrating additional Regional Leader Drones (RLDs). These RLDs could individually manage sub-fleets of Mission Drones and Honey Drones, thereby ensuring efficient coordination across an expanded operational area. This scalability approach was supported by existing works: \citet{widhalm2025community} demonstrated scalable drone operations via shared service models, and \citet{qin2023coordination} highlighted the effectiveness of decentralized, energy-aware coordination in managing large drone fleets.

The computational complexity of our HD deployment algorithm (Algorithm \ref{alg:HD placement}) is $O(|L_H| \times |L_M|)$, where $|L_H|$ is the number of honey drones and $|L_M|$ is the number of mission drones. Our network architecture's modular design inherently supports larger deployments through additional RLD coordination. Our hierarchical RLD structure is designed to support scalable deployments through its modular architecture. The algorithm complexity is primarily determined by the product of HD and MD numbers, making it suitable for moderate-scale deployments while maintaining mission effectiveness through parallel RLD coordination.

\new{Because the present study employs the same physical UAV platform and communication topology as in~\cite{wan2023deception}, certain elements of the system description and simulator configuration unavoidably overlap with our prior work.}

\subsection{Node Model} \label{subsec:node-model}

At the beginning of the mission, the drone fleet is fully charged and takes off from the GCS. Once a drone's battery is depleted, it leaves the mission team and returns to the CS for recharging. The CS and GCS are not mobile. Fig.~\ref{fig: network-model} illustrates the network model, characterized by the following node types.
\subsubsection{\textbf{Charging Station (CS)}} \label{subsubsec:charging-station}
A drone returns fully recharged before rejoining the mission team by taking $T_{C}$ time.

\subsubsection{\textbf{Ground Control Station (GCS)}} \label{subsubsec:gcs}
The GCS assigns missions in the target region and sets a maximum mission duration, $T_{M}^{\mathrm{max}}$. The GCS tracks the drone fleet, monitoring mission progress reported by RLDs, e.g., via FBCB2's BFT (Blue Force Tracker) satellite network~\cite{chevli2006blue}. However, due to the high latency, low bandwidth, and instability of satellite communication, the GCS may be unable to track the drone's locations and signal strengths in real-time. Yet, the RLD is authorized to have timely control over the UAVs' settings.

\subsubsection{\textbf{Regional Leader Drone (RLD)}} \label{subsubsec:RLD}
The RLD is a high-altitude long-endurance drone \cite{siddappaji2020role}. It possesses sufficient energy to handle heavy computations and stores sensed data over the entire mission duration, $T_M$ ($\leq T_{M}^{\mathrm{max}}$). The RLD dynamically configures the drones' trajectories and transmits them to MDs.
The RLD identifies the optimal signal strength to minimize energy consumption while accomplishing the mission. The RLD moves to the center of the target area to maximize signal coverage. When an MD is offline (i.e., compromised, exhausted in battery, or disconnected from unreliable wireless connections) or online (i.e., non-compromised with sufficient battery), the RLD adjusts the trajectory for all drones to accommodate the new conditions. The RLD can analyze attack intelligence collected by HDs and update the fleet network's configuration by modifying open ports for attacked MDs, thereby avoiding DoS attacks. Although DoS attacks can potentially compromise the RLD, we assume it is equipped with robust defense mechanisms to detect attacks and has sufficient computing power to handle a large volume of requests.  In our system, a single RLD exists and may expose a single point of failure. To deal with this, when the RLD is compromised, a backup RLD, which stands by in the GCS, will replace it. During the RLD's downtime, as it serves as a data storage unit, the mission process is paused. The DRL agent, running on the RLD, controls the signal strengths of MDs and HDs to maximize mission performance. Section~\ref{subsec: defender-model} specifies how to identify the optimal signal strength to be used by HDs and MDs.

\subsubsection{\textbf{Mission Drone (MD)}} \label{subsubsec:mission-drone}
MDs are equipped with Himax HM01B0 ULP monochrome QVGA
cameras~\cite{himaxCamera}.
They transmit sensed data to the RLD through the FANET. Each MD initially follows an assigned trajectory upon deployment. When an MD goes to recharge, it notifies the RLD offline. However, if the RLD does not receive such a signal but detects that the MD is offline, it assumes the MD has been compromised. Since an MD can go offline due to terrain, obstacles, or unreliable wireless connections, a non-compromised MD may appear compromised. When such drones reconnect, NT and FT (see Fig.~\ref{fig: two-table-UAV}) are updated accordingly. We assume the drones have basic knowledge of the target region~\cite{khiati2019air}. Each drone is loaded with an optimized trajectory to complete the mission. If a drone goes offline, the trajectory is recalculated for others based on the scanned and completed cells.

\subsubsection{\textbf{Honey Drone (HD)}} \label{subsubsec:honey-drone}
Each HD consists of two logically isolated components: a lightweight honey VM and an infrastructure VM. The honey VM exposes multiple open ports to attract DoS attacks. The infrastructure VM, running on a lightweight Linux environment, monitors the memory and log of the honey VM. Upon detecting a DoS attack based on abnormal computing consumption (e.g., CPU and memory usage), the infrastructure VM's backchannel informs the RLD of the port being used by the attacker for communication. Subsequently, the RLD reconfigures the open port of the MD to prevent further attacks. In case the decoy system in the honey VM is compromised or malfunctions, the honey VM can restore it. This honeypot can be effectively implemented on a Samsung Galaxy S2 smartphone, released in 2011~\cite{liebergeld2013nomadic}, indicating it would not overload the drone.

HDs are deployed in a greedy manner following Algorithm~\ref{alg:HD placement}, where the $d(sg_{HD})=10^{\frac{sg_{HD}+100}{\eta \cdot 10}}$ is derived from Eq. \eqref{Eq:dBm-distance}. The deployment of HDs in Algorithm~\ref{alg:HD placement} considers drones' energy concerns by limiting MDs assigned to each HD for monitoring. That is, we consider an upper bound that one HD can simultaneously protect $\tau_u$ number of MDs.  Each HD is assigned a set of MDs between $\tau_l$ and $\tau_u$ to protect and monitor. Some HDs may not be assigned any MDs and may move around to find available MDs.  In this case (i.e., line~\ref{alg:line:new_position}), the HD finds a location when $\tau_l \leq |N(l'_H)| \leq \tau_u$. We search only MDs' positions to reduce computational complexity instead of all cells.  This will lead to reducing the complexity of Algorithm~\ref{alg:HD placement} from $O(|L_H| |N_{cell}|)$ to $O(|L_H| |L_M|)$, where $|N_{cell}|$ is much higher than $|L_M|$.

\begin{algorithm}[t]
\caption{Honey Drone Deployment} \label{alg:HD placement}
\footnotesize
\begin{algorithmic}[1]
\State{$L_M \leftarrow$ A set of active and not in GCS MD locations}
\State{$L_H \leftarrow$ A set of active HD locations}

\State{$P^{H}_r \leftarrow  d(sg_{HD})$, $sg_{HD} \leftarrow DS_5$} \Comment{The signal radius/range of an HD when RLD selects ${sg}_{HD} = DS_5$ for the HD}
\State{$\mathrm{D}(x,y) \leftarrow$ The distance between two drones $x$ and $y$}
\State{$S = \emptyset$} \Comment{A set of deployed HD locations}
\State{$[\tau_{l}, \tau_{u}] \leftarrow$ The lower and upper bounds of the number of MDs}
\For{$l_H \in L_H$}
\If{$|L_M| = 0$}
\State{$S \leftarrow S \cup \{l_H\}$}
\Else{}
\State{$N(l_H)=\{{\mathrm{D}}(l_H,l_M)< P^{H}_r : l_M \in L_M\}$} \Comment{A set of MDs in the protect range of HD $l_H$}
\If{$|N(l_H)| < \tau_l$} 
\State{Find a new position $l'_H$ such that $\tau_l \leq |N(l'_H)| \leq \tau_u$} \label{alg:line:new_position}
\State{where $N(l'_H)=\{{\mathrm{D}}(l'_H,l_M)< P^{H}_r : l_M \in L_M\}$} \Comment{A set of MDs detected/protected by HD $l'_H$}
\State{$L_M \leftarrow L_M \setminus N(l'_H)$} \Comment{Remove protected MDs from set $L_M$}
\State{$S \leftarrow S \cup \{l'_H\}$} \Comment{Add deployed HD to set $S$}
\ElsIf{$\tau_l \leq |N(l_H)| \leq \tau_u $}
\State{$L_M \leftarrow L_M \setminus N(l_H)$}
\State{$S \leftarrow S \cup \{l_H\}$}
\Else{}
\State{$N'(l_H) \subseteq N(l_H)$, where $|N'(l_H)|=\tau_u$} \Comment{Select the nearest $\tau_u$ MDs from $N(l_H)$ and assign them to HD $l_H$}
\State{$L_M \leftarrow L_M \setminus N'(l_H)$}
\State{$S \leftarrow S \cup \{l_H\}$}
\EndIf
\EndIf
\EndFor
\State{Output: HD location set $S$}
\end{algorithmic}
\end{algorithm}

\subsection{Energy Model} \label{subsec:energy-model}
Our simulation (see Section~\ref{sec:experimental-setup}) considers Crazyflie 2.X quadrotor drones~\cite{bitcrazeDrone} with the ``gym-pybullet-drones'' simulator~\cite{panerati2021learning}. Both MDs and HDs consume energy as follows:
\begin{gather}
E_{MD} = E_P + E_C + \frac{E_R  \cdot DS_j}{10}, \; \;  E_{HD} = E_P +  \frac{E_R \cdot DS_j}{10},
\label{eq:energy-model}
\end{gather}
where $E_C$ represents the consumption rate of the Himax camera \cite{palossi2021fully}.
$E_P$ corresponds to the power consumed by the drone platform, including the Standard Operating Conditions (SoCs) and four motors. The estimation of $E_P$ is derived from the platform's flight time. $E_R$ represents the maximum consumption rate of the radio. As the defense strategy controls the signal level, $E_R \times \frac{DS_j}{10}$ represents the real-time energy consumed by taking a given defense strategy. 

The energy consumption model is based on real-world hardware specifications, with power values calibrated from Bitcraze Crazyflie 2.X~\cite{bitcrazeDrone} and Himax camera data~\cite{himaxCamera} and validated against empirical measurements reported in the literature~\cite{palossi2021fully}. Although not exhaustive, this model offers a reliable baseline for typical UAV operations, with future work planned to incorporate additional factors such as battery aging.

\subsection{Threat Model} \label{subsec:threat-model}
Recently, security threats and their serious impact on UAV systems have been recognized~\cite{ouiazzane2022multiagent, chen2019container, hooper2016securing, ly2021cybersecurity, mekdad2023survey}.
Particularly, the serious adverse impact of DoS attacks on UAV systems has been a serious security issue, while the DoS attacks are simple to launch but can cause serious data leakage and crashing drones~\cite{ly2021cybersecurity, mekdad2023survey}. Since UAVs often should perform real-time communications under high dynamics (e.g., node join, leave, or failure) and resource constraints, draining UAVs' energy and preventing their communications by DoS attacks can introduce a critical impact to the mission system and easily cause mission failure~\cite{hooper2016securing}. Therefore, this work primarily focuses on providing a security solution to address DoS attacks.

To estimate a drone's software vulnerability, we adopt the Common Vulnerability Scoring System (CVSS)~\cite{mell2006common}.
We model each drone's software vulnerability using the CVSS score of an Android device~\cite{cvss_score}. We represent a drone $\kappa$'s software vulnerability as a real value, $\mathrm{vul}_{\kappa} \in [0, 1]$, indicating the likelihood of a successful attack.  We do not assume attackers can capture drones' vulnerability through reconnaissance attacks, which is possible when the attackers can observe a target system for an extended period because this concerned mission system will be assigned a short-term mission. Therefore, we consider intelligent attack strategies in choosing their signal strengths in Section~\ref{subsec:attacker_model}.

We will consider DoS attackers sitting on the ground to ensure their infrastructure needs, such as high-power antennas and a stable power source. In addition, for the attackers to avoid physical detection of their presence, we do not consider attackers who are physically proximate to the UAV mission team. Each drone's presence and its connectivity with the UAV network can be detected by the signal strength level~\cite{baazaoui2023modeling}. In addition, recipients receiving such strong signals can save their energy~\cite{gachhadar2023power}.  A Denial-of-Service (DoS) attacker is more likely to target UAVs with stronger signal strengths to maximize the impact of their attacks while conserving energy. Due to the inverse relationship between signal-to-noise ratio (SNR) and packet error rate (PER)~\cite{mahmood2018cross}, stronger signals enable the attacker to deliver malicious packets more reliably and efficiently, ensuring that a higher proportion of these packets reach the UAV and effectively overwhelm its communication system~\cite{wikipedia2024stochastic}. This increased reliability enhances the overall effectiveness of the attack, leading to greater disruption of the UAV's operations. Additionally, targeting UAVs with robust signal strengths allows the attacker to use lower transmission power~\cite{bjornson2018energy, mahmood2019energy}, thereby conserving energy compared to attacking UAVs with weaker signals, which would require higher power levels to achieve the same packet delivery success. Attackers can optimize their resources by focusing on UAVs with stronger signals to sustain prolonged attacks with maximum disruptive potential.  We assume an attacker is also limited to its computing power and energy resources. Hence, we consider the attacker's budget, $\zeta$, representing the maximum number of drones to launch its attack. Larger $\zeta$ means more severe attack strength, whose impact is analyzed in Section~\ref{sec:resuts-analyses}.

While our current focus is on DoS attacks due to their prevalence and severe impact on UAV systems, the underlying principles of our HT-DRL approach can be adapted to address other cyber threats with suitable modifications in utility models and defense strategies. The modular design of our framework facilitates such extensions. Future work will extend our framework to address additional attack types such as data tampering and spoofing.

\section{Strategy Selection in HT-DRL}
\label{sec:game-framework}
A challenge in mission systems is the lack of prior knowledge about attack behaviors. Real-time defense cannot rely on instructions from a central entity due to the high delay of satellite networks.
Therefore, an adaptive mechanism is needed to learn and make decisions. DRL algorithms support autonomous interactions with the environment and learning independently without prior knowledge.
We use DRL to identify an optimal signal strength to maintain network connectivity while minimizing vulnerabilities to DoS attacks.

We formulate attack-defense interactions as a series of simulations by the attacker and defender to achieve their respective goals during mission execution. The mission's completion time is $T_M$. $T_M^{\mathrm{max}}$ is the maximum time allowed for mission completion. If $T_M > T_M^{\mathrm{max}}$, it indicates the partial completion, which means mission failure. The mission consists of multiple rounds of interaction between the attacker and defender, resulting in mission success or failure. 
The mission team of RLD, multiple MDs, and multiple HDs takes off from the GCS and continues until the mission concludes. 

In this work, two players, the attacker and defender, are given 10 strategies representing 10 different levels of signal strengths. This number was determined based on our experiments and showed a sufficient level of diverse strategies while it does not introduce too high computational complexity.  As our work adopts hypergame theory, we use the concept of a subgame each player can use to reduce its action space, lowering solution search complexity. However, when there is uncertainty, they will use a full game with all 10 strategies as the action space. We described each player's hypergame theoretic decision-making process in Sections~\ref{subsec:attacker_model} and \ref{subsec: defender-model}.

\subsection{Key Procedures of HT-DRL}
The balance between exploration and exploitation in solution search by DRL agents is critical. The dynamic $\epsilon$-greedy exploration allows initial random decision-making under a lack of knowledge while more exploitation is used as the agent accrues knowledge. However, this strategy often introduces a long convergence time to a close-to-optimal solution, particularly in scenarios without prior knowledge. Therefore, we introduce HT-DRL, which uses Hypergame Expected Utility (HEU) to guide a DRL agent to choose its best action. That is, instead of randomly exploring solutions at the beginning, the agent can use HEU-based action distributions for DRL's policy function at the beginning to resolve the cold start problem without being stuck at local optima. The hypergame outcomes are integrated into the DRL framework to guide exploration, ensuring that the agent learns under conditions that reflect real-world uncertainty. We analyze HT-DRL's performance and compare it with the performance of other baseline and existing counterparts in Section~\ref{sec:resuts-analyses}. 

\begin{figure}
\centering
\includegraphics[width=0.5\textwidth]{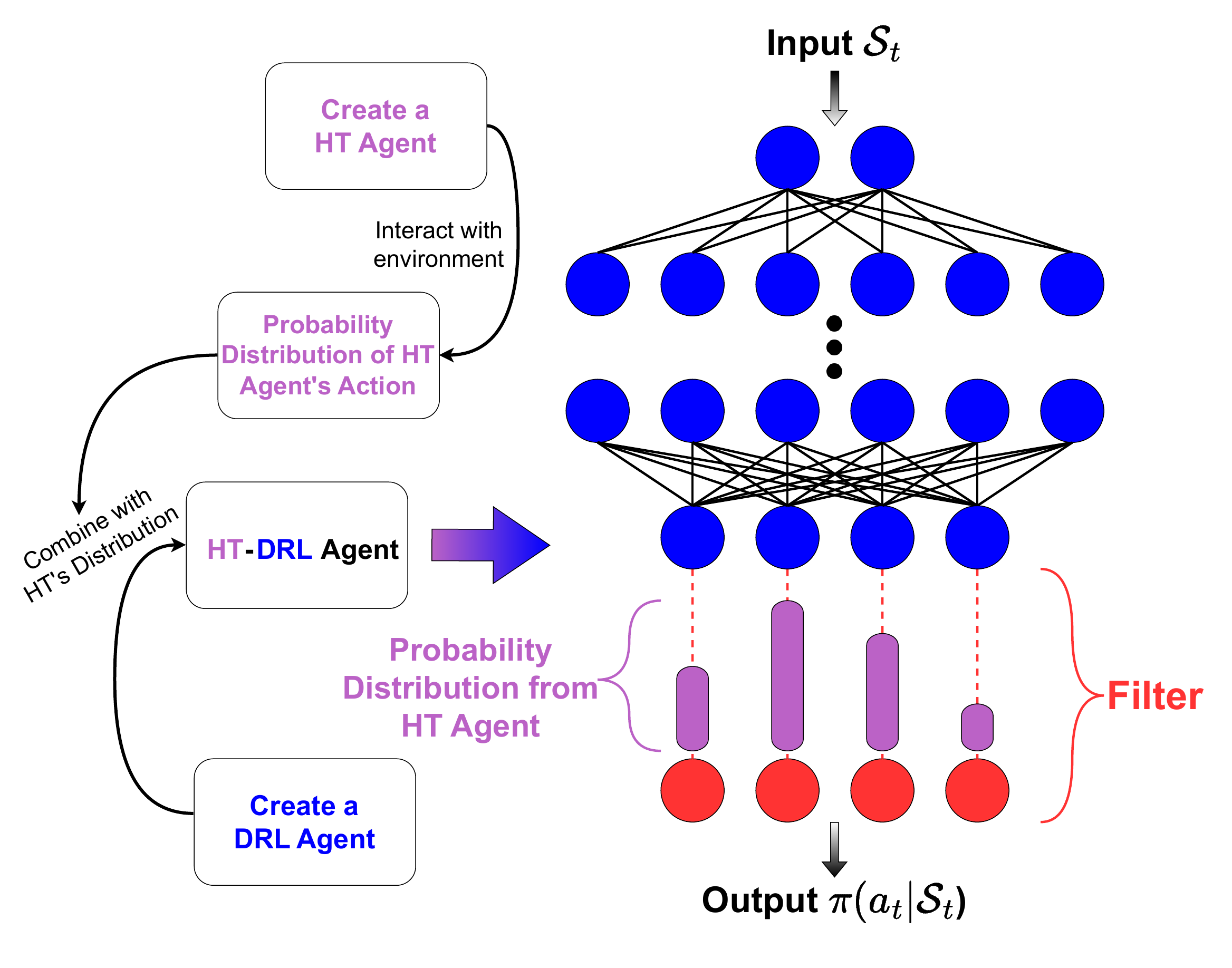}
\caption{{\bf The procedures generating the solutions by a HT-DRL agent:} $\mathcal{S}_t$ is the state at round $t$ and $\pi(a_t|\mathcal{S}_t)$ is the probability of all actions.}
\label{fig: HT-guided_DRL}
\end{figure}

HT-DRL achieves its outperformance by incorporating a unique HT-guided filter layer into the neural network and taking the following procedures: 
\begin{enumerate}
\item Utilize prior knowledge to train an HT agent to choose its action based on the HEU. Due to the characteristics of HT, only a small amount of data is required for this training.
\item Collect the HT agent's action choices to construct its action probability distribution (APD).
\item Instantiate a DRL agent using the Advantage Actor-Critic (A2C) technique.
We add a filter to the output layer to alter the action distribution by multiplying each output value by a corresponding weight.
\item Apply the APD from the pre-trained HT agent to each filter weight so that the DRL agent can perform a more strategic rather than completely randomized initial exploration.
\item Train this HT-DRL as a standard DRL for further fine-tuning parameters.
\end{enumerate}
We summarize the procedures above in Fig.~\ref{fig: HT-guided_DRL}.

\new{
\paragraph{Hypergame-Theoretic Game Formulation}
We model the attacker--defender interaction as a finite-horizon stochastic game
$\mathcal{G} = \langle P, \mathcal{S}, \{\mathcal{A}^p\}_{p \in P}, T, \{u^p\}_{p \in P} \rangle$,
where:
\begin{itemize}
  \item $P = \{A,D\}$ is the set of players: the attacker $A$ and defender $D$.
  \item $\mathcal{S}$ is the state space, containing the mission progress, drone connectivity, and scan map, e.g., $\mathcal{S}^{A/D}_t = (\mathcal{R}_{MC}^t, \mathcal{M}_{SP}^t, N_{TR}^t)$.
  \item $\mathcal{A}^A$ and $\mathcal{A}^D$ are the action spaces of attacker and defender, where each action corresponds to a discretized signal-strength range ($AS_i$ for the attacker, $DS_j$ for the defender).
  \item $T: \mathcal{S} \times \mathcal{A}^A \times \mathcal{A}^D \rightarrow \Delta(\mathcal{S})$ is the transition kernel induced by the UAV dynamics, energy consumption, and DoS effects in the simulator.
  \item $u^A$ and $u^D$ are the instantaneous utilities of the attacker and defender, derived from the gains and losses in \eqref{eq: attacker-utility} and \eqref{eq: defender-utility}.
\end{itemize}
Classical stochastic games typically assume that both players share the same game $\mathcal{G}$ and have correct beliefs about each other's strategies. In contrast, our setting is naturally modeled as a \emph{hypergame}, where each player $p \in \{A,D\}$ maintains its own perceived game $\widehat{\mathcal{G}}^p$ with a perceived utility $u^p$ and belief $C^p_{\Sigma}$ over the opponent's actions. The hypergame expected utilities (HEUs) in \eqref{Eq:A-HEU} and \eqref{Eq:D-HEU}
are computed on these perceived games and used to construct practical HEU-guided policies, which we subsequently integrate into the DRL framework. Hyper Nash Equilibrium (HNE)~\cite{kanazawa2006replicator} was analyzed in detail in our prior work~\cite{wan2023resisting}, and a full HNE analysis is beyond the scope of the present study.}

\subsection{Attacker Model} \label{subsec:attacker_model}

\subsubsection{\bf Attacker's Action Space based on Signal Strengths} \label{subsubsec:attacker_action}
The attacker observes the drones' signal strengths and selects its attack strategy accordingly. We define the attack strategy as $AS_i \in \{AS_1, \dots, AS_{10}\}$, where each action corresponds to a range of received signal strengths $[sg^l_i, sg^u_i]$ chosen by the attacker to identify the target drone set, $S_{\mathrm{target}, i}$, for the DoS attack. If $S_{\mathrm{target}, i} = \emptyset$, it implies no attack. Thus, $S_{\mathrm{target}, i}$ is given by:
\begin{equation}
S_{\mathrm{target}, i} = \{\kappa \mid sg_i^l \leq sg_{\kappa} \leq sg_i^u\},  
\end{equation}
where $\kappa$ is a drone's ID, and $sg_{\kappa}$ is the signal strength received by the attacker from drone $\kappa$. The resource-constrained attacker can target at most $\zeta$ drones.  We map the attacker's strategies into 10 signal strength ranges in $dBm$ by: 

\vspace{1mm}
\noindent $[sg^l_i, sg^u_i] \in \{$$(-100, -98.1]$, $(-98.1, -96.1]$, $(-96.1, -93.8]$, $(-93.8, -91.1], (-91.1, -87.9], (-87.9, -84.0], (-84.0, -79\\.0]$, $(-79.0$, $-72.0]$, $(-72.0, -60], (-60, 20]\}$. 
\vspace{1mm}

These ranges are formed based on Eq.~\eqref{Eq:dBm-distance}, which evenly groups drones based on the distance between the defender and attacker and converts it to the received signal strength. 

We estimate signal attenuation by~\cite{xue2017improved}:
\begin{equation}
P_{dBm}(d) = P_{dBm}(d_0) - \eta \cdot 10 \cdot \log_{10}(\frac{d}{d_0}),
\label{Eq:dBm-distance}
\end{equation}
where $\eta=4$ is a path loss exponent,
and $P_{dBm}(d)$ and $P_{dBm}(d_0)$ are the observed signal strengths at distance $d$ and $d_0$, respectively. According to the drone's specifications~\cite{bitcrazeDrone}, $P_{dBm}(d_0) = 20$ $dBm$ (decibel-milliwatts) when $d_0 = 1$ $m$ (meter). Since $-60$ $dBm$ and $-100$ $dBm$ are typical values for the strongest and lowest signal strength that a drone uses~\cite{sauter2010gsm}, we consider a signal to be strong when $d < 100$ $m$, and signal weak when $d = 1000$ $m$. Based on real-world test results~\cite{bitcrazeDrone}, the maximum control range of $1000$ $m$ well reflects a real-world scenario.

The distance, $d$, between an attacker ($A$) and a drone ($\kappa$) is calculated based on their coordinates, $(x, y, z)$:
\begin{equation}
d (A, \kappa) = \sqrt{|x^A - x^{\kappa}|^2 + |y^A - y^{\kappa}|^2 + |z^A - z^{\kappa}|^2},
\label{Eq: distance}
\end{equation}

\subsubsection{\bf Hypergame Theoretic Attack Strategy Selection} \label{subsec:hypergame-attack} This section discusses how an attacker selects its best strategy based on the hypergame expected utility (HEU). Our framework builds upon established hypergame theory to provide practical equilibrium solutions suitable for the dynamic UAV environment, where traditional game-theoretic assumptions may not fully apply due to resource constraints and real-time operational requirements.

\paragraph{Attacker's Utility Calculation}
An attacker estimates the HEU as a function of its utility. When the defender takes strategy $j$, the attacker's utility, $u^A_{ij}$, by taking strategy $i$ is:
\begin{gather}
\label{eq: attacker-utility}
    u^A_{ij} = G^A_{ij} - L^A_{ij},  \\
    G^A_{ij} = \mathrm{ai}^A_{ij} + \mathrm{dc}^A_{ij}, \; \;  L^A_{ij} = \mathrm{di}^A_{ij} + \mathrm{ac}^A_{ij},
\end{gather}
where $G^A_{ij}$ and $L^A_{ij}$ are gains and losses of the attacker $G^A_{ij}$ includes attack impact, $\mathrm{ai}^A_{ij}$, and defense cost, $\mathrm{dc}^A_{ij}$. $L^A_{ij}$ is based on defense impact, $\mathrm{di}^A_{ij}$ and attack cost, $\mathrm{ac}^A_{ij}$. 

Attack impact, $\mathrm{ai}^A_{ij}$, is computed by:
\begin{equation}
\mathrm{ai}^A_{ij} = \frac{\sum_{\kappa_j \in S^A_{\mathrm{target}, i} \mathcal{ASR'}_{\kappa_j}  \mathrm{C_{\kappa_j}}}}{\zeta},
\label{eq:attack-impact}
\end{equation}
where $S^A_{\mathrm{target}, i}$ is the target drone set when attacker selects strategy $i$, and $\zeta$ is the attack budget. $\mathcal{ASR'}_{\kappa_j}$ is the expected attack success ratio for drone $\kappa_j$. It is calculated as the ratio of successful DoS attacks to the total number of DoS attacks performed by the attacker up to round $t$. $C_{\kappa_j}$ is the criticality of drone $\kappa_j$. Since the network topology of the drone fleet is unknown to the attacker, it determines a drone's criticality based on the observed signal strength.

Attack cost, $\mathrm{ac}^A_{ij}$, is determined by the size of the target drone set, $S_{\mathrm{target}, i}$, and estimated by:
\begin{equation}
\label{eq:attack-cost}
\mathrm{ac}^A_{ij} =  e^{|S_{\mathrm{target}, i}|-\zeta}.
\end{equation}
Defense impact, $\mathrm{di}^A_{ij}$, is the opposite of the attack impact:
\begin{equation}
\mathrm{di}^A_{ij} = 1 - \mathrm{ai}^A_{ij}.
\label{eq:attacker-perceived-defense-impact}
\end{equation}

The defender takes $j$, the defense cost perceived by the attacker, $\mathrm{dc}^A_{ij}$, is calibrated by the drone's energy consumption and the impact by a compromised drone, and estimated by:
\begin{eqnarray}
\mathrm{dc}^A_{ij} = \frac{j}{\mathrm{sig}_{max}} + \mathrm{ai}_{ij}^A,
\label{eq:attacker-perceived-defense-cost}
\end{eqnarray}
where $j$ is the defense strategy perceived by attacker, and $\mathrm{sig}_{max}$ is the maximum signal (i.e., $10$ in this work).

\paragraph{Attacker's Column-Mixed Strategies (CMSs)} The probabilities of the CMSs are derived from the attacker's experience when each subgame $k$ consists of multiple defense strategies~\cite{vane2000hypergame}:
\begin{gather}
\label{Eq: CMS}
CMS^A_k = [c^A_{k1}, \dots, c^A_{km}], \; \\
\text{where} \; \; \sum_{j=1}^m c^A_{kj} = 1, \; \; c^A_{kj} = \frac{\gamma^A_{kj}}{\sum_{j \in DS_k} \gamma^A_{kj}}. \nonumber 
\end{gather}
Here $c^A_{kj}$ follows the Dirichlet distribution and $m$ is the number of the defender's strategies.  $\gamma^A_{kj}$ is the number of times the defender takes $DS_j$ when the attacker plays subgame $k$. Since the attacker cannot capture drones' true signal strength levels due to attenuation, $\gamma^A_{kj}$ and $\gamma^A_j$ are estimated based on the received signal strength levels.  

\paragraph{Attacker's Belief Contexts} The attacker's belief contexts are modeled by a subgame $k$~\cite{vane2000hypergame}. The attacker's set of probabilities taking each subgame is formulated by:
\begin{eqnarray}
P^A = [P_0^A, P_1^A, P_2^A, P_3^A], \; \; \text{where } \sum_{k=0}^3 P_k^A = 1. 
\label{eq:attack-belief-context}
\end{eqnarray}
Here $P_0^A$ is the probability of taking a full game with all possible observed signals by the attacker, where $P_1^A$, and $P_2^A$, $P_3^A$ are the probabilities of taking a game with signal strengths $(-100, -93.8]$, $(-93.8, -79.0]$, and $(-79.0, 20]$, respectively. Those ranges are designed as discussed in Section~\ref{subsubsec:attacker_action}. A subgame reduces the solution space for efficiency. In hypergame theory, the concept of a subgame (e.g., $P_1^A, P_2^A, P_3^A$) is used to reduce the cost of computing all utilities associated with each strategy. When an attacker is uncertain about the game (i.e., $P_r < g^A$, where $P_r$ is a predefined value, and $g^A$ is the attacker's perceived uncertainty), the attacker takes a full game, $P^A_0$. Otherwise, each subgame's probability, $P_i^A$, is the number of drones using a signal strength in the signal range over the total number of drones observed in a round.  

Across all subgames, the attacker's belief in the defender taking $j$ strategy is estimated by:
\begin{equation}
S_j^A =\sum_{k=0}^3 P_k^A \cdot c_{kj}, \; \text{where } \sum_{j=1}^{m} S^A_j = 1.
\label{Eq:attacker-belief}
\end{equation}
Here $P_k^A$ and $c_{kj}$ are explained in Eq.~\eqref{eq:attack-belief-context} and Eq.~\eqref{Eq: CMS}, respectively. A set of the attacker's belief in the defender taking $m$ strategy is denoted by $C^A_{\Sigma} = \{S^A_1, \dots, S^A_m\}$.

\paragraph{Attacker's Uncertainty} We model an attacker's perceived uncertainty, $g^A$, as a function of the degree of detectability toward given deception (i.e., honey drones) and the amount of successful experience in launched attacks. More specifically, we formulate this using an exponential decay function which shows the reduction in uncertainty as the attacker experiences more attack success and deception detection by: 
\begin{equation}
g^A= e^{-\lambda^A \cdot \mathrm{ad} \cdot N_{\mathrm{AS}}},
\label{eq:attacker-uncertainty}
\end{equation}
where $\lambda^A$ is a parameter for the attacker to control the range of uncertainty,  $\mathrm{ad}$ is the attacker's deception detectability, randomly selected in the range of $[0, 0.5]$, and $N_{\mathrm{AS}}$ is the number of attack successes since the mission begins.  The modeling of the attacker's perceived uncertainty accounts for its increase as the attacker detects whether the defender employs deception and successfully executes attacks. This approach is well-supported by existing literature~\cite{fugate2019artificial, huang2019dynamic}.

\paragraph{Attacker's Hypergame EU (AHEU)} The AHEU by taking strategy $AS_i$ is formulated as:
\begin{gather}
\label{Eq:A-HEU}
HEU (AS_i, g^A, C^A_\Sigma, DS^A_w) \\ = (1-g^A) \cdot EU^A(AS_i, C^A_\Sigma) + g^A \cdot EU^A(AS_i, DS^A_w), \nonumber
\end{gather}
where $g^A$ is given in Eq.~\eqref{eq:attacker-uncertainty}, $EU^A(C_\Sigma)$ is the attacker's expected utility (AEU), calculated by Eq.~\eqref{Eq: EU_C}, and $EU^A(DS_w)$ is the AEU when the defender takes $w$ strategy, producing the lowest utility to the attacker, computed by Eq.~\eqref{Eq: EU_worst}. 

The attacker's EU, $EU^A(AS_i, C_\Sigma)$, is estimated based on the attacker's belief, $S^D_j$, and utility, $u_{ij}^A$, where the attacker and defender take $i$ and $j$ strategies, respectively, and given by:
\begin{equation}
EU^A(AS_i, C^A_\Sigma) = \sum_{j=1}^{m} S^A_j \cdot u^A_{ij},
\label{Eq: EU_C}
\end{equation}
where $C^A_\Sigma$ is a set of the attacker's beliefs toward all defense strategies. $S^A_j$ is estimated in Eq.~\eqref{Eq:attacker-belief} and $u^A_{ij}$ is given in Eq.~\eqref{eq: attacker-utility}.  $EU^A(AS_i, C^A_\Sigma)$ is considered when the attacker is certain about the game. However, when the attacker is uncertain about the game (i.e., $g^A$), it considers the worst situation and uses the following AEU:
\begin{equation}
EU^A(AS_i, DS^A_w) = m \cdot S^A_w \cdot u^A_{iw},
\label{Eq: EU_worst}
\end{equation}
where the defender takes strategy $w$ with the lowest EU to the attacker, and $m$ is the number of defense strategies.

\subsubsection{\bf Attack Strategy Selection in DRL} \label{subsubsec:att-DRL} We employ the DRL strategy selection algorithm for the attacker. Using a single DRL agent, the attacker identifies the optimal attack strategy $AS_i$ that maximizes its accumulated reward, $G^A$, by:
\begin{itemize}
\item \textbf{State} ($\mathcal{S}_t^A$): $\mathcal{S}_t^A = (N_{TR}^t)$, where $N_{TR}^t$ is the number of drones in each signal strength range at round $t$.

\item \textbf{Action Set} ($\mathcal{A}^A$): $\mathcal{A}^A = \{a_1, \dots, a_i, \dots, a_n\}$, where $a_i$ is $AS_i$ that determines the set of target drones, $S_{\mathrm{target}, i}$. Action $i$ the attacker DRL agent takes at round $t$ is represented by $a_i^t$. Each action in the set is aligned with the attacker strategies as discussed in Section \ref{sec:game-framework}.

\item \textbf{Reward Function} ($\mathcal{R}_t^A ({a_i^t})$): $\mathcal{R}_t^A ({a_i^t}) = \mathcal{N}_{MNC}^t$, where $\mathcal{N}_{MNC}^t$ represents the number of mission tasks not completed in round $t$.  The attacker's DRL agent aims to maximize its accumulated reward, $G^A = \sum^\infty_{t=0} {\gamma^A}^t \cdot \mathcal{R}_t^A$, where $\gamma^A$ is the decay factor.
\end{itemize}
These designs are also employed by the DRL agent in our proposed HT-DRL. Specifically, HT-DRL begins by utilizing HT during the initial stages of the learning process and subsequently transitions to a DRL approach for solution optimization, as described in Fig.~\ref{fig: HT-guided_DRL}.

\begin{figure*}
\centering
\includegraphics[width= \textwidth]{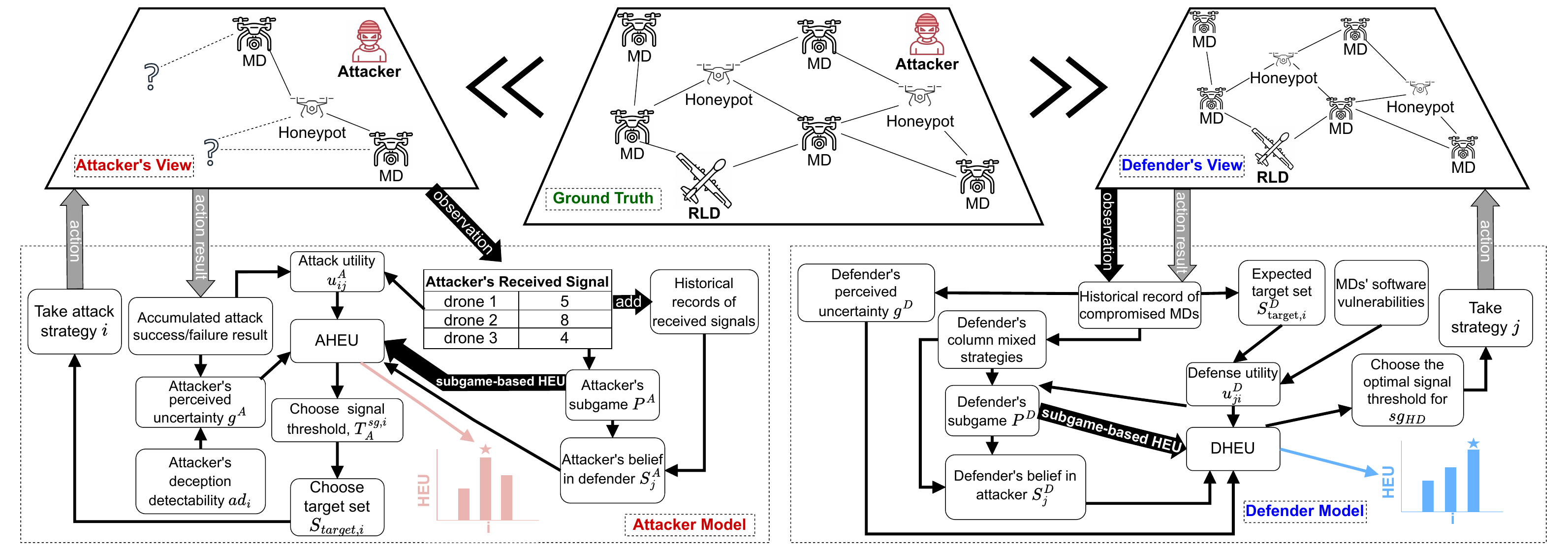}
\caption{Hypergame-Theoretic Strategy Selection Methods by the Attacker and Defender.}
\label{fig: HEU-decision}
\end{figure*}

\subsection{Defender Model}
\label{subsec: defender-model}

\subsubsection{\bf Defender's Action Space based on Signal Strength} \label{subsubsec: drone-signal-strength}

We use a defense strategy, $DS_j \in \{DS_1, \dots, DS_{10}\}$, to adjust the signal strength of HDs, $sg_{HD}$. MDs' signal strength levels are determined by $sg_{MD} = sg_{HD}-\rho$, where $\rho$ is a predefined integer to give a higher signal strength to HDs (see Table~\ref{tab:parameter}) than MDs. We evenly split the signal transmission range from 100m to 1000m based on Eq.~\eqref{Eq:dBm-distance} and map $sg_{HD} \in \{-20, -7.9, -0.9, 4.0, 7.9, 11.1, 13.8, 16.1, 18.1, 20\}$. We leverage DRL to identify the optimal defense strategy, $sg_{HD}$, for adjusting MDs' and HDs' signal strength levels.

\subsubsection{\bf Hypergame Theoretic Defense Strategy Selection} \label{subsec:hypergame-defense} The defender will estimate its HEU and choose defense strategy ($DS_j$) to determine $sg_{HD}$ and $ sg_{MD} = sg_{HD} - \rho$. 

\paragraph{Defender's Utility Calculation}
The defender's utility by taking strategy $j$ when the attacker takes strategy $i$ is:
\begin{gather}
\label{eq: defender-utility}
u^D_{ji} = G^D_{ji} - L^D_{ji},  \\
G^D_{ji} = \mathrm{di}^D_{ji} + \mathrm{ac}^D_{ji}, \; \;  L^D_{ji} = \mathrm{ai}^D_{ji} + \mathrm{dc}^D_{ji},
\end{gather}
where $G^D_{ji}$ and $L^D_{ji}$ are the defender's gains and losses.

Defense impact, $\mathrm{di}^D_{ji}$, is measured by:
\begin{equation}
\mathrm{di}^D_{ji} = 1 - \frac{\sum_{\kappa \in S^D_{\mathrm{target}, i}} \mathrm{vul}_{\kappa}}{\zeta} + \frac{N'_{connect, j}}{N_{drone}} ,
\label{eq:defense-impact}
\end{equation}
where $\mathrm{vul}_\kappa$ refers to the vulnerability of drone $\kappa$ as a real number in $[0, 1]$, as in Section~\ref{subsec:threat-model}. The number of target drones perceived by the defender in $S^D_{\mathrm{target}, i}$ is based on experience. The defender keeps track of which drones are targeted when the attacker selects strategy $i$. The $\zeta$ is the attack budget. $N'_{connect, j}$ is the expected number of connected drones after selecting defense strategy $j$, and $N_{drone}$ is the total number of drones initially assigned to the mission team. Defense cost, $\mathrm{dc}^D_{ji}$, is:
\begin{equation}
\mathrm{dc}^D_{ji} = e^{j-\mathrm{sig}_{max}},
\label{eq:defense-cost}
\end{equation}
where $j$ is the defense strategy and $\mathrm{sig}_{max}$ refers to the maximum signal (i.e., 10).

The attack impact, $\mathrm{ai}^D_i$, is given by: 
\begin{equation}
\mathrm{ai}^D_{ji} = 1 - \mathrm{di}^D_{ji}
\label{eq:defender-perceived-attack-impact}
\end{equation}
where $\mathrm{vul}_{\kappa_j}$ is used as the probability for drone $\kappa_j$ to be compromised, as discussed in Section~\ref{subsec:threat-model}. 

The attack cost, $\mathrm{ac}^D_{ji}$, is estimated as: 
\begin{equation}
\label{eq:defender-perceived-attack-cost}
\mathrm{ac}^D_{ji} = 
\frac{|S^D_{\mathrm{target}, i}|}{\zeta},
\end{equation}
where $|S^D_{\mathrm{target}, i}|$ is based on the defender's prediction towards attack strategy $i$.

\paragraph{Defender's Column-Mixed Strategies (CMSs)} The probability of CMS when the defender plays $k$-th subgame is obtained by:
\begin{equation}
CMS^D_k = [c^D_{k1}, \dots, c^D_{kn}], \; \text{where} \sum_{i=1}^n c^D_{ki} = 1,
\label{eq:defender-CMS}
\end{equation}
where $c^D_{ki}$ follows the Dirichlet distribution with $c^D_{ki} = \frac{\gamma^D_{ki}}{\sum_{j \in AS_k} \gamma^D_{ki}}$ and $n$ is the count of the attacker's strategies. Here $\gamma^D_{ki}$ is the number of times the attacker takes $AS_i$ based on the defender's observations in subgame $k$ during the observation window from $t=0$ to $t-1$ at round $t$.

\paragraph{Defender's Belief Contexts} The defender's beliefs are formulated by: 
\begin{gather}
P^D = [P_0^D, P_1^D, P_2^D, P_3^D], \; \; \text{where  } \sum_{k=0}^3 P_k^D = 1. \label{eq:defender-belief-contexts}
\end{gather}
$P_0^D$ is the probability of taking a full game with all strategies and $P_1^D$, $P_2^D$, and $P_3^D$ are the probabilities of the defender taking subgame 1, 2, and 3 with $\{1, 2, 3\}$, $\{4, 5, 6\}$, and $\{7, 8, 10\}$, respectively. When the defender is uncertain about the game (i.e., $P_r < g^D$ where $P_r$ is a random number with uniform distribution in $[0, 1]$ and $g^D$ is estimated by Eq.~\eqref{eq:defender-uncertainty}), it takes the full game, $P_0^D$. Otherwise (i.e., $P_r \geq g^D$), it takes a subgame with probability $P_k^D$, estimated by:
\begin{gather}
P_k^D = \sum_{j \in B_k} \sum_{i=1}^n c^D_{ki} \cdot u_{ji}^D,
\label{eq:P_k^D}
\end{gather}
where $B_k$ is the set of strategies available to the defender in subgame $k$ and $i$ is an attack strategy. The defender's belief in the attacker taking $i$ strategy is estimated by:
\begin{equation}
S^D_i =\sum_{k=0}^3 P^D_k \cdot c^D_{ki}, \; \text{where } \sum_{i=0}^{9} S^D_i = 1,
\label{Eq:defender-belief}
\end{equation}
where the attacker has 10 strategies, representing integer signal strength in $[0, 9]$. The defender's belief on attacker's $n$ strategies is denoted by $C^D_{\Sigma} = \{S^D_1, \dots, S^D_n\}$.
\begin{figure*}
\centering
\includegraphics[width= \textwidth]{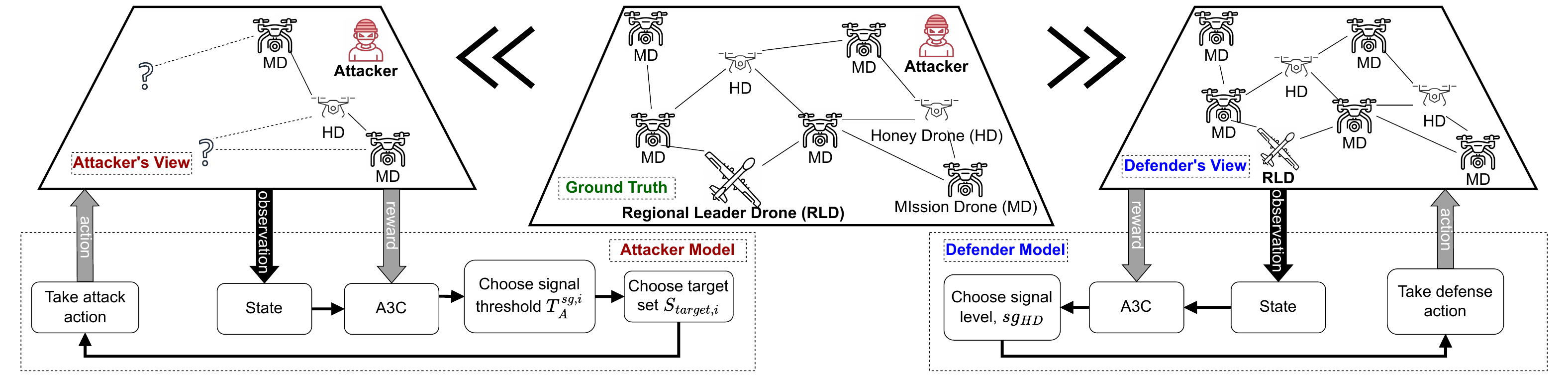}
\caption{\new{Conceptual model of the proposed honey-drone mission system in which both the DoS attacker and the defender employ Deep Reinforcement Learning to select their optimal signal strengths. This figure is adapted from our prior design~\cite{wan2023deception}.}}
\label{fig: DRL-decision}
\vspace{-5mm}
\end{figure*}
\paragraph{Defender's Uncertainty}
The defender's perceived uncertainty is denoted by $g^D$. The defender's uncertainty reduces as more attack alerts are received in honeypots. These alerts give the defender insights into attack patterns. Accordingly, $g^D$ is estimated by:
\begin{equation}
g^D = e^{-\lambda^D \cdot N_{\mathrm{alert}}},
\label{eq:defender-uncertainty}
\end{equation}
where $\lambda^D$ is a parameter for the defender to control the uncertainty and $N_{\mathrm{alert}}$ refers to the number of alerts provided by honeypots about DoS attacks.

\paragraph{Defender's Hypergame Expected Utility (DHEU)} The DHEU of each defense strategy $DS_j$ is calculated by:
\begin{gather}
\label{Eq:D-HEU}
HEU(DS_j, g^D, C^D_\Sigma, AS^D_w) \\ = (1-g^D) \cdot EU^D(DS_j, C^D_\Sigma) + 
    g^D \cdot EU^D(DS_j, AS^D_w), \nonumber 
\end{gather}
where $g^D$ is the uncertainty perceived by the defender and calculated by Eq.~\eqref{eq:defender-uncertainty}. $EU^D(DS_j, C^D_\Sigma)$ is the expected utility of the strategy calculated by Eq.~\eqref{Eq: defender-EU_C}. $EU^D(DS_j, AS^D_w)$ is the expected utility when the attacker selects the strategy that gives the defender the worst result and is estimated by Eq.~\eqref{Eq: defender-EU_worst}.

A defender's EU, $EU^D(DS_j, C^D_\Sigma)$, is estimated based on its beliefs about attacker moves, $S^D_i$, and its utility, $u^D_{ji}$, where the defender and attacker take $j$ and $i$, respectively. The defender's EU is formulated by:
\begin{equation}
EU^D(DS_j, C^D_\Sigma) = \sum_{i=1}^{n} S^D_i \cdot u^D_{ji},
\label{Eq: defender-EU_C}
\end{equation}
where $C^D_\Sigma$ is a set of the defender's beliefs toward attack strategies in all possible subgames, where $S^D_i$ is estimated by Eq.~\eqref{Eq:defender-belief} and $u^D_{ji}$ by Eq.~\eqref{eq: defender-utility}. When the defender is uncertain about the game, it estimates the expected utility under the worst case:
\begin{equation}
EU^D(DS_j, AS^D_w) = n \cdot S^D_w \cdot u^D_{wj},
\label{Eq: defender-EU_worst}
\end{equation}
where $AS^D_w$ is attacker strategy, $w$, providing the lowest expected utility to the defender and $n$ is the total number of attack strategies.

\subsubsection{\bf Defense Strategy Selection in DRL} \label{subsubsec:def-DRL}
The defender's DRL agent optimizes the signal strength of the HDs to maximize its total accumulated reward, $G^D$. This optimization procedure involves the following components:
\begin{itemize}
\item \textbf{State} ($\mathcal{S}_t^D$): The state at round $t$, $\mathcal{S}_t^D$, is a tuple composed of the mission completion ratio and the scan progress map, defined as 
$\mathcal{S}_t^D= (\mathcal{R}_{MC}^t, \mathcal{M}_{SP}^t)$.
Here, $\mathcal{R}_{MC}^t$ represents the ratio of completed mission tasks at round $t$. It is a real number between 0 (indicating no tasks have been completed) and 1 (indicating all tasks have been completed). $\mathcal{M}_{SP}^t$ is a map indicating the scan progress for each cell in the target area at round $t$. Each cell value in this map reflects the level of scanning progress, providing a detailed snapshot of the surveillance status of the target area.

\item \textbf{Action Set} ($\mathcal{A}^D$): The action set is denoted by $\mathcal{A}^D=\{a_1, \cdots, a_j, \cdots, a_m\}$ where each action $a_j$ is a defense strategy $DS_j$ indicating the signal strength of the HDs. For MDs, we follow in Section~\ref{subsubsec: drone-signal-strength}. The action $j$ selected by the defender in round $t$ is denoted by $a_j^t$. Each action in the set is aligned with the defender strategies as discussed in Section \ref{sec:game-framework}.

\item \textbf{Reward Function} ($\mathcal{R}^D_t (a_j^t)$): The immediate reward for the defender upon executing action $a_j^t$ is given by $\mathcal{R}^D_t (a_j^t) = \mathcal{N}_{MC}^t$. In this equation, $\mathcal{N}_{MC}^t$ represents the number of mission tasks completed in round $t$.  The defender's DRL agent will maximize its accumulated reward, $G^D = \sum^\infty_{t=0} {\gamma^D}^t \cdot \mathcal{R}_t^D$, where $\gamma^D$ is the decay factor.
\end{itemize}
These designs are also employed by the DRL agent in our proposed HT-DRL. The procedure of HT-DRL is described in Fig.~\ref{fig: HT-guided_DRL}. Fig.~\ref{fig: HEU-decision} shows how the attacker and defender take their best strategies based on the estimated HEU. Fig.~\ref{fig: DRL-decision} describes the high-level description of how the attacker and defender use their DRL agents to take strategies.

Regarding the transition functions and the simulation environment, our approach employs a model-free DRL methodology, which does not require explicit knowledge of the environment's transition dynamics. Instead, the DRL agents learn optimal policies through continuous interaction with a simulated environment that accurately reflects the operational dynamics of the UAV fleet and the adversarial nature of DoS attacks. The simulation environment, detailed in Section IV, encompasses key factors such as drone mobility, signal attenuation, and energy consumption, providing a realistic setting for the agents to explore and adapt their strategies. By leveraging a model-free approach, we ensure that the agents can effectively learn and optimize their strategies in response to evolving attack patterns without predefined transition models. This design choice reinforces the practicality and adaptability of our proposed HT-DRL framework, demonstrating its capability to intelligently deceive attackers and maintain mission integrity in dynamic and uncertain environments.

\begin{table}[t]
\centering
\caption{\sc \centering Key Design Parameters and Default Values}
\vspace{-2mm}
\small 
\begin{tabular}{|P{0.9cm}|p{5cm}|P{1.5cm}|}
\hline
Symbol & \multicolumn{1}{c}{Meaning} & Default \\
\hline
\noline
$\rho$ & Signal strength decrement interval & 5 \\
\noline
$T_{C}$ & Time duration of battery being charged & 30 \\ 
\noline
$T_M^{\mathrm{max}}$ & Maximum mission duration & 150 \\
\noline
$[\tau_{l}, \tau_{u}]$ & The maximum number of MDs that an HD can protect simultaneously & $[2, 4]$ \\
\noline
$E_P$ & Energy consumption rate by a drone & 7,900 mW \\
\noline
$E_C$ &Energy consumption rate by a camera & 4 mW \\
\noline
$\zeta$ & Maximum number of targeted drones by the attacker in a single round & 5 \\
\hline
\end{tabular}
\label{tab:parameter}
\vspace{-7mm}
\end{table}

\begin{figure*}[t]
\centering
\subfloat{\includegraphics[height=0.04\textwidth]{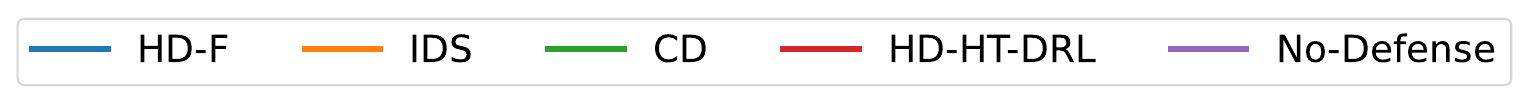}}
\hfil

\vspace{-3mm}
\setcounter{subfigure}{0}
\subfloat[$\mathcal{R}_{MC}$ under fixed attack.]{\includegraphics[width=0.25\textwidth]{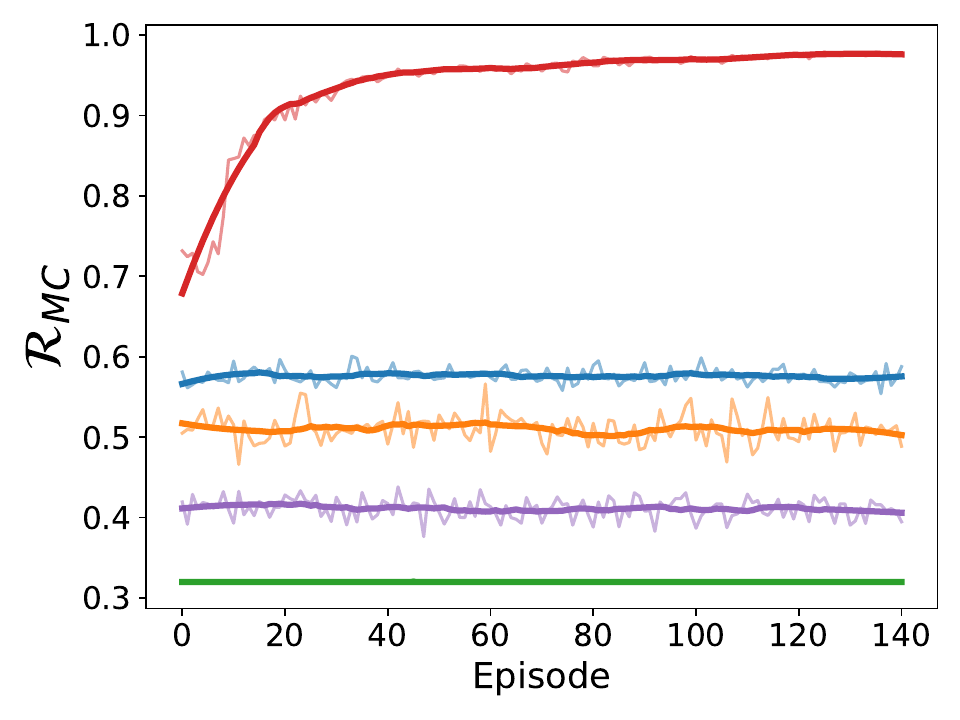}}
\hfil
\subfloat[$\mathcal{R}_{MC}$ under HT attack.]{\includegraphics[width=0.25\textwidth]{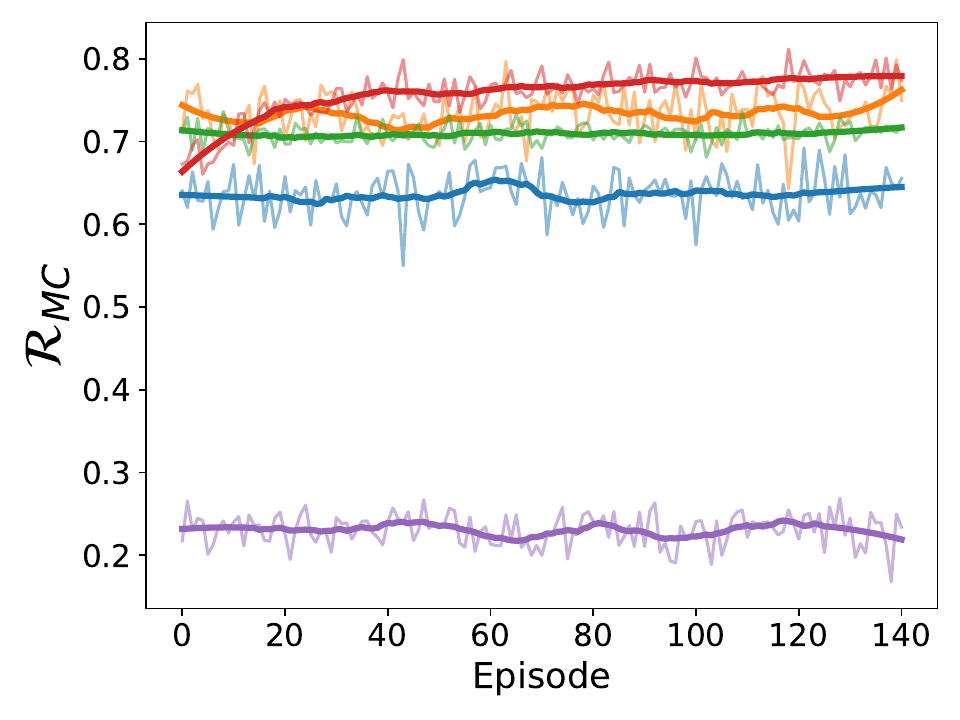}}
\hfil
\subfloat[$\mathcal{R}_{MC}$ under DRL attack.]{\includegraphics[width=0.25\textwidth]{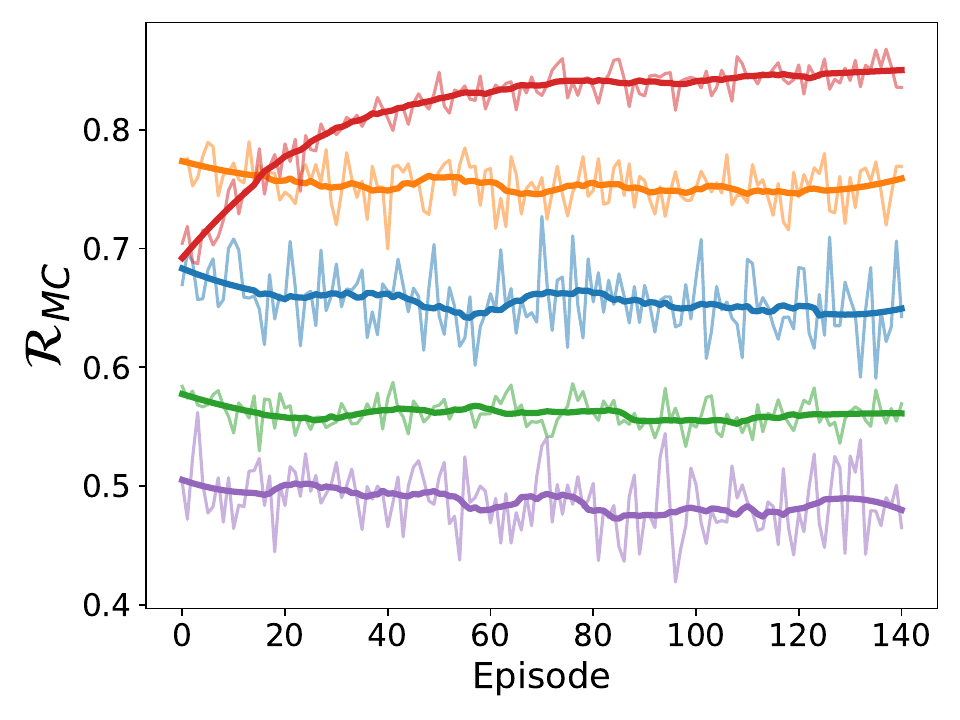}}
\hfil
\subfloat[$\mathcal{R}_{MC}$ under HT-DRL attack.]{\includegraphics[width=0.25\textwidth]{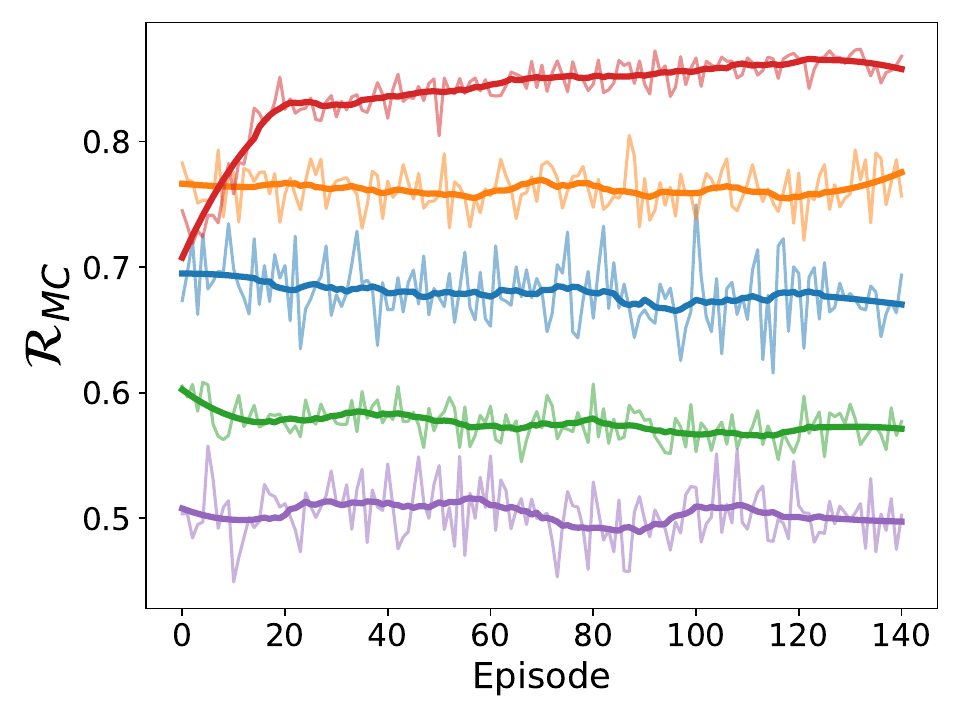}}
\hfil
\caption{Performance comparison of the considered defense approaches with respect to the ratio of completed mission tasks ($\mathcal{R}_{MC}$), given an attack strategy.}
\label{fig: Performance analysis-compare existing: Ratio of Completed Mission Tasks}
\end{figure*}

\begin{figure*}[t]
\centering
\subfloat{\includegraphics[height=0.04\textwidth]{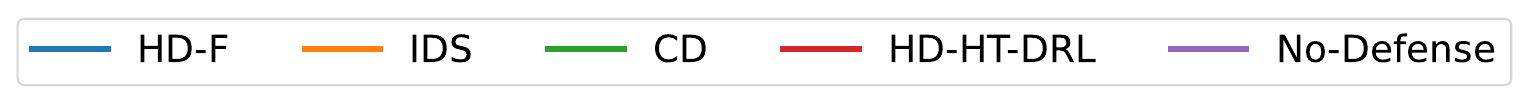}}
\hfil

\vspace{-3mm}
\setcounter{subfigure}{0}
\subfloat[$\mathcal{EC}$ under fixed attack.]{\includegraphics[width=0.24\textwidth]{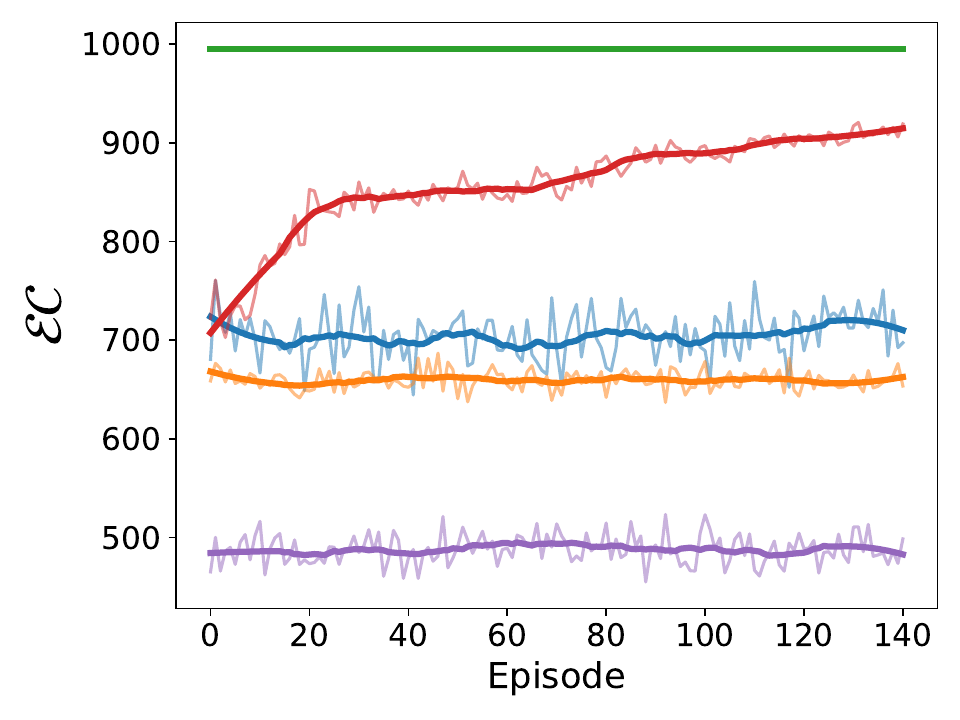} \label{fig: Attacker fixed-Drones' Energy Consumption-CompareExist}}
\hfil
\subfloat[$\mathcal{EC}$ under HT attack.]{\includegraphics[width=0.24\textwidth]{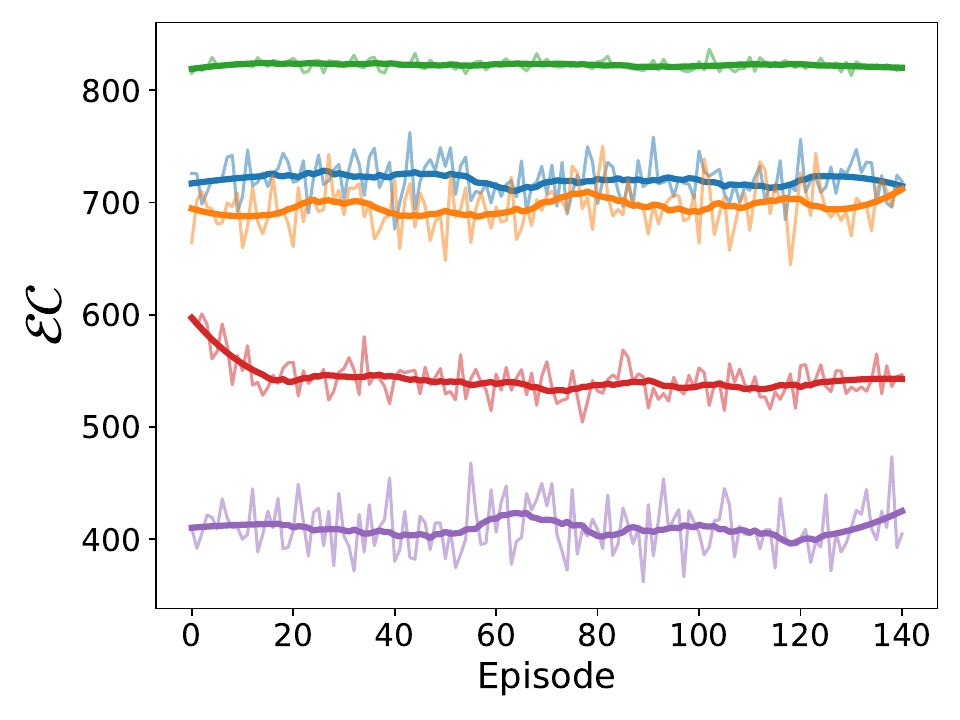}}
\hfil
\subfloat[$\mathcal{EC}$ under DRL attack.]{\includegraphics[width=0.24\textwidth]{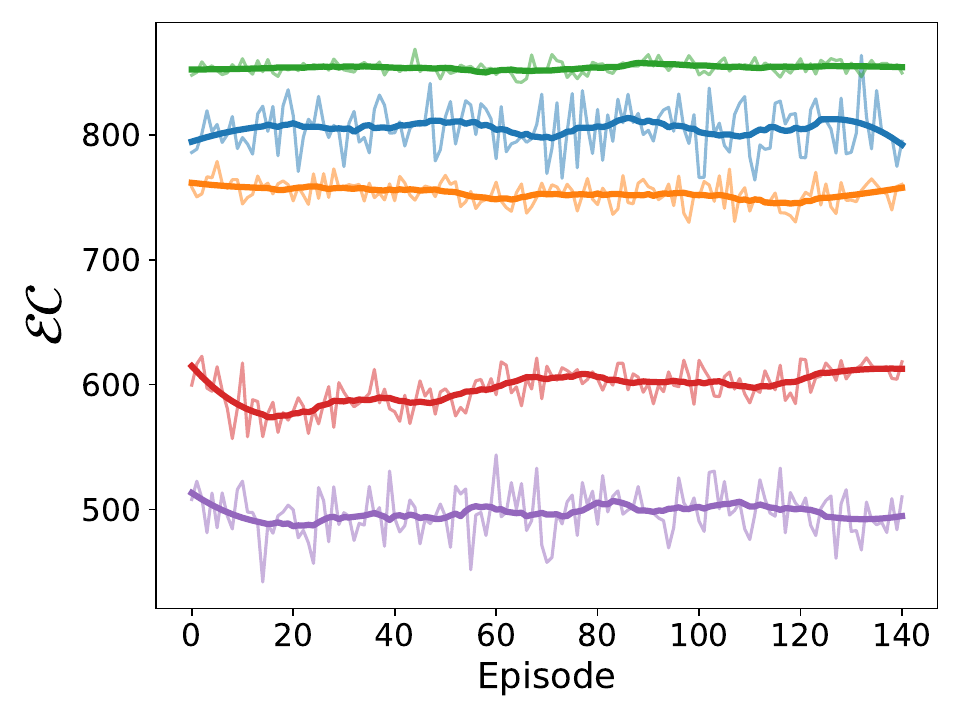}}
\hfil
\subfloat[$\mathcal{EC}$ under HT-DRL attack.]{\includegraphics[width=0.24\textwidth]{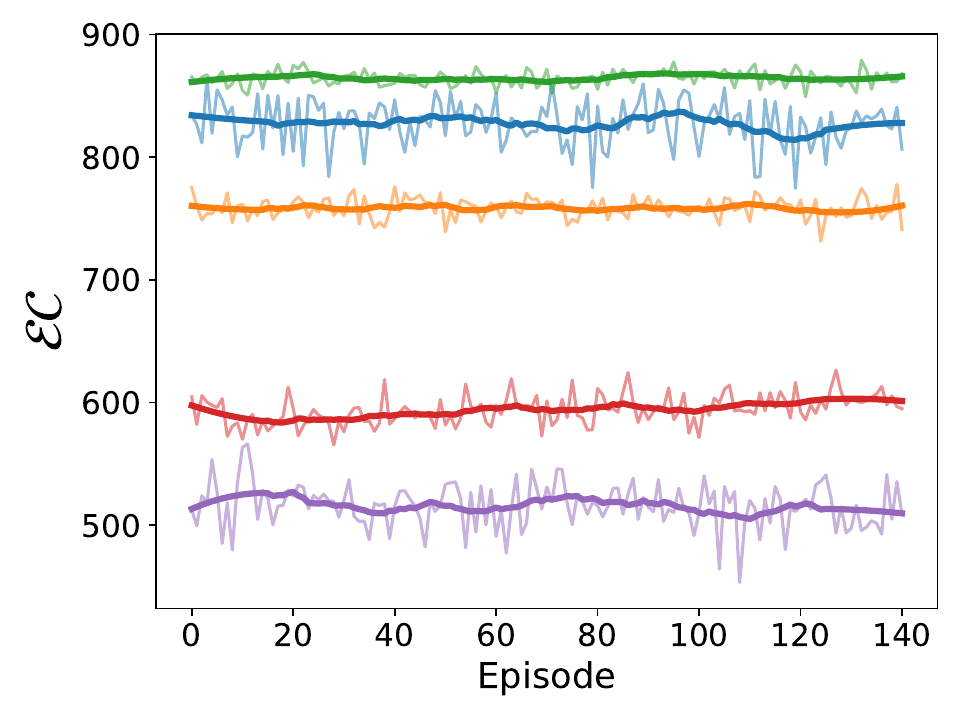}}
\hfil
\caption{Performance analysis of HD-HT-DRL, IDS, CD, and fixed defense, given an attack strategy with respect to energy consumption ($\mathcal{EC}$).} \label{fig: Performance analysis-compare existing: Drones' Energy Consumption}
\vspace{-5mm}
\end{figure*}

\section{Experimental Setup} \label{sec:experimental-setup}
\subsection{Simulation Environment Setup} \label{subsec:simulation-environment-setup}
We evaluated the efficacy of our proposed mission system through extensive experiments in a simulated environment based on PyTorch deep learning library
and the NetworkX graph library.
Our simulation was executed on a high-performance computing cluster with 128-thread AMD EPYC 7702 CPUs.
Our simulation environment reflects operational UAV dynamics, including realistic mobility, signal attenuation, and energy usage patterns derived from actual hardware specifications. This approach ensures that our simulation outcomes closely approximate those expected in live deployments.
Our experiments utilized the A2C algorithm.
This synchronous approach maintains an ideal balance between efficiency and stability, unlike its asynchronous variant, A3C, which often introduces unnecessary complexity and unpredictability through multiple parallel workers.

We allocated a 750m $\times$ 750m target area, subdivided into 25 cells of 150m $\times$ 150m each. The GCS and CS were stationed outside the target area. The drone fleet was composed of 15 MDs and 5 HDs.  A drone was classified as having crashed if its altitude dropped below 0.1 meters. Several factors can trigger such a crash: navigation errors induced by DoS attacks, inter-drone collisions, loss of balance due to improper operations, and battery depletion.
We generate the trajectory of MDs with OR-Tools library~\cite{orTools}, a Python-based efficient, close-to-optimal path-finding algorithm~\cite{voudouris1996partial}. Using this algorithm, the drone fleet can optimally utilize a subset of available MDs to accomplish a mission if the number of MD exceeds the requirements.

In situations where energy depletion occurred mid-mission, the remaining drones at the GCS were deployed to replace the energy-depleted ones. However, if no drones were available, the mission team had to pause operations and wait for a recharged drone. Additionally, if a drone crashes, RLD will reactivate the path-finding algorithm to assign a new drone from GCS. If no replacement drones were present at the GCS during such an event, the mission performance could be adversely impacted.  We considered an energy consumption model as detailed in Section~\ref{subsec:energy-model}.  


For high experimental validity, we conducted 100 runs for each scenario. The mean values of the relevant results were used for further analysis. We summarize key design parameters of the simulation environment and their default values based on real drone specifications and behaviors in Table I. We adopted the Advantage Actor-Critic (A2C) algorithm for the deep learning agent, a synchronous approach that effectively balances efficiency and stability. We set the reward decay factor ($\gamma$) at $0.99$ and initialized the learning rate at $0.0005$ with a subsequent decay factor of $0.9$. The neural networks employed LeakyReLU activation functions for all hidden layers, promoting non-linearity and mitigating the vanishing gradient problem. Network parameters were initialized using the Xavier normal distribution to maintain appropriate variance across layers, facilitating faster convergence. Further, we incorporated layer normalization and a dropout rate of 0.01 for each layer to enhance training stability and prevent overfitting. The policy and value networks of the A2C are structured as multilayer perceptrons with internal layers arranged as (128, 128, 64, 32). To ensure stable training, we implemented gradient clipping, which helps prevent excessively large updates that could destabilize the learning process. We did not employ a target network during training, as the synchronous nature of A2C obviates the need for one. We used a memory buffer size of $10,000$ and a mini-batch size of $32$, applying prioritized experience replay~\cite{schaul2015prioritized} to efficiently learn from key experiences. Since the DRL agent cannot be trained until the memory is full, all the results are generated after the memory population. 

While our HT-DRL framework is method-agnostic and can be integrated with other policy gradient methods such as PPO or SAC, in this work, we employ A2C due to its demonstrated stability and convergence in our simulation environment. The key novelty of our approach lies in the hypergame-theoretic guidance mechanism, which mitigates the cold-start problem independently of the underlying DRL backbone. Future work will explore comparative evaluations of HT-guided PPO, SAC, and related algorithms.

We made the source code available at~\cite{wan2023sourcecode}.

\begin{figure*}[t]
\centering
\subfloat{\includegraphics[height=0.04\textwidth]{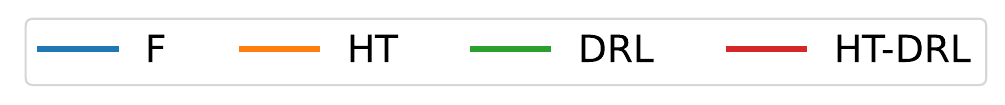}}
\hfil

\vspace{-3mm}
\setcounter{subfigure}{0}
\subfloat[$\mathcal{R}_{MC}$ under fixed attack. ]{\includegraphics[width=0.25\textwidth]{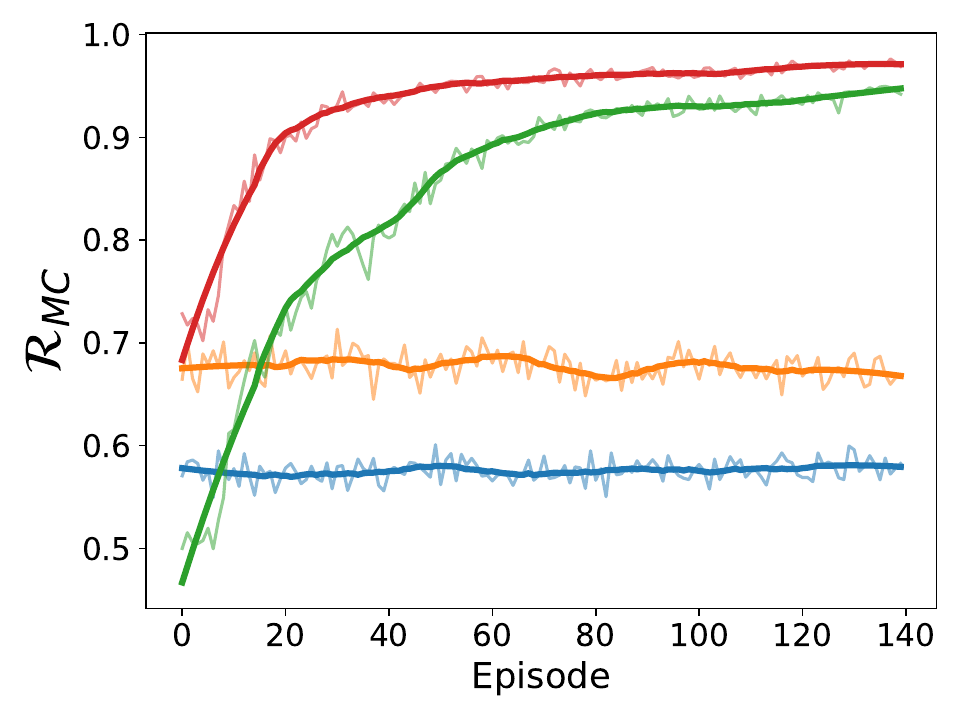}\label{fig: Attacker Fixed-Ratio of Completed Mission Tasks}}
\hfil
\subfloat[$\mathcal{R}_{MC}$ under HT attack.]{\includegraphics[width=0.25\textwidth]{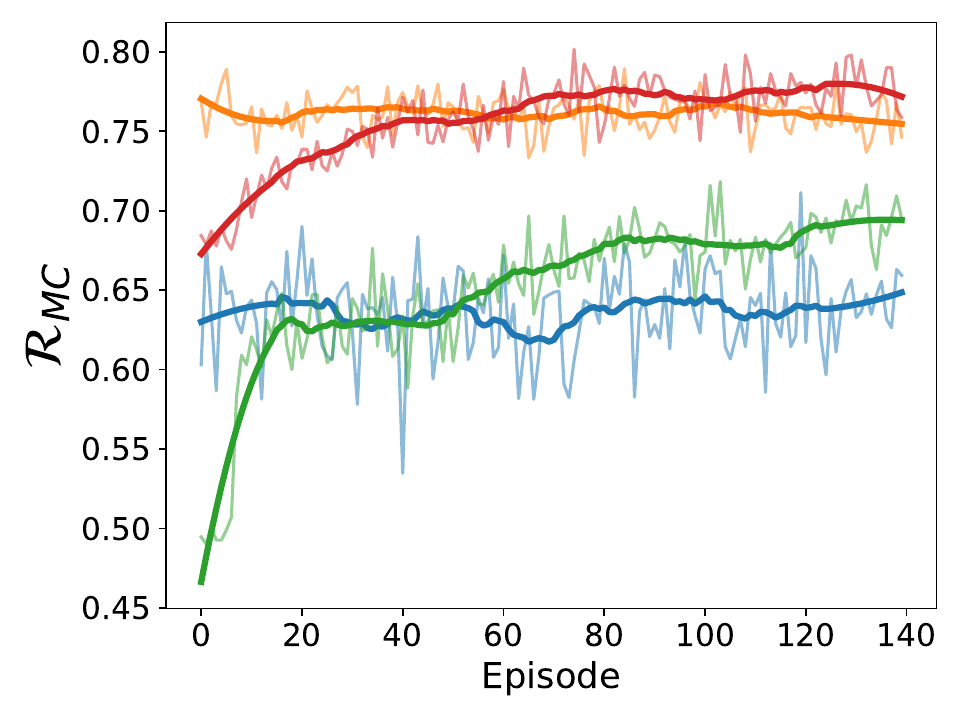}\label{fig: Attacker HT-Ratio of Completed Mission Tasks}}
\hfil
\subfloat[$\mathcal{R}_{MC}$ under DRL attack.]{\includegraphics[width=0.25\textwidth]{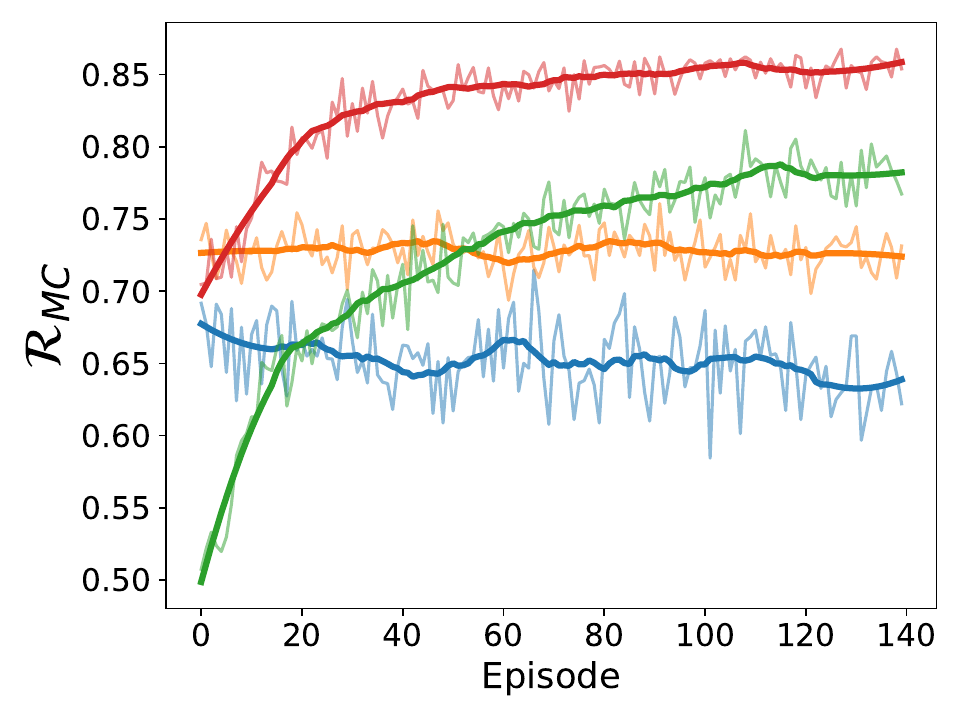}\label{fig: Attacker DRL-Ratio of Completed Mission Tasks}}
\hfil
\subfloat[$\mathcal{R}_{MC}$ under HT-DRL attack.]{\includegraphics[width=0.25\textwidth]{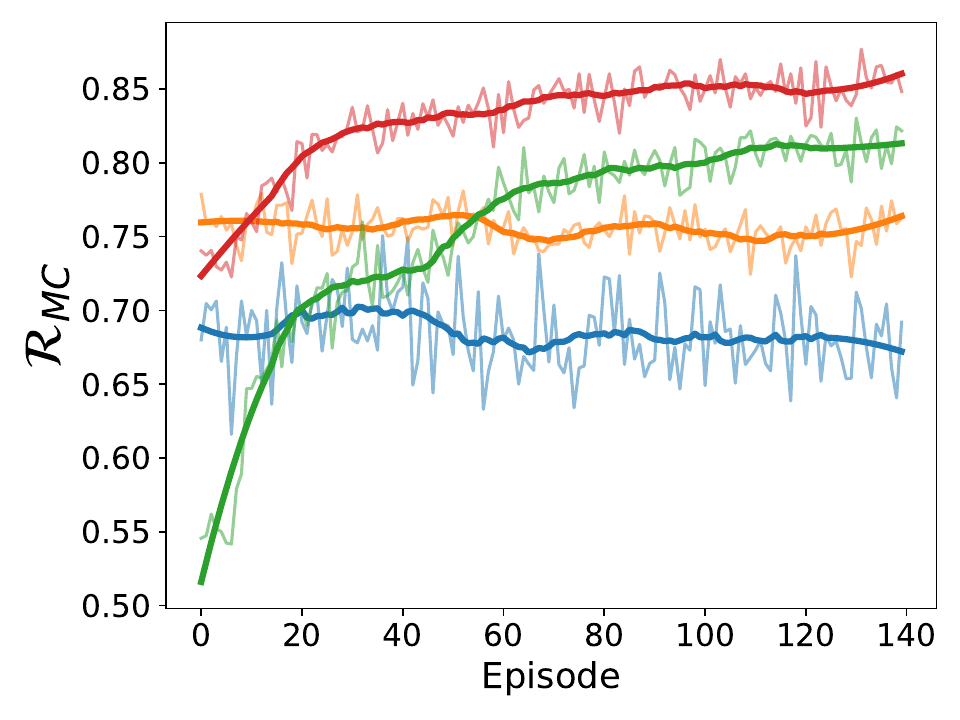}\label{fig: Attacker HT-DRL-Ratio of Completed Mission Tasks}}
\hfil
\caption{Performance analysis of honey drone defense strategies with respect to the ratio of completed mission tasks ($\mathcal{R}_{MC}$), given an attack strategy.}
\label{fig: Performance analysis: Ratio of Completed Mission Tasks}
\vspace{-3mm}
\end{figure*}

\begin{figure*}[t]
\centering
\subfloat{\includegraphics[height=0.04\textwidth]{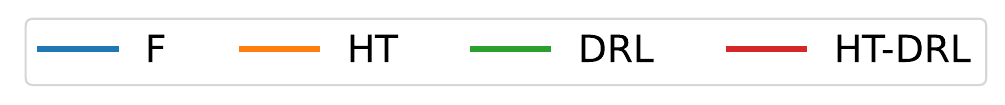}}
\hfil

\vspace{-3mm}
\setcounter{subfigure}{0}
\subfloat[$\mathcal{EC}$ under fixed attack.]{\includegraphics[width=0.25\textwidth]{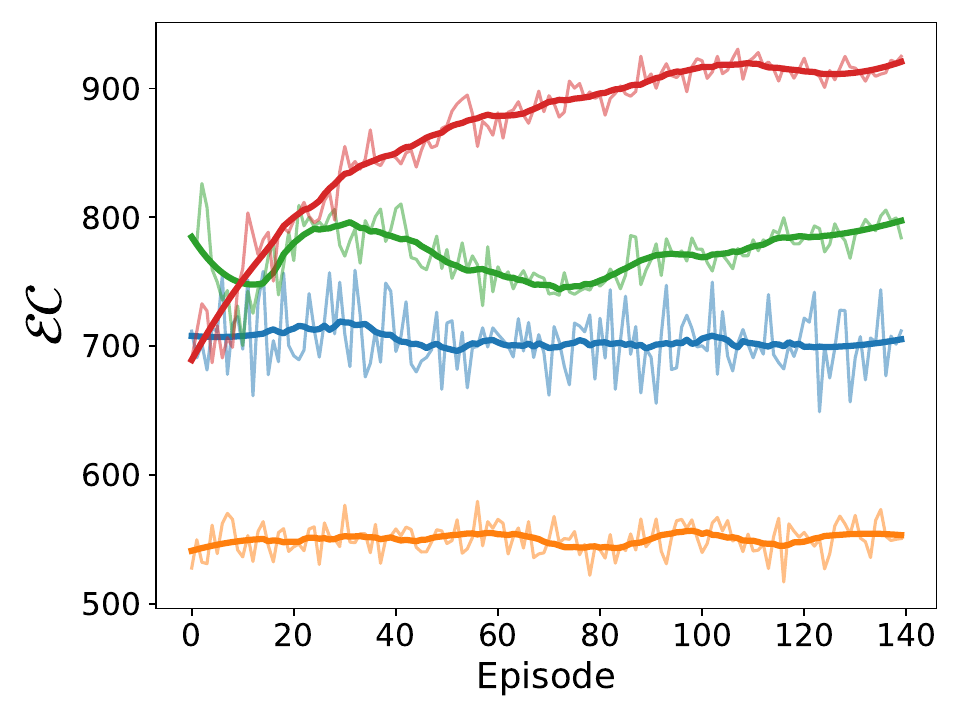}\label{fig: Attacker Fixed-Drones' Energy Consumption}}
\hfil
\subfloat[$\mathcal{EC}$ under HT attack.]{\includegraphics[width=0.25\textwidth]{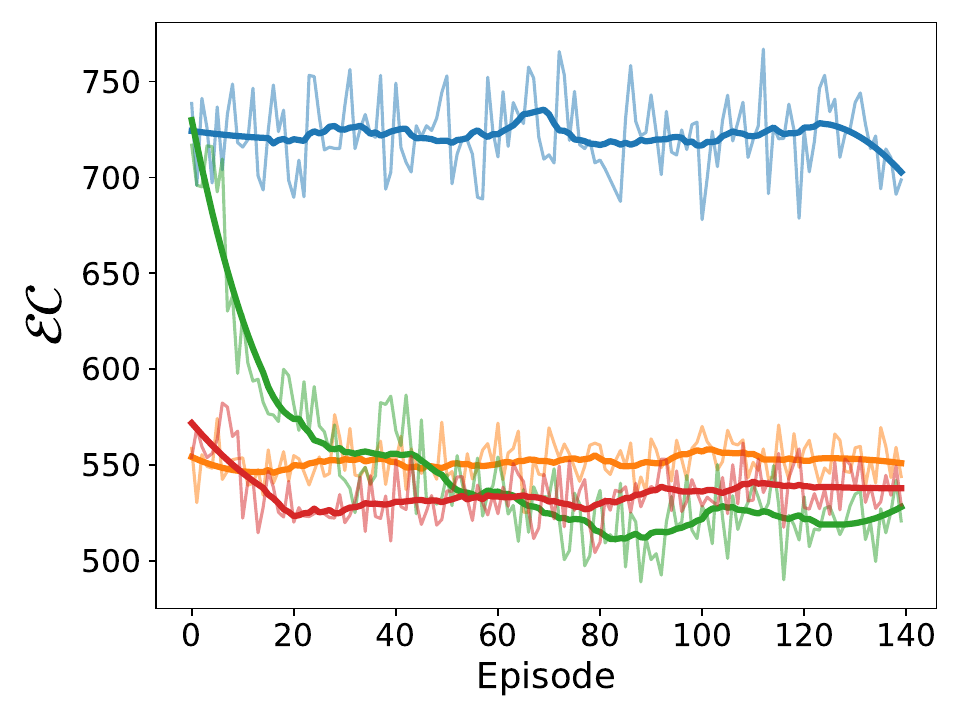}\label{fig: Attacker HT-Drones' Energy Consumption}}
\hfil
\subfloat[$\mathcal{EC}$ under DRL attack.]{\includegraphics[width=0.25\textwidth]{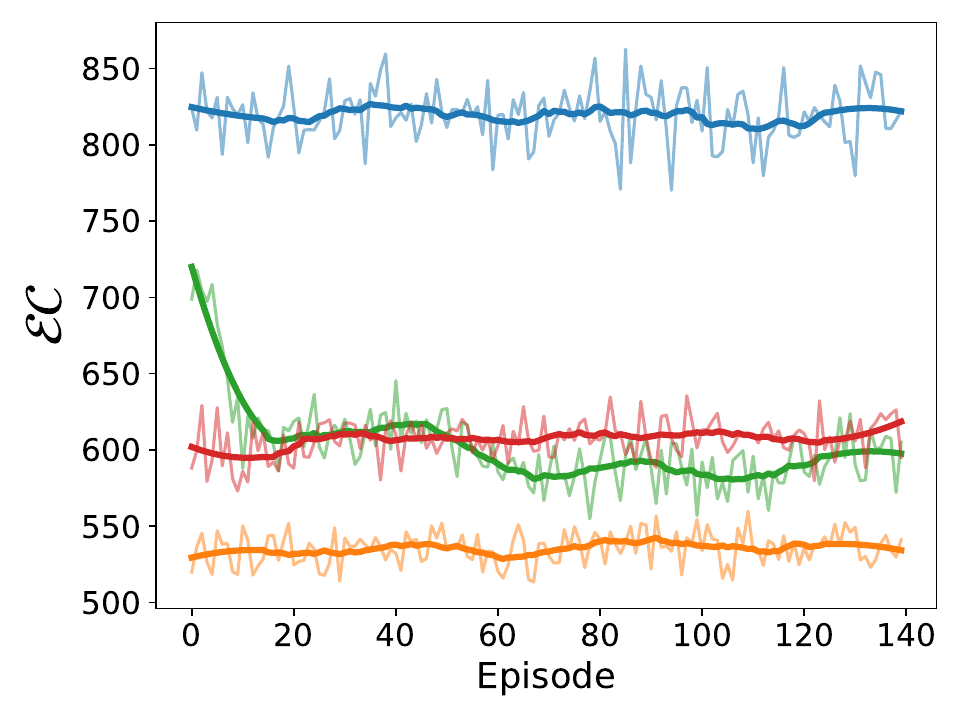}\label{fig: Attacker DRL-Drones' Energy Consumption}}
\hfil
\subfloat[$\mathcal{EC}$ under HT-DRL attack.]{\includegraphics[width=0.25\textwidth]{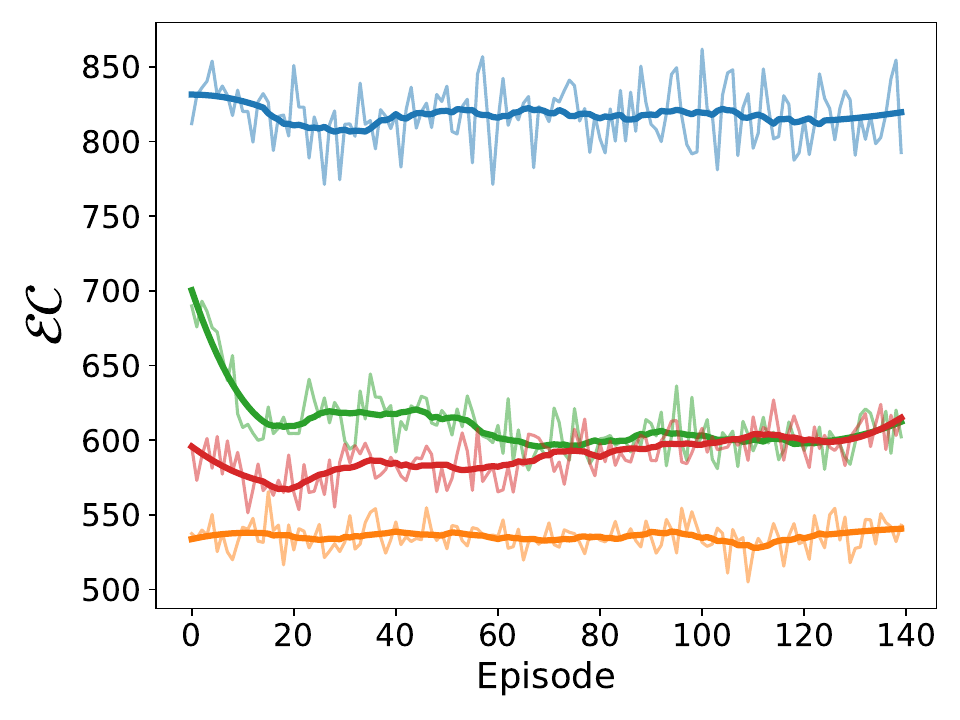}\label{fig: Attacker HT-DRL-Drones' Energy Consumption}}
\hfil
\caption{Performance analysis of honey drone defense strategies with respect to energy consumption ($\mathcal{EC}$), given an attack strategy.} \label{fig: Performance analysis: Drones' Energy Consumption}
\vspace{-5mm}
\end{figure*}

\subsection{Metrics} \label{subsec:metrics}
We use the following metrics to evaluate the proposed approach and the existing counterparts:
\begin{itemize}
\item {\bf Ratio of Completed Mission Tasks ($\mathcal{R}_{MC}$)} is the ratio of cells completed to all assigned cells. 
\item {\bf Energy Consumption ($\mathcal{EC}$)} calculates the total energy consumed by all drones, including HDs and MDs.
\item {\bf Number of Active, Connected Drones ($\mathcal{N}_{AC}$)} is the number of non-compromised drones in the network.
\item {\bf Accumulated Reward ($G^A$ or $G^D$)} by the attacker or defender are given in Sections~\ref{subsubsec:att-DRL} and~\ref{subsubsec:def-DRL}, respectively. 
\end{itemize}
Due to space limitations, we show the experimental results using $\mathcal{N}_{AC}$ and $\mathcal{G}^A$/$\mathcal{G}^D$ in the supplement document.

\subsection{Comparing Schemes} \label{subsec:comparing-schemes}

For the comparative comparison of the proposed HT-DRL with the state-of-the-art (SOTA) non-HD counterparts in Section~\ref{subsec:different-defenses-analysis}, we consider two HD-based approaches, including HD-F using a fixed signal strength level and HD-HT-DRL using dynamic signal strength levels identified by HT-DRL.  We consider two SOTA non-HD-based approaches, using an intrusion detection system (IDS)~\cite{condomines2019network} and ContainerDrone (CD)~\cite{chen2019container} technique, along with a baseline mode with No Defense in handling DoS attacks. We also conduct the ablation study in our proposed HT-DRL by investigating the effect of using the fixed signal strength level, HT, and DRL in Section~\ref{subsec:hd-analysis}. 

\section{Numerical Results and Analyses} \label{sec:resuts-analyses}

\new{Due to space constraints, additional ablation studies, sensitivity analyses, and supplementary metric evaluations are provided in the supplemental material.}

\subsection{Comparative Performance Analysis with Non-HD Defenses}
\label{subsec:different-defenses-analysis}

We compare HD defenses (HD-F and HD-HT-DRL) with other defenses (i.e., IDS, CD), and no defense with a fixed signal under a given attack strategy.

\subsubsection{\bf Ratio of Mission Completion} Fig.~\ref{fig: Performance analysis-compare existing: Ratio of Completed Mission Tasks} shows the performance of different defenses against DoS attacks in terms of $\mathcal{R}_{MC}$.
(1) HD-HT-DRL outperforms all. HD-F performs well only using HDs. This shows the effectiveness of defensive deception. The trends are well aligned with $\mathcal{N}_{AC}$ results shown in Fig. 1 in Appendix A of the supplement document.  (2) The effect of the HT attack is stronger than other attacks based on its lower $\mathcal{R}_{MC}$. This shows that the strategic HT attack has a more adverse impact on mission performance.


\subsubsection{\bf Energy Consumption} Fig.~\ref{fig: Performance analysis-compare existing: Drones' Energy Consumption} demonstrates the performance of different types of defenses against DoS attacks in terms of $\mathcal{EC}$. The key observations are: (1) The drone fleet protected by CD had the highest $\mathcal{EC}$, meaning the CD can protect drones but lowers mission performance by stopping its operation. Such `zombie drones' hover on the target area for the maximum mission duration but do not perform mission operations. (2) As seen in Fig.~\ref{fig: Performance analysis: Drones' Energy Consumption}, except Fig.~\ref{fig: Attacker fixed-Drones' Energy Consumption-CompareExist}, HD-HT-DRL consumes less energy than other schemes for the same reason. More active, connected drones lead to higher $\mathcal{R}_{MC}$ and naturally consume more energy. However, HD-HT-DRL consumes much less than CD and HD-F, despite its high performance in $\mathcal{R}_{MC}$.  We also show the results using the metrics of accumulated reward and connected drones in Appendix A of the supplement document.

\subsection{Performance Analysis of Honey Drone Defenses} \label{subsec:hd-analysis}
In this section, we analyze the performance of HD-based defenses under various attacks in terms of the ratio of mission completion ($\mathcal{R}_{MC}$) and energy consumption ($\mathcal{EC}$). 

\subsubsection{\bf Ratio of Mission Completion} \label{subsubsec: Ratio of Mission Completion}
Fig.~\ref{fig: Performance analysis: Ratio of Completed Mission Tasks} compares the performance of HD defenses, in terms of $\mathcal{R}_{MC}$, given an attack strategy. The key observations are: (1) Overall we observe the outperformance of HT-DRL over other HD defenses because the DRL agent in HT-DRL can explore solutions from its local optima identified by hypergame-theoretic, strategic decision-making without random exploration. We explain this process in Fig. 12 of the Appendix, demonstrating the agent using hypergame theory and later using RL to explore optimal solutions further. HT-DRL also leverages the benefit of using A2C as a DRL algorithm in exploring global optimal solutions adaptable to complex, non-stationary environments.  Fig.~13 in the Appendix shows the distinct probability distributions in selecting strategies for the trained HT-DRL and DRL models. We can consistently observe the similar trends in Fig.~\ref{fig: Performance analysis: Ratio of Completed Mission Tasks} and sensitivity analyses shown in Fig. 7 of the Appendix.

\subsubsection{\bf Energy Consumption} Fig.~\ref{fig: Performance analysis: Drones' Energy Consumption} compares the performance of diverse HD-based defenses, in $\mathcal{EC}$, given an attack strategy. We observe: (1) Except Fig.~\ref{fig: Attacker Fixed-Drones' Energy Consumption}, HT-DRL shows acceptable energy consumption compared to its highest $\mathcal{R}_{MC}$. Energy consumption relates to how many drones are active in the given network. Hence, more active, connected drones ($\mathcal{N}_{AC}$; see Appendix B of the supplement document 
) 
are aligned with higher $\mathcal{R}_{MC}$, resulting in higher $\mathcal{EC}$. However, despite high performance in $\mathcal{N}_{AC}$ and $\mathcal{R}_{MC}$, HT-DRL's energy consumption is relatively low, meaning that HT-DRL considers energy efficiency in selecting its best strategy.  (2) In Fig.~\ref{fig: Attacker Fixed-Drones' Energy Consumption}, the highest $\mathcal{EC}$ is HT-DRL, showing much higher $\mathcal{R}_{MC}$ (i.e., over 95\%) under fixed attack than under other attacks in Figs.~\ref{fig: Attacker HT-Drones' Energy Consumption}--\ref{fig: Attacker HT-DRL-Drones' Energy Consumption}.


\section{Related Work} \label{sec:related-work}

\subsection{Defenses Against DoS Attacks in UAVs}

\citet{chen2019container} discussed a defense mechanism to deal with DoS attacks in real-time UAV systems using containers, called \textit{ContainerDrone} (CD), to prevent excessive resource usage for protecting critical resources (CPU, memory, and communication channel) from DoS attacks. Hence, if a certain level of resource utilization is reached, it stops providing services.  
\citet{sedjelmaci2017hierarchical} proposed a hierarchical detection and response system to enhance security against DoS attacks, including attacks on the GPS module, in UAV networks. 
\citet{ouiazzane2022multiagent} proposed an IDS that utilizes the decision tree model to protect the multi-agent UAV system from DoS attacks.
\citet{gudla2018defense} proposed a moving target defense (MTD) technique to protect the Parrot AR drone from DoS attacks by shuffling communication channels.
\citet{feng2020application} proposed a reinforcement learning approach that uses a multi-objective reward function to adaptively defend against application-layer DDoS attacks by balancing between minimizing false positives during low system load and maximizing attack mitigation during high load.
\citet{li2023transfer} proposed a transfer double deep Q-network-based DDoS detection method for the Internet of Vehicles that leverages traffic flow similarity between base stations to speed up DRL training for newly added base stations.

\textbf{Limitations}: Although these works offer valuable defenses against DoS attacks in UAV systems, they may be limited when facing intelligent attackers empowered by machine learning or deep learning. For instance, the Moving Target Defense (MTD) technique \cite{gudla2018defense} is susceptible to pattern recognition algorithms, potentially allowing attackers to anticipate communication channel shuffling.  While \citet{feng2020application} showed promising DDoS defense using reinforcement learning, their work was limited to traditional server environments. Further, it did not consider the unique challenges of UAV systems, such as energy constraints, mobility, and dynamic network topology changes.  While \citet{li2023transfer} addressed the long training time issue of DRL through transfer learning, it still relies on collecting historical traffic data from existing base stations to calculate similarities, which may not be feasible in mission-critical UAV systems where immediate defense is needed and historical data collection opportunities are limited. Additionally, their approach focuses specifically on DDoS detection rather than comprehensive defensive deception strategies.


\subsection{Game Theory-based Defensive Deception}
\citet{xiao2018attacker} examined the framing effect of an advanced persistent threat (APT) attacker on its detection. They employed {\em Cumulative Prospect Theory} to consider players' subjective, uncertainty-aware decision-making for obtaining Nash Equilibrium (NE) solutions. \citet{basak2019identifying} deployed honeypots using a multistage Stackelberg game and evaluated the accuracy of detecting the attackers collected by the honeypots.  \citet{pibil2012game} used honeypots as a defensive deception technique to lure an attacker aiming to choose real servers. They formulated this as a zero-sum game with imperfect and incomplete information.

Recently, Hypergame Theory (HT) has been used to develop defensive deception techniques.  \citet{cho2019modeling} developed a cyber deception hypergame (CDHG) using HT to examine the best strategy selection by an attacker and defender by developing the Stochastic Petri Nets models with the CDHG.  \citet{wan2021foureye} used HT to model the attack-defense interactions where an attacker and defender identify their best strategy using hypergame expected utility (HEU). Further, they investigated the effect of multiple attacks when the defender can choose a set of defense strategies based on HT~\cite{wan2023resisting}.  \citet{anwar2022honeypot} considered HT to develop intelligent honeypot strategies using either low-interaction or high-interaction honeypots for efficient and effective defensive deception.

\textbf{Limitations:} However, the existing game-theoretic approaches~\cite{xiao2018attacker, basak2019identifying, pibil2012game} have not considered the attacker's and defender's perceived uncertainty in estimating their utilities for selecting the best strategies. In addition, the works using hypergame theory (HT)~\cite{cho2019modeling, wan2021foureye, wan2023resisting, anwar2022honeypot} has mainly applied HT in static or low-mobility settings, such as enterprise or mobile cloud networks, which do not have the concerns of limited resources such as energy or wireless bandwidths. Further, they have not integrated game theory with DRL which can enhance the solution quality. 

\subsection{DRL-based Defensive Deception (DD)} \label{subsec:drl-dd-related-work}

DRL has also been used to develop defensive deception strategies.  \citet{li2022defensive} used DRL to generate an optimal DD strategy. They formulated a utility function to model the underlying threats related to common vulnerabilities in the virtual machine to deal with reconnaissance attacks.  \citet{li2022optimal} proposed a DRL-based DD strategy selection approach to generate optimal placement strategies for the decoys and deceptive routing.  \citet{huang2022reinforcement} examined the vulnerability of systems where RL is used to choose the best defense strategy among MTD, DD, and assistive human security technologies that aim to mitigate human-related vulnerabilities. Similarly, \citet{huang2019deceptive} investigated RL's vulnerability under malicious falsification of cost signals based on the relationship between the falsified cost and the Q-factors. \citet{charpentier2022deep} used DRL (e.g., Deep Q-Network or DQN) to identify the best defense strategy for four types of MTD and DD strategies. \citet{olowononi2022deep} studied how to deploy Intelligent Reflective Surfaces (IRS) in UAVs to strengthen wireless communications for battlefield contexts. They developed data-driven power allocation in communication channels using RL for obfuscating the attack surface by luring jammers into designated channels to mitigate DoS attacks.

\textbf{Limitations}: The existing DRL-based defensive deception methods~\cite{li2022defensive, li2022optimal, huang2022reinforcement, huang2019deceptive, charpentier2022deep, olowononi2022deep} often requires extensive training and data to demonstrate its outperformance. However, they did not try to reduce the training time as the long training time is not allowed in resource-constrained environments.


\subsection{DRL Integrated with Game Theory}

\citet{zheng2022stackelberg} presented a game-theoretic interpretation of actor-critic (AC) algorithms, modeled as a two-player general-sum Stackelberg game, where the actor is the leader, and the critic is the follower.  \citet{cao2022game} proposed a game-theoretic inverse RL to learn the parameters of the dynamic system and individual cost function of multistage games from the demonstrated sequences of system states. 

\textbf{Limitations}: Both studies~\cite{zheng2022stackelberg, cao2022game} did not address the long convergence time to learn in DRL using the integrated approach of DRL and game theory. Furthermore, these existing works did not explore the applicability of their methods in domains like cybersecurity or UAV systems.

\new{In summary, existing studies on hypergame-based cyber defense, DRL-based UAV defense, and HD deployment share three key limitations. First, most DRL-based defenses treat the attacker as part of a stationary environment and do not model asymmetric perception or misperception between attacker and defender. Second, DRL-based defenders generally begin from uninformed random exploration, resulting in slow and unstable learning under highly non-stationary DoS/DDoS attack behaviors. Third, HD deployment strategies and energy-consumption models are frequently simplified or static, limiting their applicability to realistic, energy-constrained multi-UAV surveillance missions.}

\new{Our proposed HT-DRL framework is designed primarily to address the first and second limitations. By constructing a hypergame-based threat model and computing HEU-induced action probabilities, we explicitly account for perception asymmetry between attacker and defender and leverage it to guide the defender’s early action selection. This mechanism mitigates cold-start effects and reduces inefficient exploration in the presence of adaptive DoS/DDoS attackers. In addition, our extended HD deployment and energy-consumption model partially addresses the third limitation by incorporating heterogeneous HD/MD roles and per-UAV energy costs. More detailed hardware-level energy modeling and broader resource constraints remain important directions for future work.}



\section{Conclusion, Limitations, \& Future Work} \label{sec:conclusion}

Denial-of-Service (DoS) attacks on drones have become increasingly prevalent, representing a significant threat to UAV systems in real-world scenarios.
Our defensive deception strategy using honey drones presented a practical solution to countermeasure DoS attacks effectively and efficiently. We integrated hypergame theory with deep reinforcement learning to provide a solution for enhancing the resilience and adaptability of the proposed cyber deception using honey drones, which can be well-suited for complex environments and real-world applications.  In particular, we developed a hybrid approach integrating hypergame theory (HT) with deep reinforcement learning (DRL), namely HT-DRL, to avoid a cold start problem in DRL process of converging to close-to-optimal solutions.  The key idea is how to intelligently select optimal signal strengths to minimize security vulnerabilities to DoS attacks while maximizing mission performance with acceptable energy consumption. 

\subsection{Key Findings}
Via extensive experimental analysis, we obtained the following {\bf key findings}: (1) Honey drone (HD) approaches outperformed all considered counterparts, particularly when attack strategies are highly intelligent by using game theory or DRL or both. Our proposed approach, HT-DRL, showed outperformance in mission performance and energy consumption. (2) Except for the case under a fixed attack strategy (F), HT-DRL (HD-HT-DRL) showed lower or comparable energy consumption while achieving high mission performance. This demonstrated the energy efficiency of HT-DRL. (3) When the attack uses F, the HT-DRL showed the best mission performance, compared to when other attack strategies. This is well explained by the trends observed in the number of active, connected drones (shown in the supplement document), consuming more energy.

\subsection{Limitations}
The presented work has the following {\bf limitations} which can be addressed in our future work. First, a key characteristic of hypergame theory is its capability to deal with an agent's perceived uncertainty. Although the defense system can estimate its own uncertainty, the estimation of the attacker's uncertainty is based on our assumption that the system knows when it is exposed to the attacker and for how long. However, this may not be known in practice. In real-world deployments, this limitation can be addressed through practical approaches based on observed attack patterns or past records from similar scenarios. Second, our proposed HT-DRL leverages hypergame expected utility to tackle the cold-start problem. However, it requires a small amount of prior knowledge to train the HT agent. Hence, when no prior knowledge is available, the effectiveness of HT-DRL is reduced. Lastly, we assumed a static mission scenario. However, in real-world operations, mission parameters such as target area boundaries and surveillance time requirements may change dynamically. This limitation can be addressed in future work by developing adaptive mechanisms to adjust drone parameters in real-time when mission parameters change.

We acknowledge that a well-trained hypergame theory agent and a static mission scenario may not fully reflect the dynamic nature of real-world environments. Future work will explore adaptive techniques to address these challenges.

While our simulation-based DRL approach shows promising results, the derived parameters and strategies are not directly transferable to real-world scenarios due to the inherent complexities of real-world environments. Nevertheless, this work provides high-level guidelines for leveraging DRL and hypergame-theoretic DRL in cyber deception drone scenarios, offering valuable insights for future real-world applications.

The proposed techniques have not been validated with physical drones. Such an experiment could provide a more comprehensive evaluation of the techniques, taking into account the complexities and uncertainties that arise in real-world scenarios.

\subsection{Future Research}
Considering all these limitations in mind, we have a plan for {\bf future research} as follows: (1) addressing additional types of attacks; (2) considering {\em transfer learning}
to mitigate the cold start in addition to HT with respect to the above methods; and (3) developing more realistic measures of agents' changing uncertainty reflecting real-time scenarios. 
(4) Implementing and evaluating our deception strategies using actual UAV hardware and environments to assess performance.
(5) Incorporating strategies specifically designed to mitigate data manipulation and spoofing attacks, leveraging the flexibility of our HT-DRL approach to enhance the overall security and robustness of mission-critical UAV operations.
An important direction for future research is extending our framework to multi-attacker and multi-defender scenarios. This would require developing n-player hypergame formulations and corresponding multi-agent reinforcement learning algorithms. The current single attacker-defender model provides the foundational framework that can be extended through coalition game theory and distributed multi-agent reinforcement learning (MARL) approaches.
\new{(6) conducting a systematic sensitivity analysis of the utility trade-off weights $(\rho, \zeta)$ and the uncertainty parameters $(\lambda_A, \lambda_D)$ to quantify how their values influence both mission coverage (RMC) and energy consumption (EC) under diverse attacker behaviors.}

\section*{Acknowledgment}
This work is partly supported by the Army Research Office and Army Research Laboratory under Grant Contract Numbers W911NF-20-2-0140, W911NF-19-2-0150, W911NF-17-1-0370, and W911NF-23-2-0012 and NSF award 2107450. 
\bibliographystyle{IEEEtranN}
\bibliography{ref}

\vskip 0pt plus -1fil
\vspace{-5mm}
\begin{IEEEbiography}
[{\includegraphics[width=1in,height=1.25in,clip,keepaspectratio]{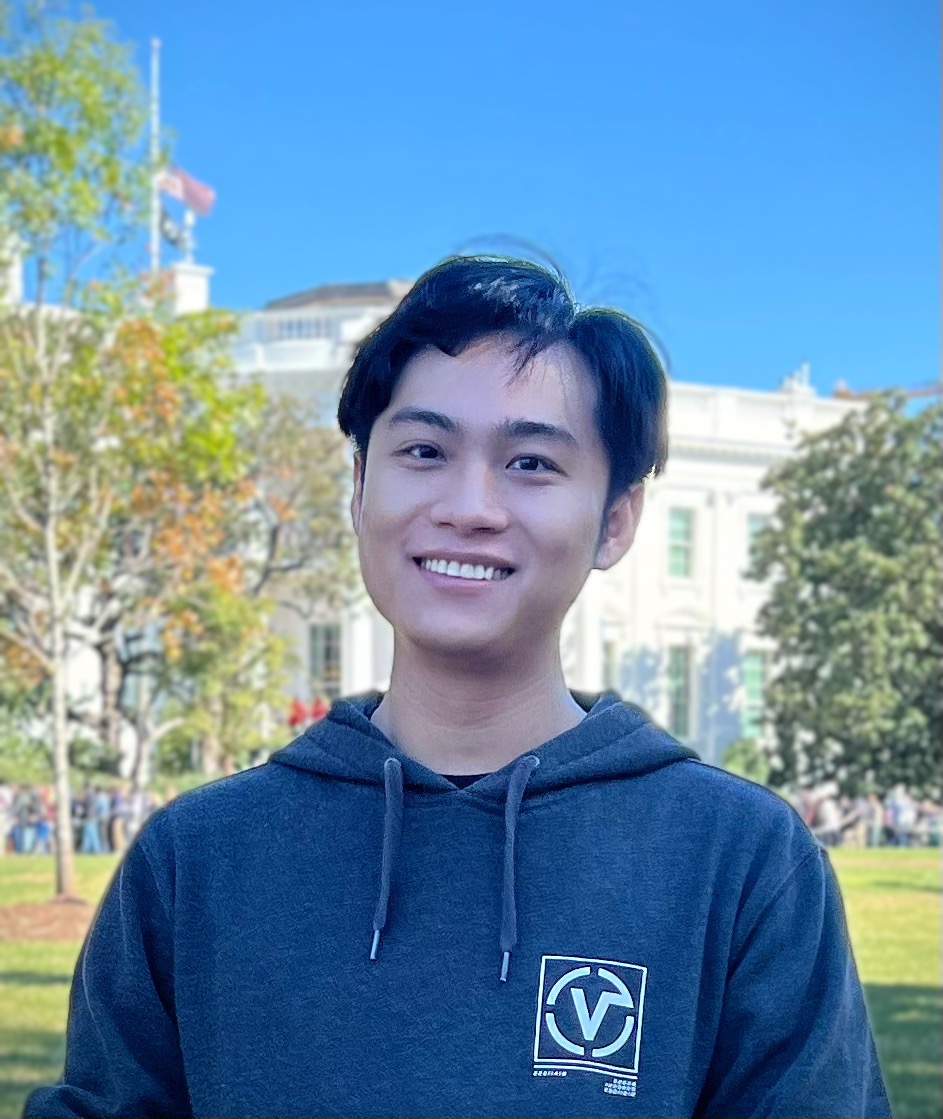}}]{Zelin Wan} is an AI Research Scientist at Bobyard in San Francisco, CA. He earned his Ph.D. in Computer Science from Virginia Tech in 2025, following an M.S. in Computer Science from Virginia Tech in 2021 and a B.S. in Computer Science from the University of Arizona in 2019. His research interests include game theory, deep learning, cybersecurity, and network science. He has been nominated for the Joseph Frank Hunkler Memorial Scholarship, received the College of Engineering Graduate Student Publication Fellowship, and was selected for the department’s Best Dissertation Award for 2024–2025 CS Ph.D. students. He is a student member of the ACM.
\end{IEEEbiography}

\vspace{-10mm}

\begin{IEEEbiography}
[{\includegraphics[width=1in,height=1.25in,clip,keepaspectratio]{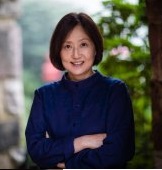}}]{Jin-Hee Cho} (M’09; SM’14) is an Associate Professor in the Department of Computer Science at Virginia Tech, where she has served since August 2018, and the director of the Trustworthy Cyberspace Lab. Prior to joining Virginia Tech, she was a computer scientist at the U.S. DEVCOM Army Research Laboratory (US DEVCOM ARL) in Adelphi, MD, beginning in 2009. Dr. Cho has published extensively in leading journals and conferences in the areas of cybersecurity, decision-making under uncertainty, and network science. She has received multiple best paper awards, including IEEE TrustCom 2009, BRIMS 2013, IEEE GLOBECOM 2017, ARL’s Publication Award in 2017, and IEEE CogSima 2018. She was awarded the 2015 IEEE Communications Society William R. Bennett Prize in Communications Networking and the 2023 IEEE ComSoc Network Operations \& Management (CNOM) Test of Time Paper Award. In 2013, Dr. Cho received the Presidential Early Career Award for Scientists and Engineers (PECASE), the highest honor bestowed by the U.S. government on early-career researchers. She was also recognized with the 2022 Faculty Fellow Award from the College of Engineering (CoE) and named a 2025–2026 CoE Dean’s Fellow at Virginia Tech. In addition, she was awarded the Stephen and Cherye Tyndall Moore Junior Faculty Fellowship, which provides endowed research support. Dr. Cho earned her Ph.D. in Computer Science from Virginia Tech in 2008. She currently serves as an associate editor for the IEEE Transactions on Network and Service Management, IEEE Transactions on Services Computing, and The Computer Journal. She is a Senior Member of IEEE and a member of ACM.
\end{IEEEbiography}

\vskip 0pt plus -1fil

\begin{IEEEbiography}
[{\includegraphics[width=1in,height=1.25in,clip,keepaspectratio]{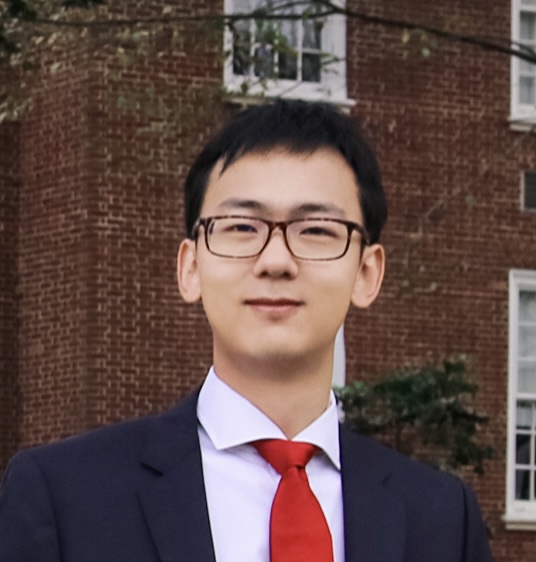}}]{Mu Zhu} is a postdoctoral researcher at the Computer Network Information Center of the Chinese Academy of Sciences as of 2024. His research focuses on cybersecurity and defensive deception, with a particular interest in enhancing these fields through game theory and machine learning. He earned his Ph.D. in Computer Science from North Carolina State University in 2024, his M.S. in Computer Engineering from the University of Delaware in 2017, and his B.S. in Electronic Engineering from Zhengzhou University, Henan, China, in 2014.
\end{IEEEbiography}

\vskip 0pt plus -1fil

\begin{IEEEbiography}
[{\includegraphics[width=1in,height=1.5in,clip,keepaspectratio]{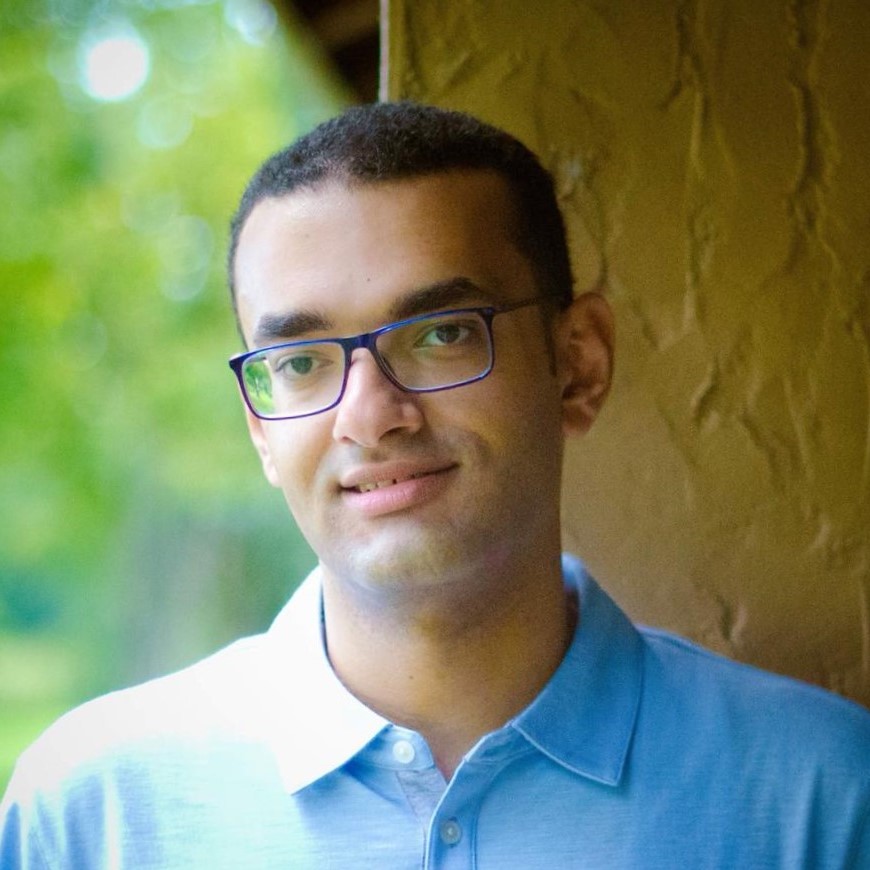}}]{Ahmed H. Anwar} (S’09; M’19) is currently a postdoctoral research scientist in the U.S. Army Research Lab in Adelphi, MD since 2019. His research interests include Network Security, Algorithmic Game Theory and Machine Learning. Dr. Anwar earned his PhD degree in Electrical Engineering from the University of Central Florida in 2019. Before that he worked as a research assistant in Nile University, Egypt, and Qatar University between 2011 and 2013. He received his BSc. degree (with highest honors) in electrical engineering from Alexandria University, Alexandria, Egypt, in 2011 and the MSc. degree in wireless Information Technology from Nile University, Egypt, in 2013.
\end{IEEEbiography}

\vspace{-10mm}
\begin{IEEEbiography}
[{\includegraphics[width=1in,height=1.25in,clip,keepaspectratio]{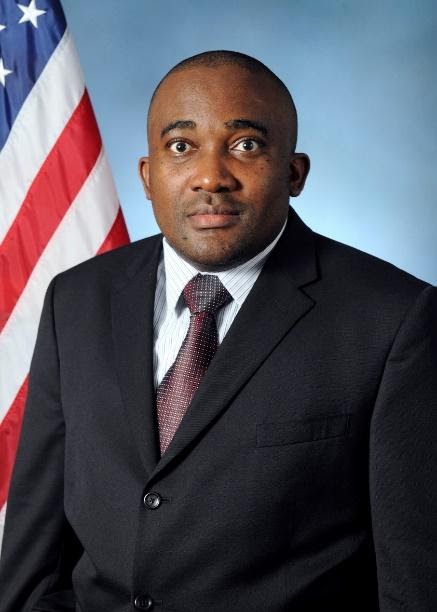}}]{Charles A. Kamhoua} is a Senior Electronics Engineer at the Network Security Branch of the ARL in Adelphi, MD, where he conducts and directs basic research on applying game theory to cybersecurity. Before joining ARL, he spent six years as a researcher at AFRL, Rome, NY, and held visiting positions at the University of Oxford and Harvard University. He has co-authored over 200 peer-reviewed papers, earning five best paper awards, and co-edited four Wiley-IEEE Press books: Game Theory and Machine Learning for Cyber Security, Modeling and Design of Secure Internet of Things, Blockchain for Distributed System Security, and Assured Cloud Computing. His scholarship and leadership have been recognized with awards such as the 2020 Sigma Xi Young Investigator Award and the 2019 US Army Civilian Service Commendation Medal. He earned a B.S. in electronics from the University of Douala (ENSET), Cameroon, in 1999, an M.S. in Telecommunication and Networking from Florida International University (FIU) in 2008, and a Ph.D. in Electrical Engineering from FIU in 2011. He is a senior member of ACM and IEEE.
\end{IEEEbiography}

\vskip 0pt plus -1fil

\begin{IEEEbiography}
[{\includegraphics[width=1in,height=1.25in,clip,keepaspectratio]{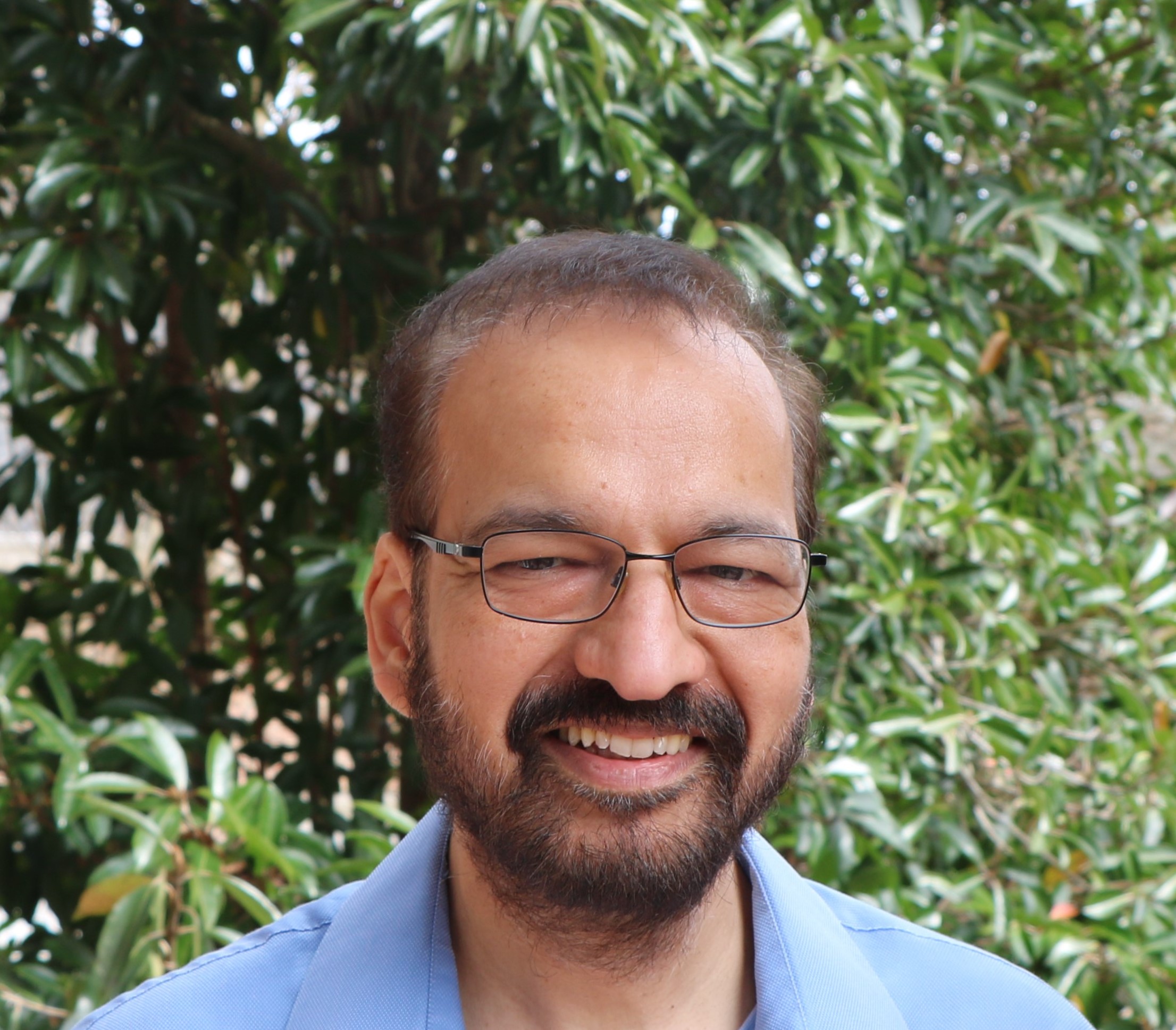}}]{Munindar P. Singh}(F'09)
is the SAS Institute Distinguished Professor and an Alumni Distinguished Graduate Professor in the Department of Computer Science at North Carolina State University. Munindar's research interests include AI and multiagent systems and their applications, including in cybersecurity.
Munindar is a Fellow of AAAI (Association for the Advancement of Artificial Intelligence), AAAS (American Association for the Advancement of Science), ACM (Association for Computing Machinery), and IEEE (Institute of Electrical and Electronics Engineers), and was elected a foreign member of Academia Europaea. He has won the ACM/SIGAI Autonomous Agents Research Award, the IEEE TCSVC Research Innovation Award, and the IFAAMAS Influential Paper Award.  He won NC State University's Outstanding Graduate Faculty Mentor Award as well as the Outstanding Research Achievement Award (twice). He was selected as an Alumni Distinguished Graduate Professor and elected to NCSU's Research Leadership Academy. He also won NC State's Outstanding Faculty Mentor Award. 
Munindar was the editor-in-chief of the ACM Transactions on Internet Technology (2012--2018) and IEEE Internet Computing (1999--2002). 
Munindar served on the founding board of directors of IFAAMAS, the International Foundation for Autonomous Agents and MultiAgent Systems. 
Thirty-one students have received PhD degrees and forty-one students MS degrees under Munindar's direction.
\end{IEEEbiography}

\end{document}


\title{Supplement Document: Cyber Deception for Mission Surveillance via Hypergame-Theoretic Deep Reinforcement Learning}

\author{Zelin Wan, Jin-Hee Cho,~\IEEEmembership{Senior Member, IEEE}, Mu Zhu, Ahmed H. Anwar, and Charles Kamhoua,~\IEEEmembership{Senior Member, IEEE}, Munindar P. Singh,~\IEEEmembership{IEEE Fellow}
\IEEEcompsocitemizethanks{\IEEEcompsocthanksitem This research was partly sponsored by the Army Research Laboratory and was accomplished under Cooperative Agreement Number W911NF-19-2-0150 and W911NF-23-2-0012. In addition, this research is also partly supported by the Army Research Office under Grant Contract Numbers W911NF-20-2-0140 and W911NF-17-1-0370. The views and conclusions contained in this document are those of the authors and should not be interpreted as representing the official policies, either expressed or implied, of the Army Research Laboratory or the U.S. Government. The U.S. Government is authorized to reproduce and distribute reprints for Government purposes notwithstanding any copyright notation herein ({\em Corresponding author: Zelin Wan}). Zelin Wan and Jin-Hee Cho are with the Department of Computer Science, Virginia Tech, Falls Church, VA, USA. Email: \{zelin, jicho\}@vt.edu. Mu Zhu and Munindar P. Singh are with the Department of Computer Science, North Carolina State University, Raleigh, NC 27695. Email: \{mzhu5, mpsingh\}@ncsu.edu. Ahmed H. Anwar and Charles A. Kamhoua are with the US Army Research Laboratory, Adelphi, MD, USA. Email: a.h.anwar@knights.ucf.edu; charles.a.kamhoua.civ@mail.mil.}}



\maketitle

\appendices

\section{HT-DRL Training Algorithm}
\label{sec:ht-drl-algorithm}

\new{This section provides the formal algorithmic presentation of the HT-DRL training procedure referenced in the main paper.}

\begin{algorithm}[h]
\caption{\new{HT-DRL Training Procedure}}
\begin{algorithmic}[1]
\State \textbf{Input:} Environment, HT parameters, RL hyperparameters
\State \textbf{Output:} Trained HT-DRL agent
\State Initialize HT agent with utility functions
\State Generate action probability distribution (APD) using HEU
\State Initialize A2C agent and apply APD weights to output layer
\For{episode = 1 to max\_episodes}
    \State Interact with environment using filtered policy
    \State Update A2C networks with gradients
\EndFor
\State \textbf{return} Trained HT-DRL agent
\end{algorithmic}
\end{algorithm}

\section{Comparative Performance Analysis with Other Defenses}
\label{sec:other-defense-analysis-appendix}

In this section, we explain the performance analyses of the defense mechanisms considered in this work, particularly in terms of honey drone (HD)-based vs. non-HD based.  Specifically, we will compared the performance of HD-F, IDS, CD, HD-HT-DRL, and the No-Defense schemes (see Section IV.A of the main paper, discussing Comparing Schemes) in terms of the number of active, connected drones ($\mathcal{N}_{AC}$) and accumulated rewards by the attacker and defender ($\mathcal{G}^A$ and $\mathcal{G}^D$), which are detailed in Section IV.B of the main paper.

\subsubsection{\bf Number of Active, Connected Drones}
Fig.~\ref{fig: Performance analysis-compare existing: Number of Active, Connected Drones} compares the performance of HD-based defenses, including F, HT, DRL, and HT-DRL, in terms of $\mathcal{N}_{AC}$, given an attack strategy. Our key observations are as follows: (1) HD-based defense techniques, specifically HD-F and HD-HT-DRL, outperform other defense techniques. This implies that honey drones are effective at luring attacker while simultaneously serving as a relay to maintain the connectivity of the drone fleet. (2) HD-HT-DRL exhibits the highest performance across different schemes, demonstrating that the synergistic, hybrid approach combining game theory and DRL can contribute to fast and autonomous learning for optimal solutions. This also can best utilize the merit of the honey drones and subsequently increases $\mathcal{N}_{AC}$, which leads to higher mission performance as shown in Fig. 5
of the main section in this paper.

\subsubsection{\bf Accumulated Rewards}
Figs.~\ref{fig: Performance analysis-compare existing: Accumulated Reward of Attacker} and \ref{fig: Performance analysis-compare existing: Accumulated Reward of Defender} compare the performance in terms of $\mathcal{G}^A$ and $\mathcal{G}^D$, respectively, under a given attack strategy. The figures show the following results: (1) No-Defense strategy results in the highest $\mathcal{G}^A$ and lowest $\mathcal{G}^D$ across all strategies because the system does not have any defense against DoS attacks. (2) In Figs.~\ref{fig: Attacker Fixed-Accumulated Reward of Attacker-CompareExist} and \ref{fig: Attacker Fixed-Accumulated Reward of Defender-CompareExist}, we observed the CD shows the highest $\mathcal{G}^A$ and lowest $\mathcal{G}^D$ among all. This is because CD prevents mission drones from being compromised but also disables them from performing the mission due to its protection mechanism. These drones can be analogous to `zombies' -- they cannot be compromised but are also unproductive. As a result, the defender cannot complete new tasks, resulting in a low $\mathcal{G}^D$. From the attacker's perspective, as many cells remain unscanned, the attacker enjoys a high immediate reward. Since those 'zombie' drones are considered active, the mission does not end due to a lack of available drones. A prolonged mission duration thus allows the attacker to accumulate a higher reward, $\mathcal{G}^A$.

\begin{figure*}[t]
\centering
\subfloat{\includegraphics[height=0.04\textwidth]{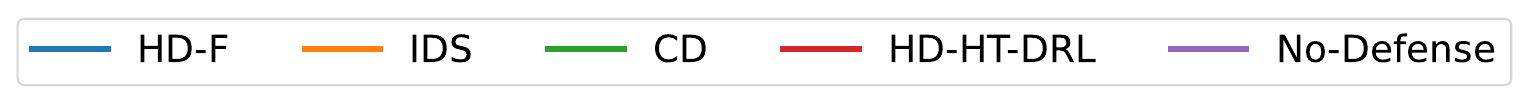}}
\hfil

\vspace{-3mm}
\setcounter{subfigure}{0}
\subfloat[$\mathcal{N}_{AC}$ under fixed attack.]{\includegraphics[width=0.25\textwidth]{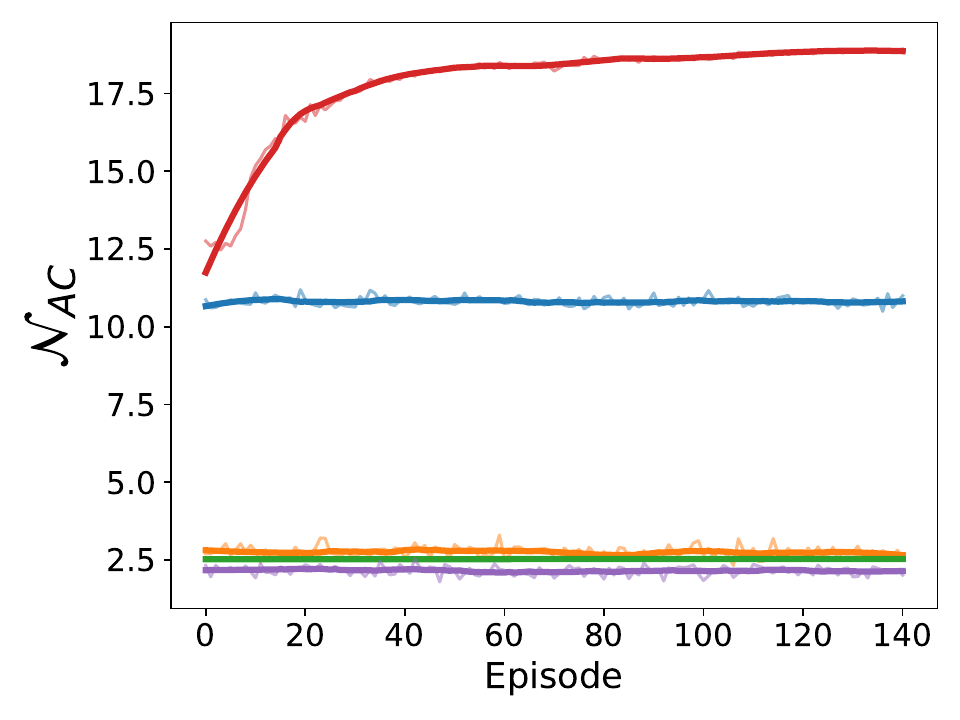}}
\hfil
\subfloat[$\mathcal{N}_{AC}$ under HT attack.]{\includegraphics[width=0.25\textwidth]{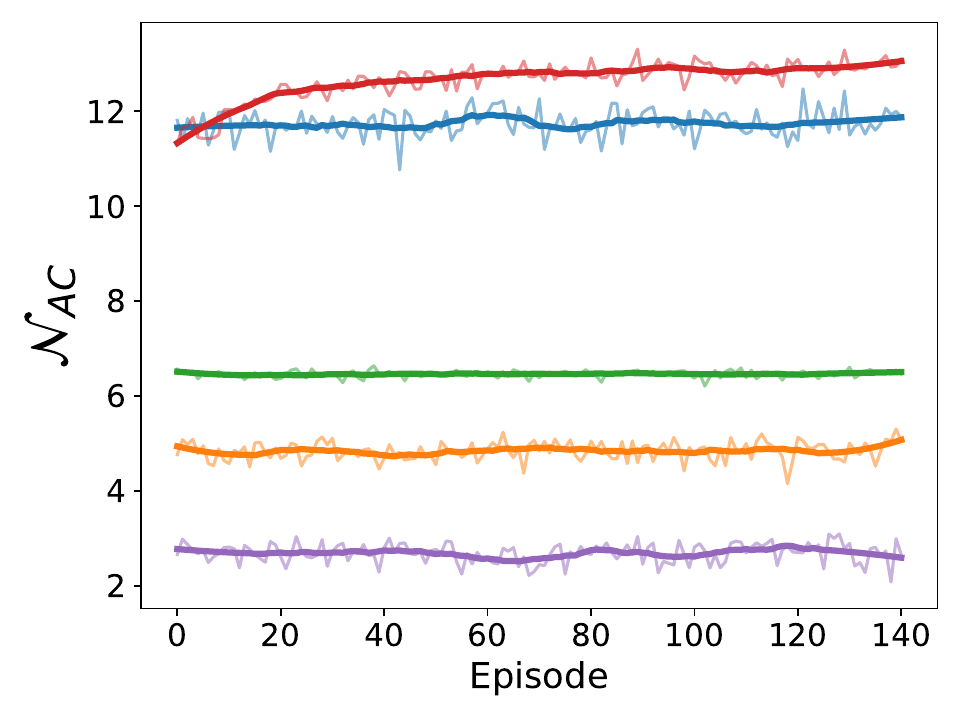}}
\hfil
\subfloat[$\mathcal{N}_{AC}$ under DRL attack.]{\includegraphics[width=0.25\textwidth]{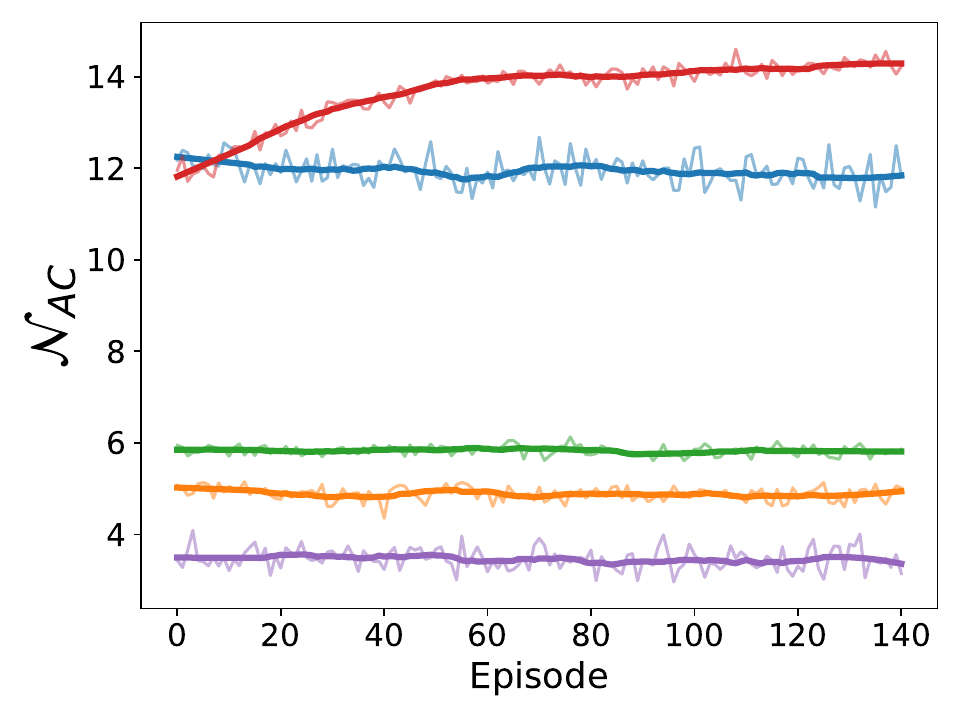}\label{fig: Attacker DRL-Number of Active, Connected Drones-CompareExist}}
\hfil
\subfloat[$\mathcal{N}_{AC}$ under HT-DRL attack.]{\includegraphics[width=0.25\textwidth]{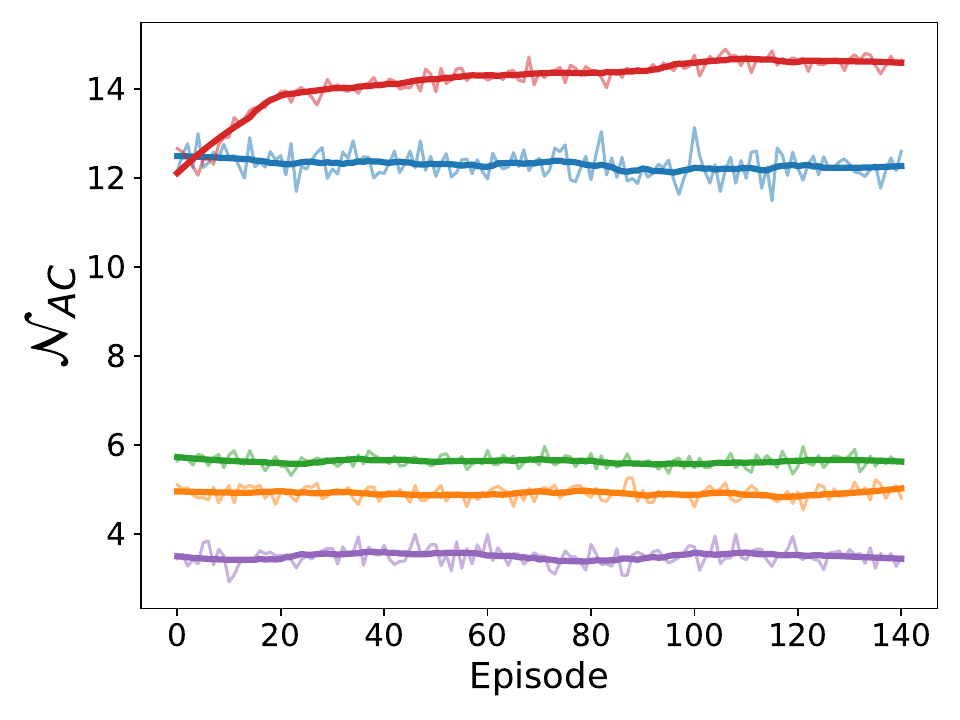}}
\hfil
\caption{Performance analysis of HD-HT-DRL, IDS, CD, and fixed defense, given an attack strategy, with respect to the number of active, connected drones ($\mathcal{N}_{AC}$).}\label{fig: Performance analysis-compare existing: Number of Active, Connected Drones}
\vspace{-2mm}
\end{figure*}

\begin{figure*}[t]
\centering
\subfloat{\includegraphics[height=0.04\textwidth]{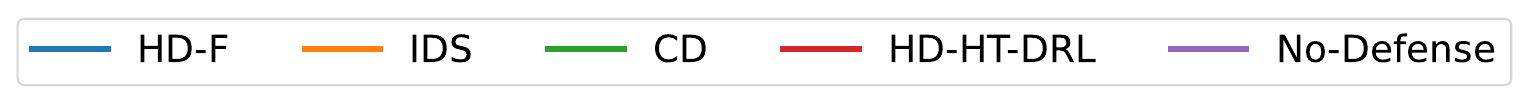}}
\hfil

\vspace{-3mm}
\setcounter{subfigure}{0}
\subfloat[$\mathcal{G}^A$ under fixed attack.]{\includegraphics[width=0.25\textwidth]{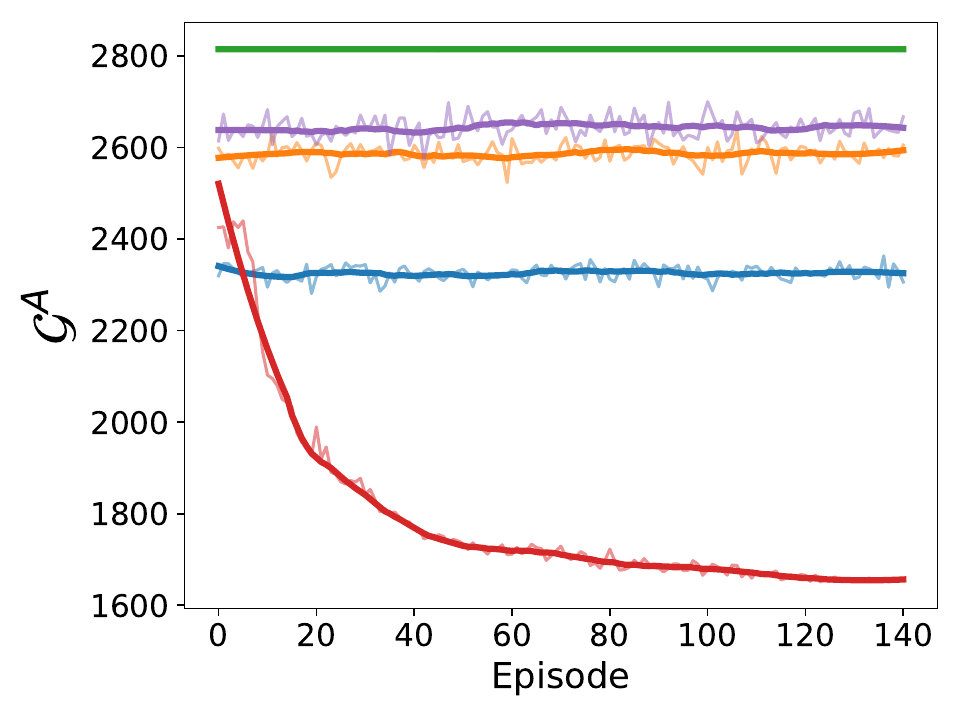}\label{fig: Attacker Fixed-Accumulated Reward of Attacker-CompareExist}}
\hfil
\subfloat[$\mathcal{G}^A$ under HT attack.]{\includegraphics[width=0.25\textwidth]{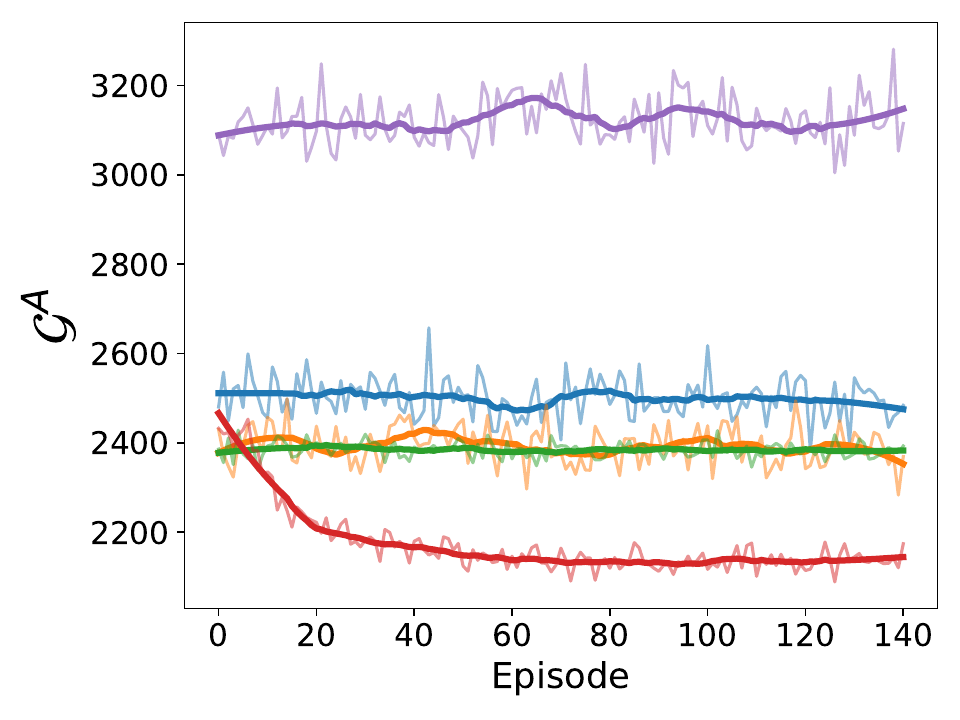}}
\hfil
\subfloat[$\mathcal{G}^A$ under DRL attack.]{\includegraphics[width=0.25\textwidth]{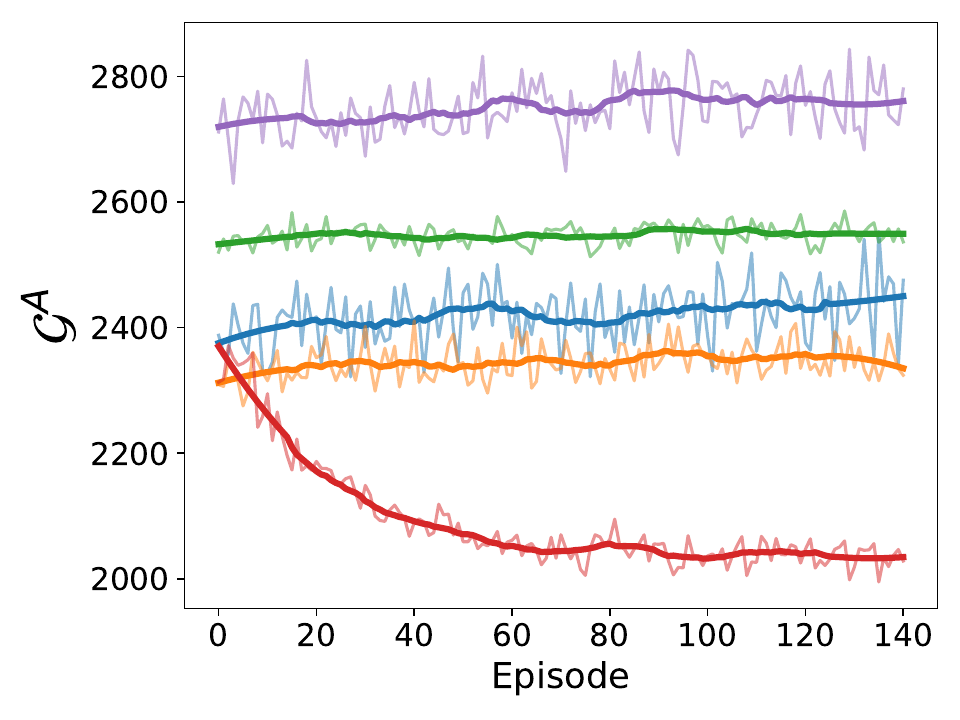}}
\hfil
\subfloat[$\mathcal{G}^A$ under HT-DRL attack.]{\includegraphics[width=0.25\textwidth]{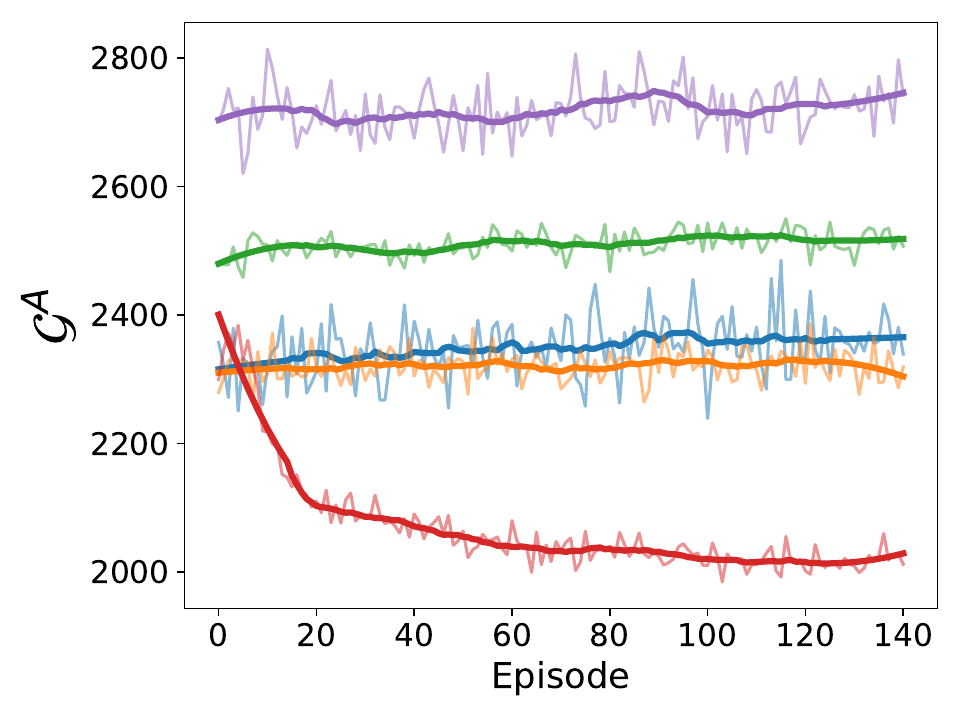}}
\hfil
\caption{Performance analysis of HD-HT-DRL, IDS, CD, and fixed defense, given an attack strategy, with respect to the attacker's accumulated reward ($\mathcal{G}^A$).}\label{fig: Performance analysis-compare existing: Accumulated Reward of Attacker}
\vspace{-2mm}
\end{figure*}

\begin{figure*}[t]
\centering
\subfloat{\includegraphics[height=0.04\textwidth]{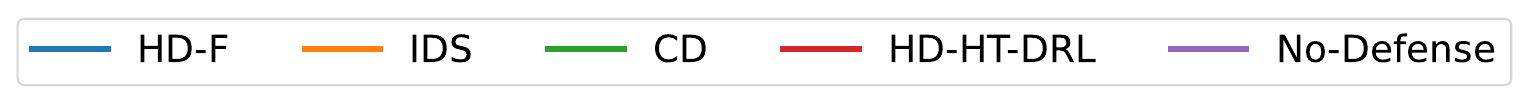}}
\hfil

\vspace{-3mm}
\setcounter{subfigure}{0}
\subfloat[$\mathcal{G}^D$ under fixed attack.]{\includegraphics[width=0.25\textwidth]{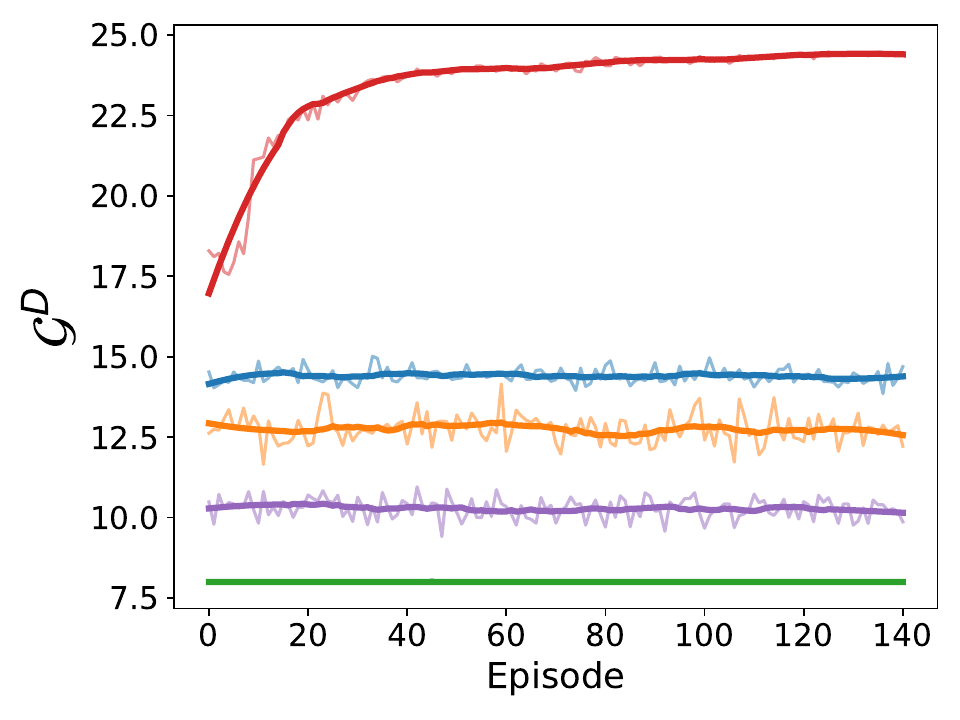}\label{fig: Attacker Fixed-Accumulated Reward of Defender-CompareExist}}
\hfil
\subfloat[$\mathcal{G}^D$ under HT attack.]{\includegraphics[width=0.25\textwidth]{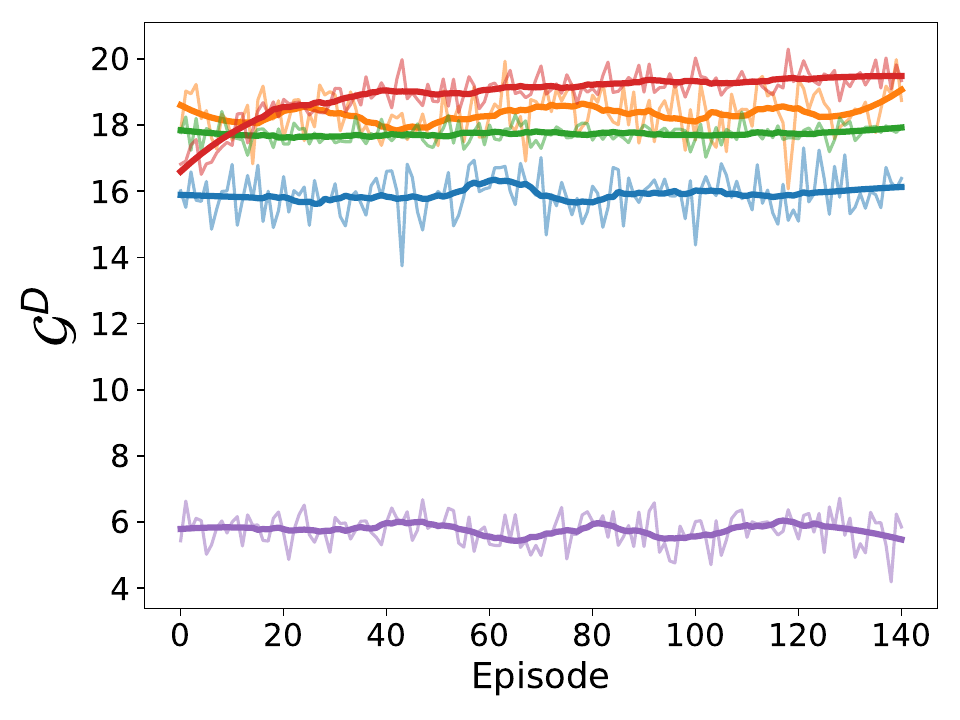}}
\hfil
\subfloat[$\mathcal{G}^D$ under DRL attack.]{\includegraphics[width=0.25\textwidth]{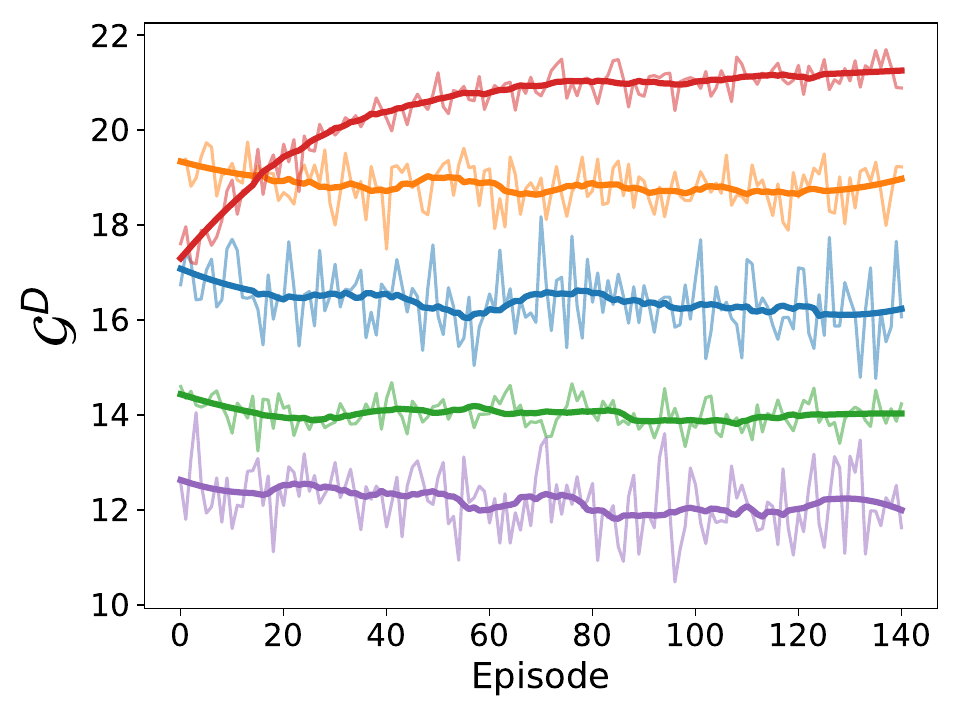}}
\hfil
\subfloat[$\mathcal{G}^D$ under HT-DRL attack.]{\includegraphics[width=0.25\textwidth]{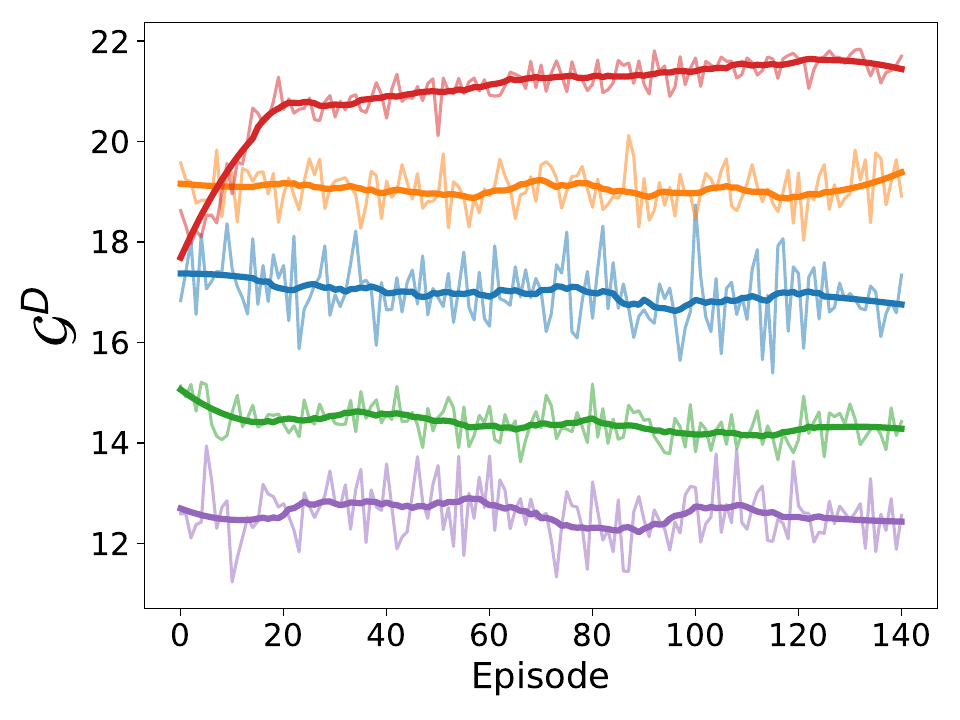}}
\hfil
\caption{Performance analysis of HD-HT-DRL, IDS, CD, and fixed defense, given an attack strategy, with respect to the defender's accumulated reward ($\mathcal{G}^D$).}\label{fig: Performance analysis-compare existing: Accumulated Reward of Defender}
\vspace{-2mm}
\end{figure*}

\section{Performance Analysis of Honey Drone-based Defenses}
\label{sec:hd-analysis-appendix}
This appendix discusses the performance of HD-based defenses, including F, HT, DRL, and HT-DRL, under the number of active, connected drones ($\mathcal{N}_{AC}$) and accumulated rewards by the attacker ($\mathcal{G}^A$) or defender ($\mathcal{G}^D$).

\subsubsection{\bf Number of Active, Connected Drones}
Fig.~\ref{fig: Performance analysis: number of active, connected drones} showcases the comparative performance in terms of $\mathcal{N}_{AC}$. We found: (1) The results in Fig.~\ref{fig: Performance analysis: number of active, connected drones} are well aligned with the results in Fig. 7 in the main paper,
where the high mission performance is achieved by more active, connected nodes, showing the performance order of HT-DRL $\geq$ DRL $\geq$ HT $\geq$ F, given a sufficiently large number of episodes (e.g., after $>60$). 
(2) HT-DRL exhibits superior performance over the other HD-based defenses with the same reasons discussed with Fig. 7
in Section V-B1
of the main paper.

\subsubsection{\bf Accumulated Reward}
Figs.~\ref{fig: Performance analysis: Accumulated Reward of Attacker} and \ref{fig: Performance analysis: Accumulated Reward of Defender} present a comparative study of HD-based defenses, specifically F, HT, DRL, and HT-DRL, in terms of $\mathcal{G}^A$ and $\mathcal{G}^D$, respectively, under a given attack strategy.  From the results shown in those figures, we observed: (1) The performance order is reversed when comparing Figs.~\ref{fig: Performance analysis: Accumulated Reward of Attacker} and \ref{fig: Performance analysis: Accumulated Reward of Defender}.  This is mainly because the immediate reward for the attacker corresponds to the number of uncompleted mission tasks in round $t$, whereas the defender's immediate reward relates to the number of completed mission tasks in the same round. This reward design effectively creates a zero-sum-like game scenario, resulting in the observed reverse order of rewards. (2) A special case is observed where DRL has a low $\mathcal{G}^A$ in Fig.~\ref{fig: Attacker HT-Accumulated Reward of Attacker} and also low $\mathcal{G}^D$ in Fig.~\ref{fig: Attacker HT-Accumulated Reward of Defender}. This may be because HT-based attacker outperforms and compromises most drones, leading to early termination of the mission.  Since both $\mathcal{G}^A$ and $\mathcal{G}^D$ represent accumulated rewards, a reduced mission duration may lead to a lower performance under the accumulated rewards.

\begin{figure*}[t]
\centering
\subfloat{\includegraphics[height=0.04\textwidth]{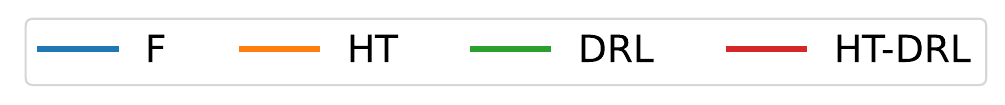}}
\hfil

\vspace{-3mm}
\setcounter{subfigure}{0}
\subfloat[$\mathcal{N}_{AC}$ under fixed attack.]{\includegraphics[width=0.25\textwidth]{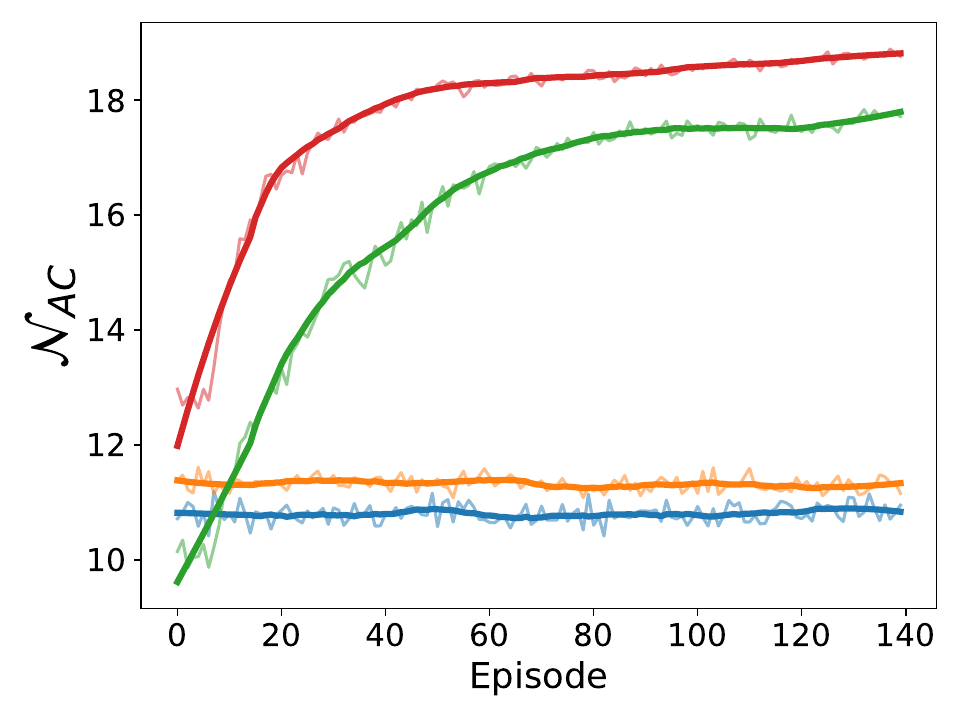}}
\hfil
\subfloat[$\mathcal{N}_{AC}$ under HT attack.]{\includegraphics[width=0.25\textwidth]{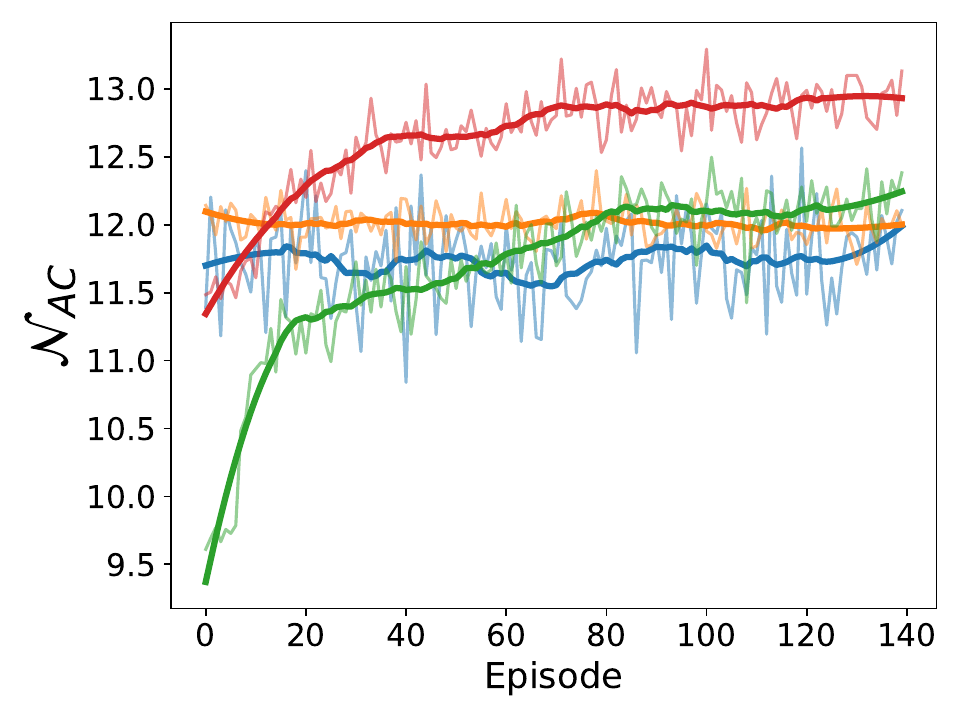}}
\subfloat[$\mathcal{N}_{AC}$ under DRL attack.]{\includegraphics[width=0.25\textwidth]{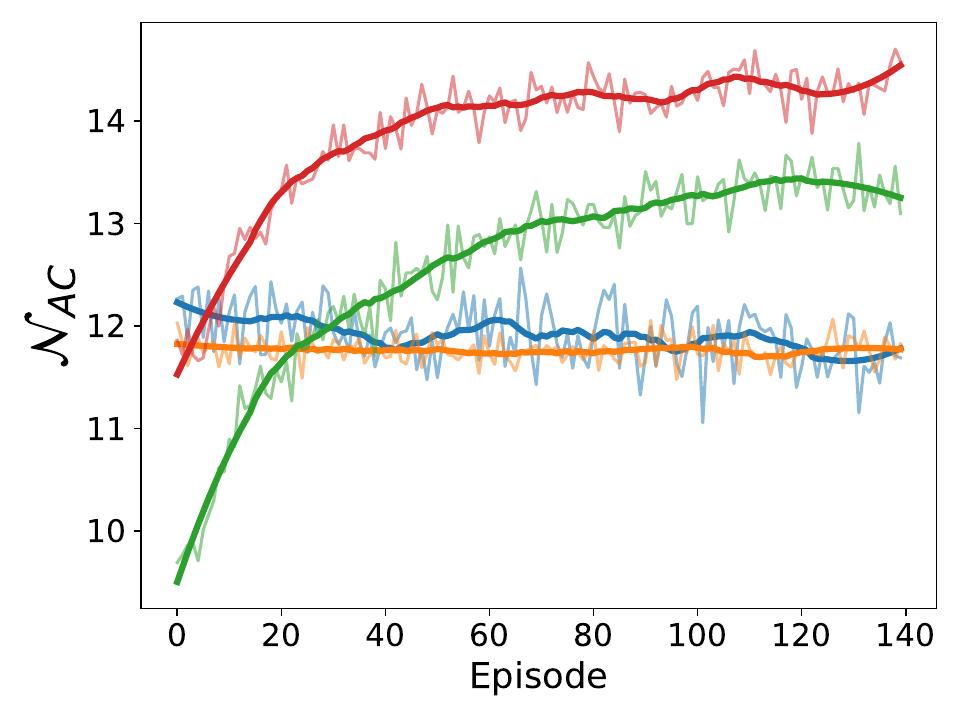}\label{fig: Attacker DRL-Number of Active Connected Drone}}
\subfloat[$\mathcal{N}_{AC}$ under HT-DRL attack.]{\includegraphics[width=0.25\textwidth]{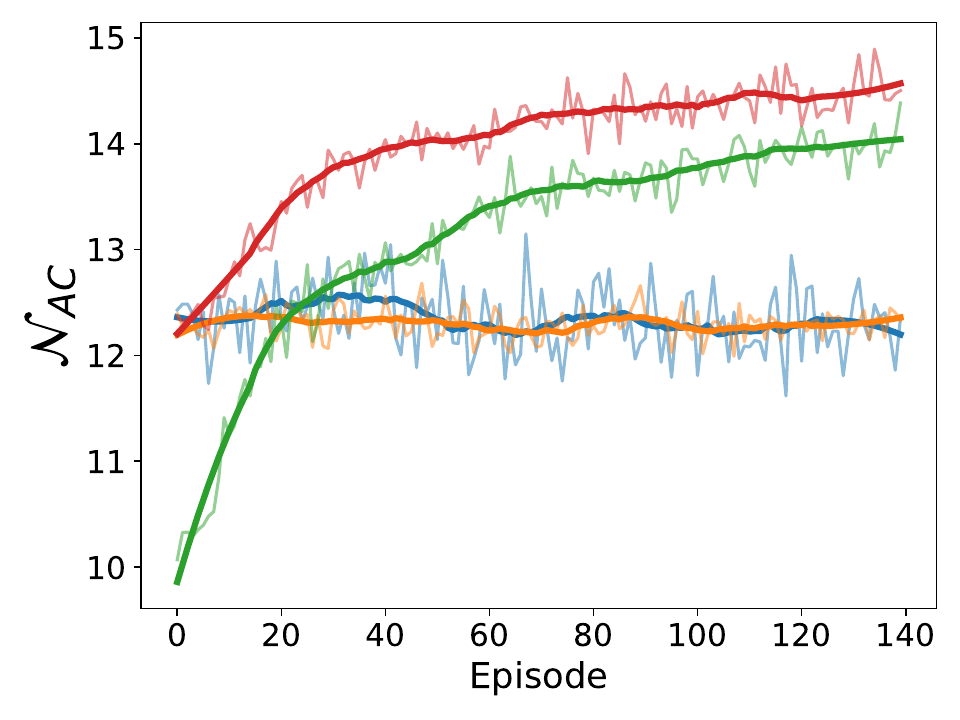}}
\hfil
\caption{Performance analysis of honey done-based defenses with respect to the number of active, connected drones ($\mathcal{N}_{AC}$), under a given attack.}\label{fig: Performance analysis: number of active, connected drones}
\vspace{-2mm}
\end{figure*}

\begin{figure*}[t]
\centering
\subfloat{\includegraphics[height=0.04\textwidth]{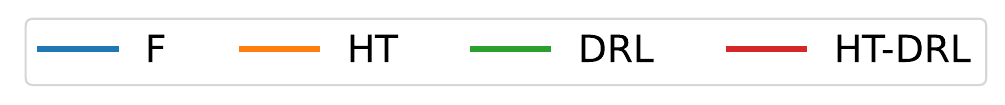}}
\hfil

\vspace{-3mm}
\setcounter{subfigure}{0}
\subfloat[$\mathcal{G}^A$ under fixed attack.]{\includegraphics[width=0.25\textwidth]{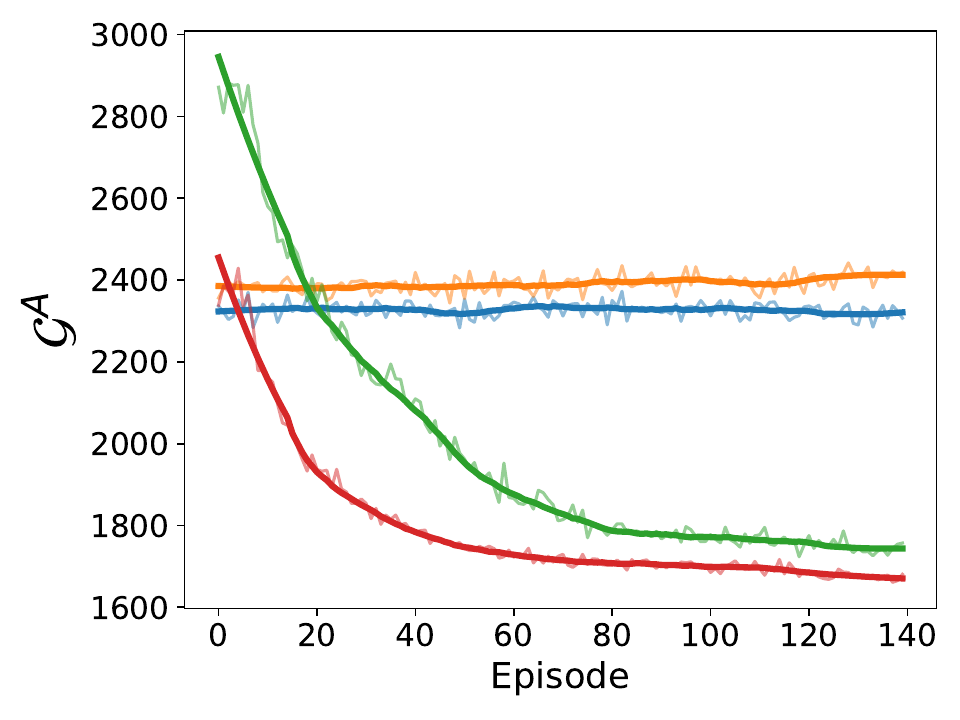}\label{fig: Attacker Fixed-Accumulated Reward of Attacker}}
\hfil
\subfloat[$\mathcal{G}^A$ under HT attack.]{\includegraphics[width=0.25\textwidth]{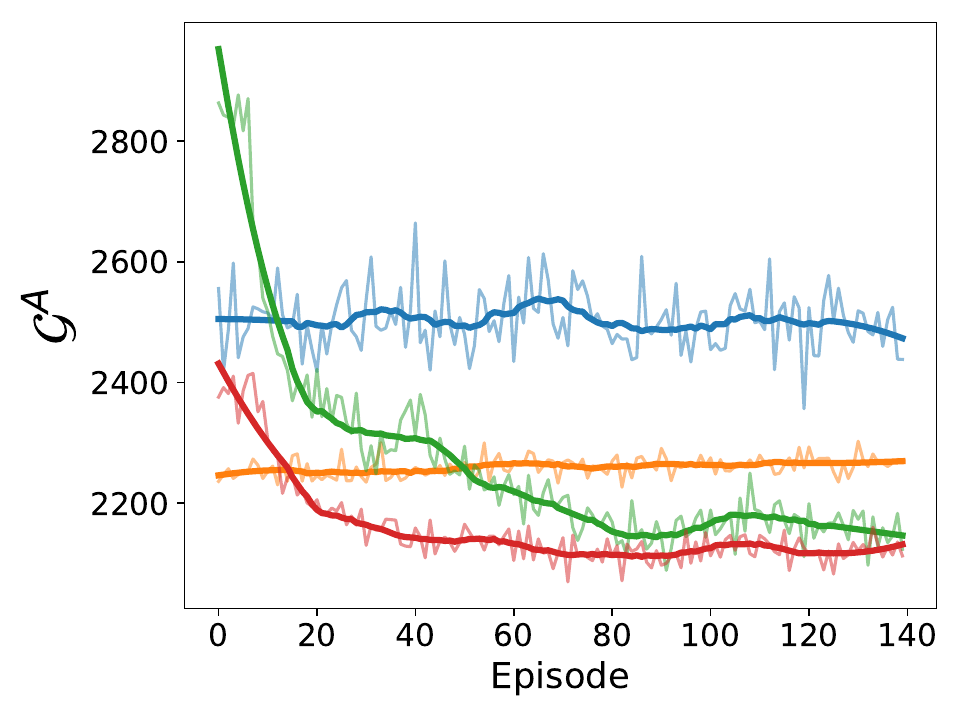}\label{fig: Attacker HT-Accumulated Reward of Attacker}}
\hfil
\subfloat[$\mathcal{G}^A$ under DRL attack.]{\includegraphics[width=0.25\textwidth]{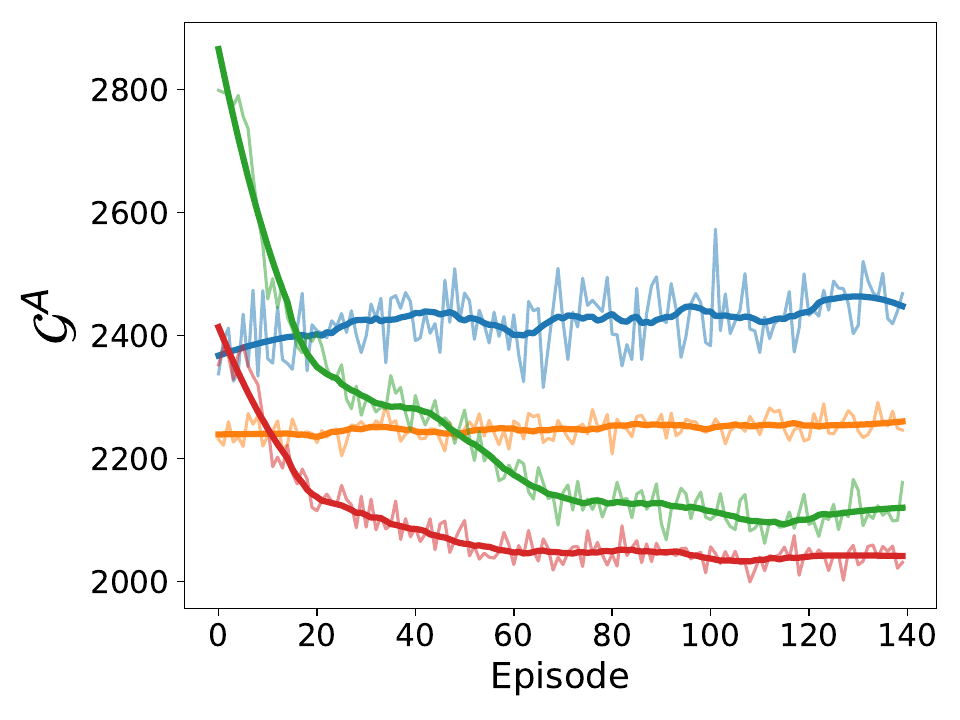}\label{Attacker DRL-Accumulated Reward of Attacker}}
\hfil
\subfloat[$\mathcal{G}^A$ under HT-DRL attack.]{\includegraphics[width=0.25\textwidth]{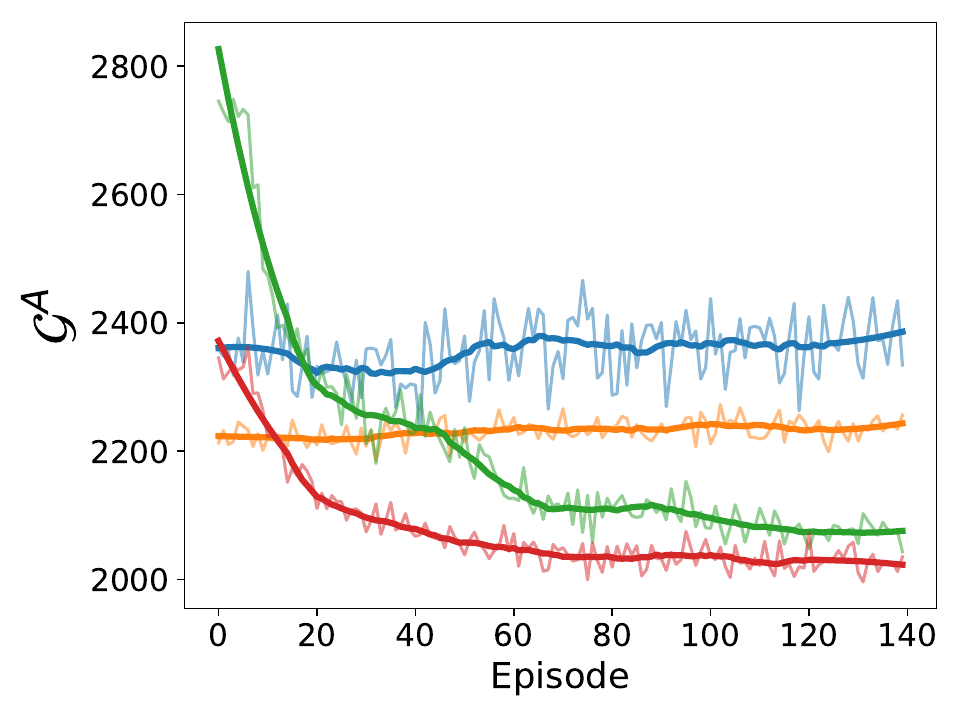}\label{Attacker HT-DRL-Accumulated Reward of Attacker}}
\hfil
\caption{Performance analysis of honey done-based defenses with respect to the attacker's accumulated reward ($\mathcal{G}^A$), under a given attack. }\label{fig: Performance analysis: Accumulated Reward of Attacker}
\vspace{-2mm}
\end{figure*}

\begin{figure*}[t]
\centering
\subfloat{\includegraphics[height=0.04\textwidth]{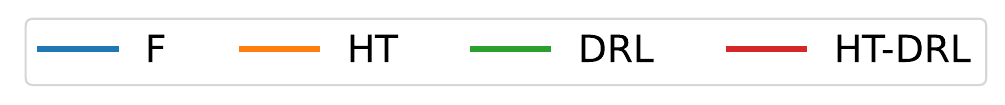}}
\hfil

\vspace{-3mm}
\setcounter{subfigure}{0}
\subfloat[$\mathcal{G}^D$ under fixed attack.]{\includegraphics[width=0.25\textwidth]{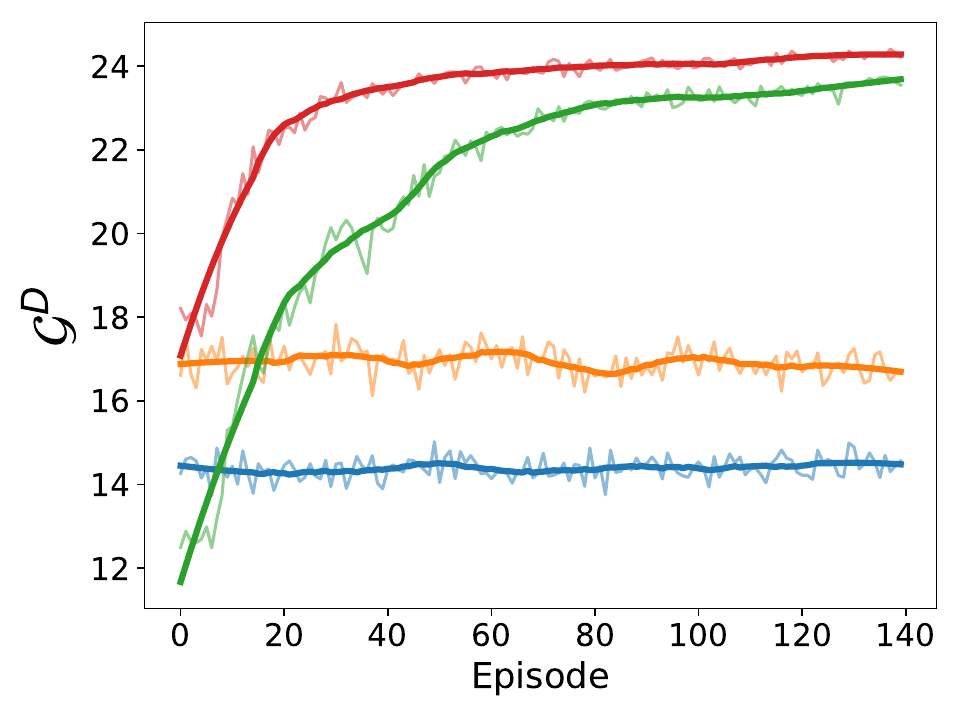}\label{fig: Attacker Fixed-Accumulated Reward of Defender}}
\hfil
\subfloat[$\mathcal{G}^D$ under HT attack.]{\includegraphics[width=0.25\textwidth]{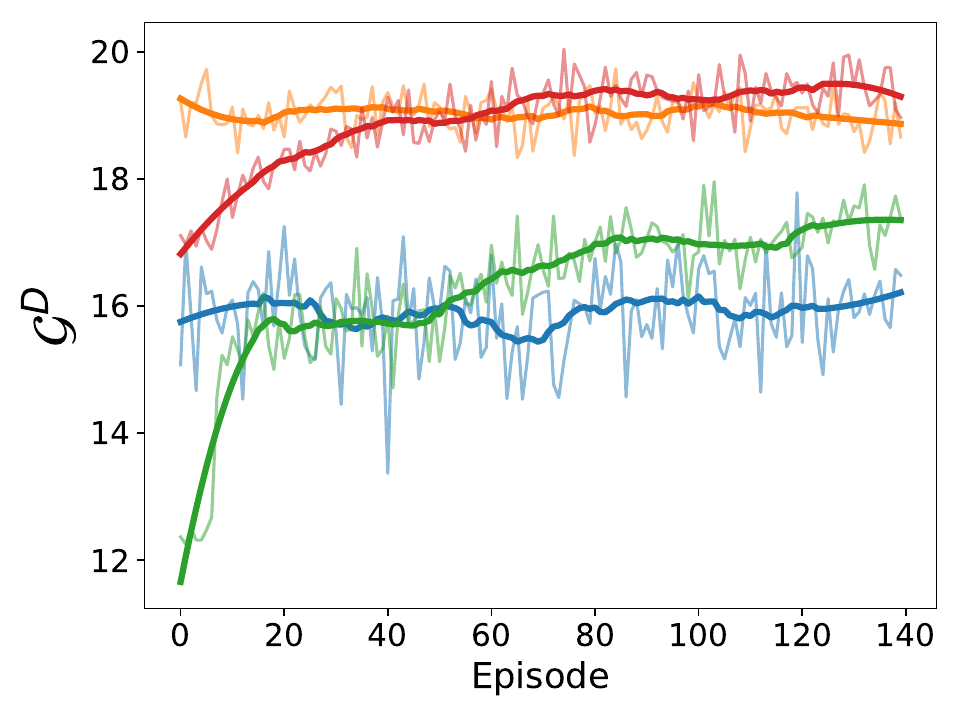}\label{fig: Attacker HT-Accumulated Reward of Defender}}
\hfil
\subfloat[$\mathcal{G}^D$ under DRL attack.]{\includegraphics[width=0.25\textwidth]{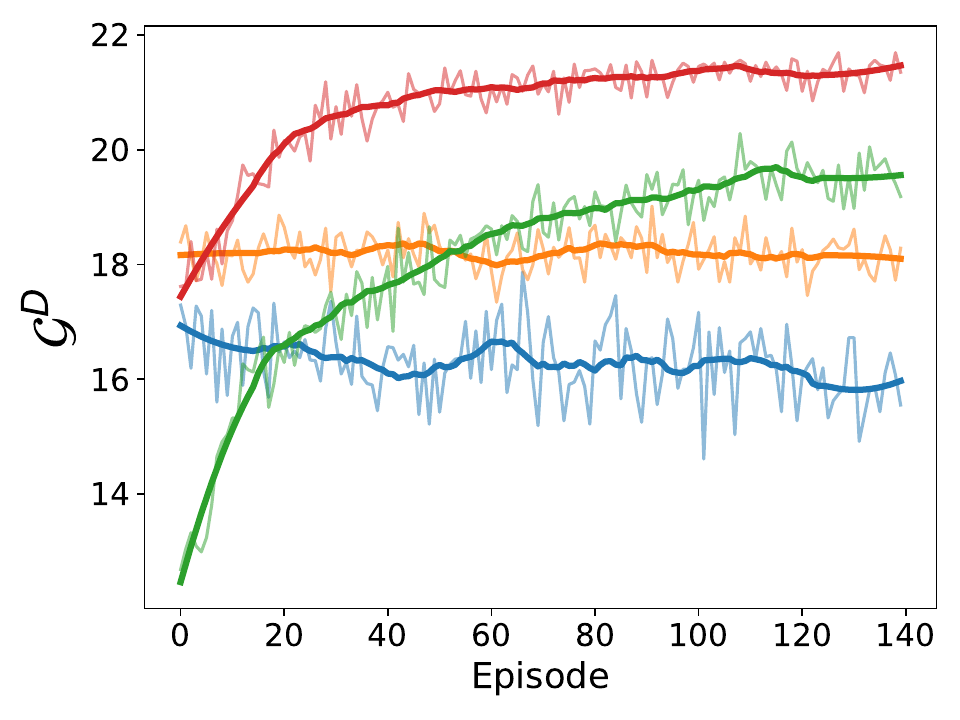}\label{fig: Attacker DRL-Accumulated Reward of Defender}}
\hfil
\subfloat[$\mathcal{G}^D$ under HT-DRL attack.]{\includegraphics[width=0.25\textwidth]{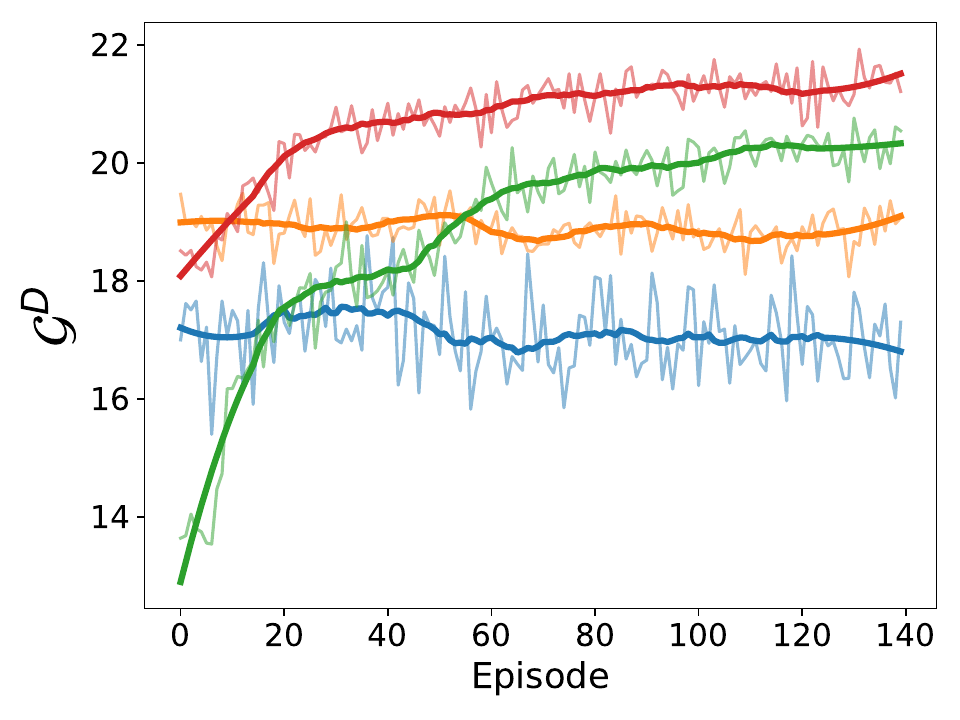}\label{fig: Attacker HT-DRL-Accumulated Reward of Defender}}
\hfil
\caption{Performance analysis of honey done-based defenses with respect to the defender's accumulated reward ($\mathcal{G}^D$), under a given attack.}\label{fig: Performance analysis: Accumulated Reward of Defender}
\vspace{-2mm}
\end{figure*}

\section{Sensitivity Analysis: Effect of Varying Attack Budget ($\zeta$)} \label{subsec:sensitivity-analysis}

This section investigates the impact of varying the attack budget ($\zeta$) on system performance based on the metrics in Section IV-B.
The results are presented based on the ones that converged at the 140th episode.

\subsubsection{\bf Ratio of Mission Completion} \label{Appendix: Sensitivity Analysis, Ratio of Mission Completion}
Fig.~\ref{fig: SenseAnalysis-Attack Budget-Ratio of Completed Mission} represents how a different level of the attack budget ($\zeta$, ranged from 1 to 9) impacts the ratio of mission completion ($\mathcal{R}_{MC}$).  From the results, our findings are: (1) Overall, as the attack budget increases, $\mathcal{R}_{MC}$ decreases because a higher attack budget empowers the attacker to launch more attacks in each step. Consequently, fewer mission drones are available, resulting in a diminished $\mathcal{R}_{MC}$. 
(2) However, we observed that after $\zeta=5$, the defense performance is mostly flat and can maintain its high performance. This is mainly because even if the attacker has higher budget, it may not find sufficient drones that have sufficient signal strength to perform attack.  (3) As we observed in Section V-B1,
HT attack is more effect in lowering down the performance of RL-based defenses due to the merit of strategic decision making by hypergame theory.


\subsubsection{\bf Energy Consumption} \label{Appendix: Sensitivity Analysis, Energy Consumption}
Fig.~\ref{fig: SenseAnalysis-Attack Budget-Energy Consumption} shows the effect of varying the attack budget ($\zeta$) on energy consumption ($\mathcal{EC}$).  We found the following from the results:  (1) $\mathcal{EC}$ decreases with higher $\zeta$ because more compromised (crashed) nodes can lead to fewer active nodes, resulting in low mission performance.  (2) However, HT-DRL consumes relatively less compared to its outperformance in mission performance because of its intelligent energy-aware strategy selection.

\subsubsection{\bf Number of Active, Connected Drones}
\begin{figure*}[t]
\centering
\subfloat{\includegraphics[height=0.04\textwidth]{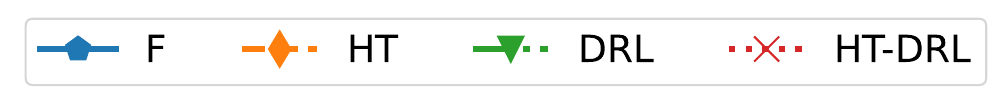}}
\hfil

\vspace{-3mm}
\setcounter{subfigure}{0}
\subfloat[$\mathcal{R}_{MC}$ under fixed attack.]{\includegraphics[width=0.25\textwidth]{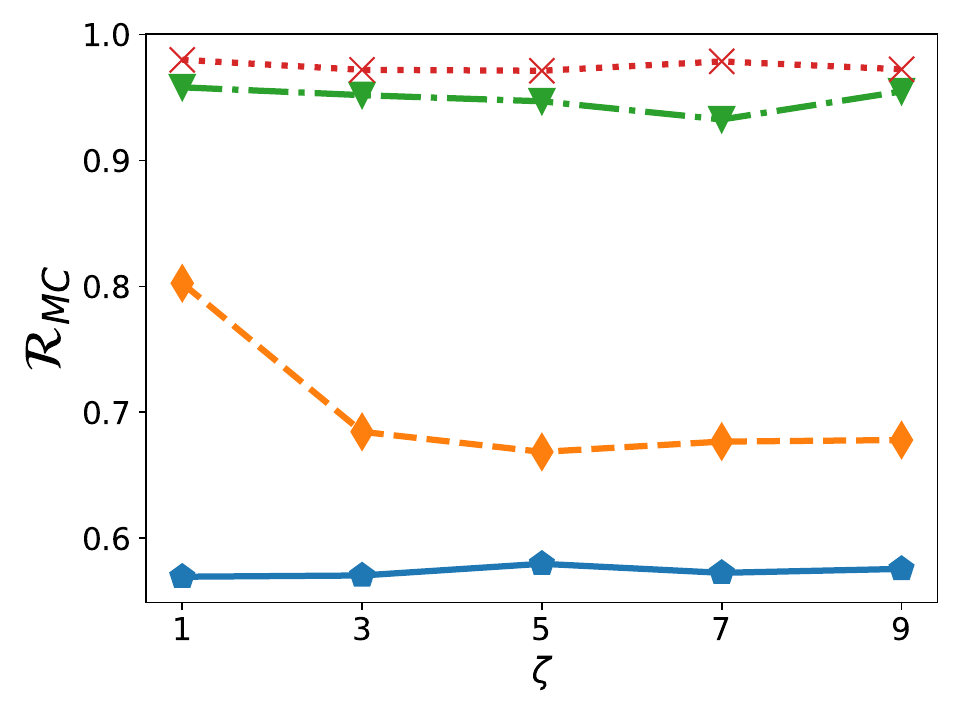}\label{fig: SenseAnalysis-Attack Budget-Attacker Fixed-Ratio of Completed Mission Tasks}}
\hfil
\subfloat[$\mathcal{R}_{MC}$ under HT attack.]{\includegraphics[width=0.25\textwidth]{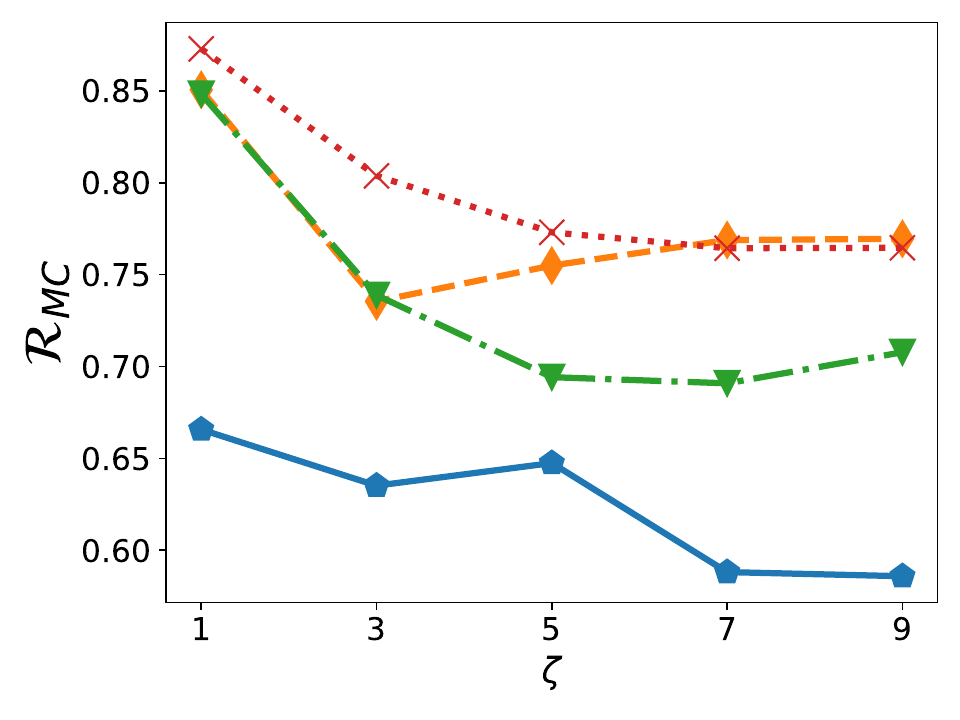}\label{fig: SenseAnalysis-Attack Budget-Attacker HT-Ratio of Completed Mission Tasks}}
\hfil
\subfloat[$\mathcal{R}_{MC}$ under DRL attack.]{\includegraphics[width=0.25\textwidth]{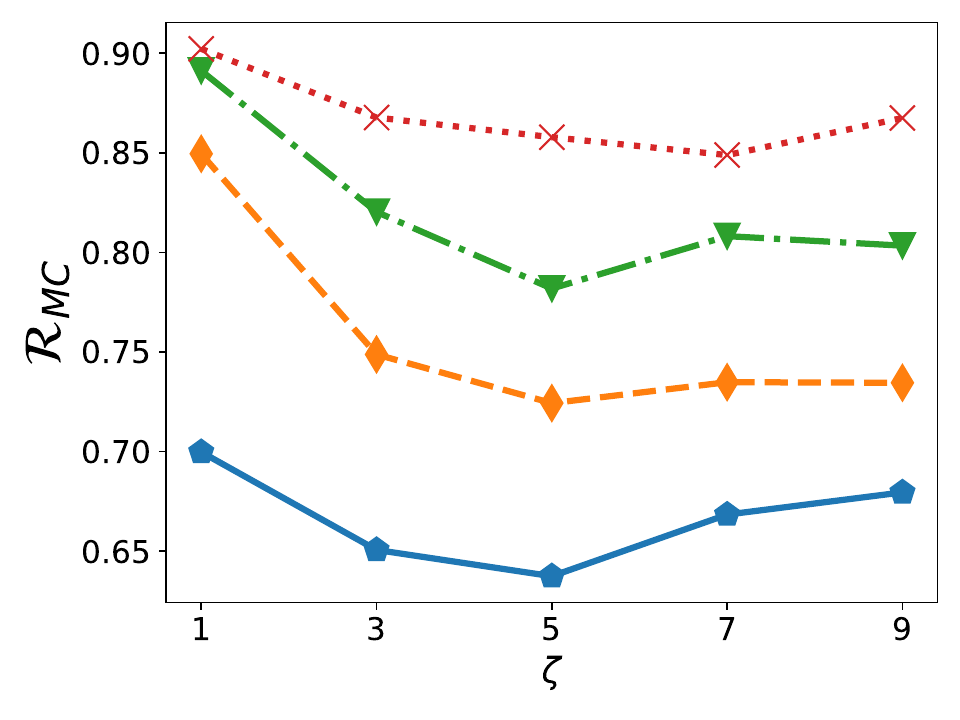}\label{fig: SenseAnalysis-Attack Budget-Attacker DRL-Ratio of Completed Mission Tasks}}
\hfil
\subfloat[$\mathcal{R}_{MC}$ under HT-DRL attack.]{\includegraphics[width=0.25\textwidth]{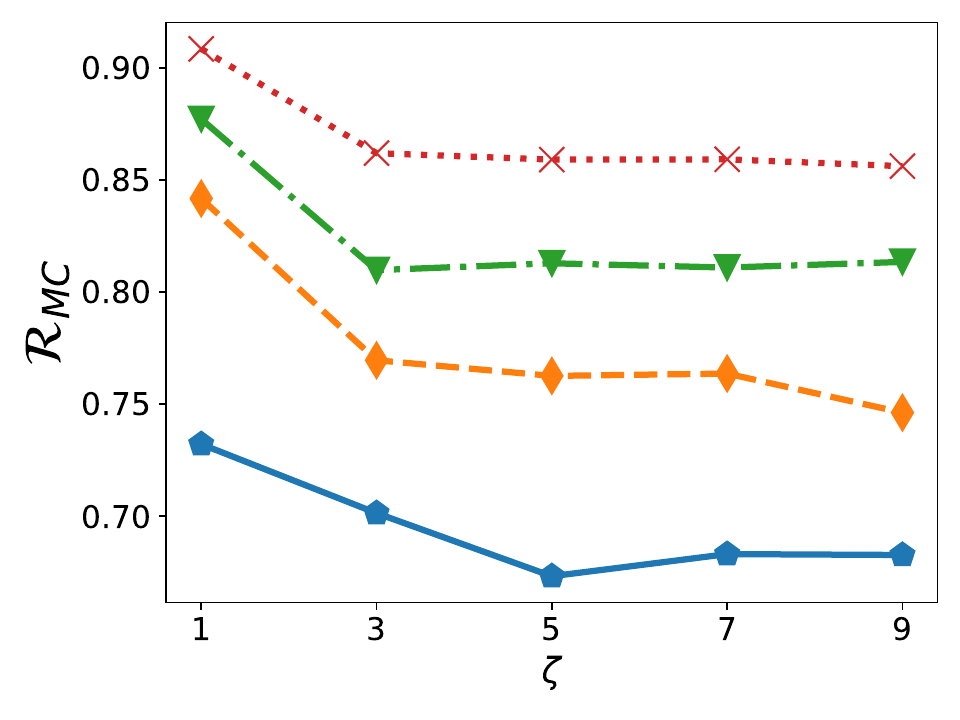}\label{fig: SenseAnalysis-Attack Budget-Attacker HT-DRL-Ratio of Completed Mission Tasks}}
\hfil
\caption{Effect of varying the attack budget ($\zeta$) on the ratio of completed mission tasks ($\mathcal{R}_{MC}$).}
\label{fig: SenseAnalysis-Attack Budget-Ratio of Completed Mission}
\end{figure*}

\begin{figure*}[t]
\centering
\subfloat{\includegraphics[height=0.04\textwidth]{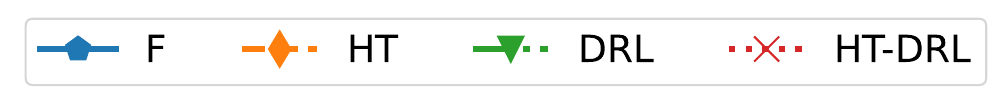}}
\hfil

\vspace{-3mm}
\setcounter{subfigure}{0}
\subfloat[$\mathcal{EC}$ under fixed attack.]{\includegraphics[width=0.25\textwidth]{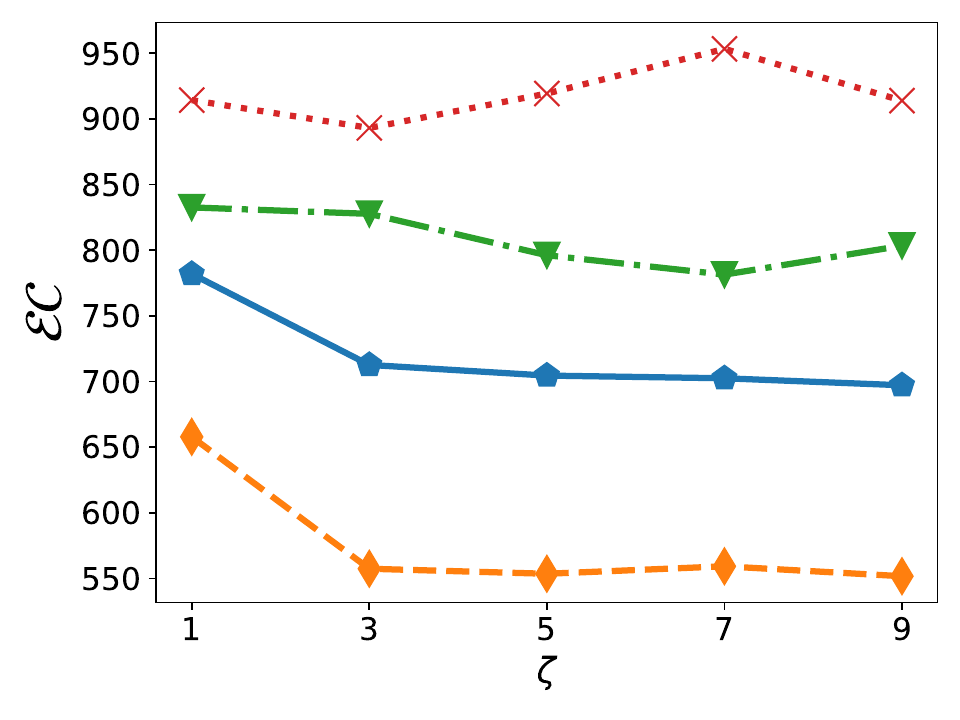}}
\hfil
\subfloat[$\mathcal{EC}$ under HT attack.]{\includegraphics[width=0.25\textwidth]{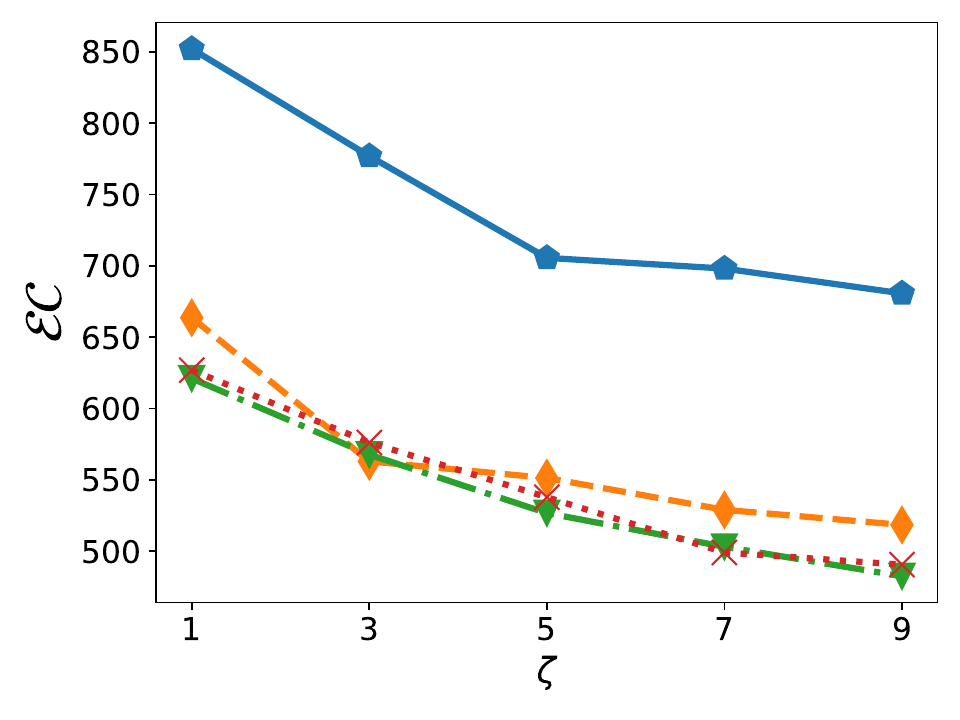}}
\hfil
\subfloat[$\mathcal{EC}$ under DRL attack.]{\includegraphics[width=0.25\textwidth]{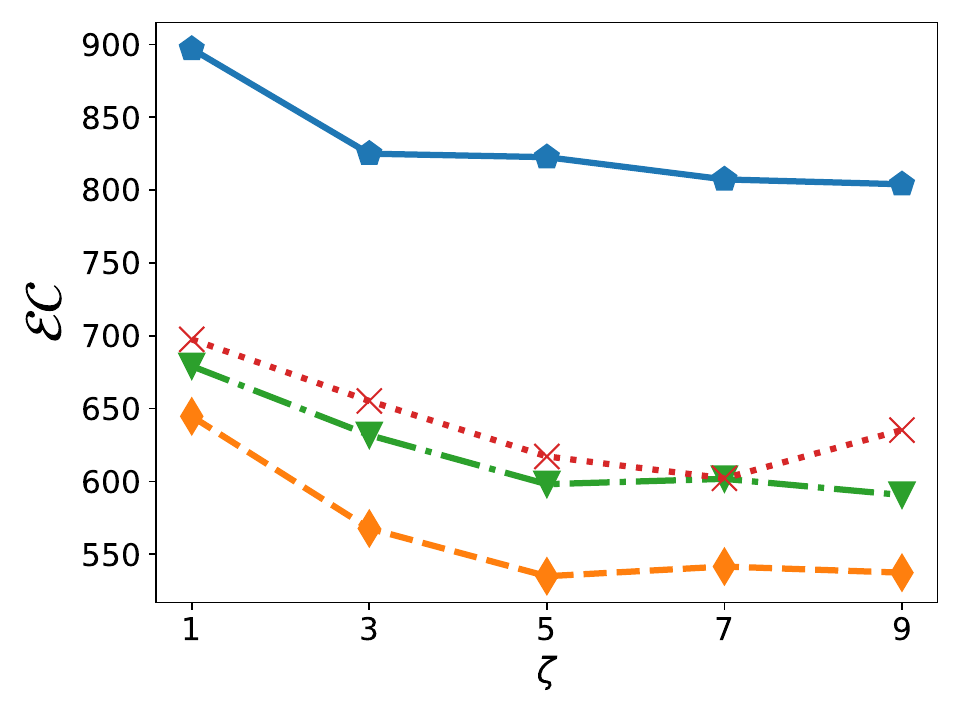}}
\hfil
\subfloat[$\mathcal{EC}$ under HT-DRL attack.]{\includegraphics[width=0.25\textwidth]{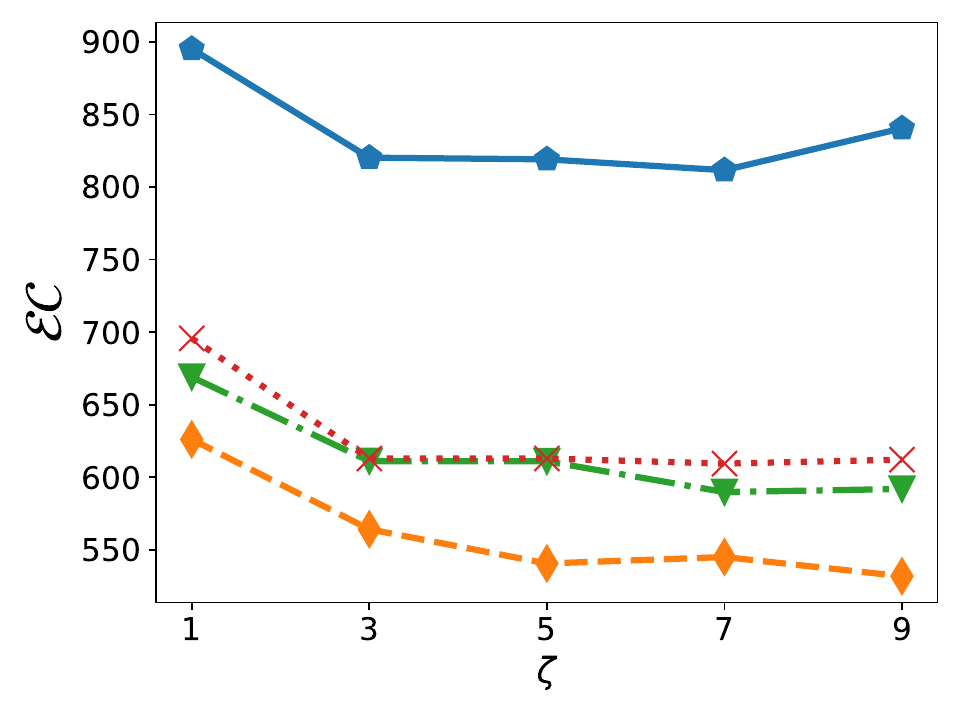}}
\hfil
\caption{Effect of varying the attack budget ($\zeta$) on energy consumption ($\mathcal{EC}$).}
\label{fig: SenseAnalysis-Attack Budget-Energy Consumption}
\end{figure*}

\begin{figure*}[t]
\centering
\subfloat{\includegraphics[height=0.04\textwidth]{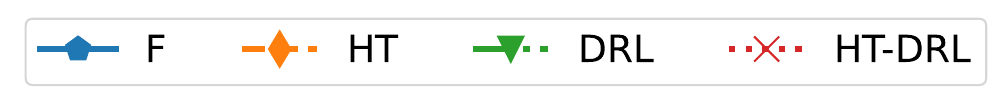}}
\hfil

\vspace{-3mm}
\setcounter{subfigure}{0}
\subfloat[$\mathcal{N}_{AC}$ under fixed attack.]{\includegraphics[width=0.25\textwidth]{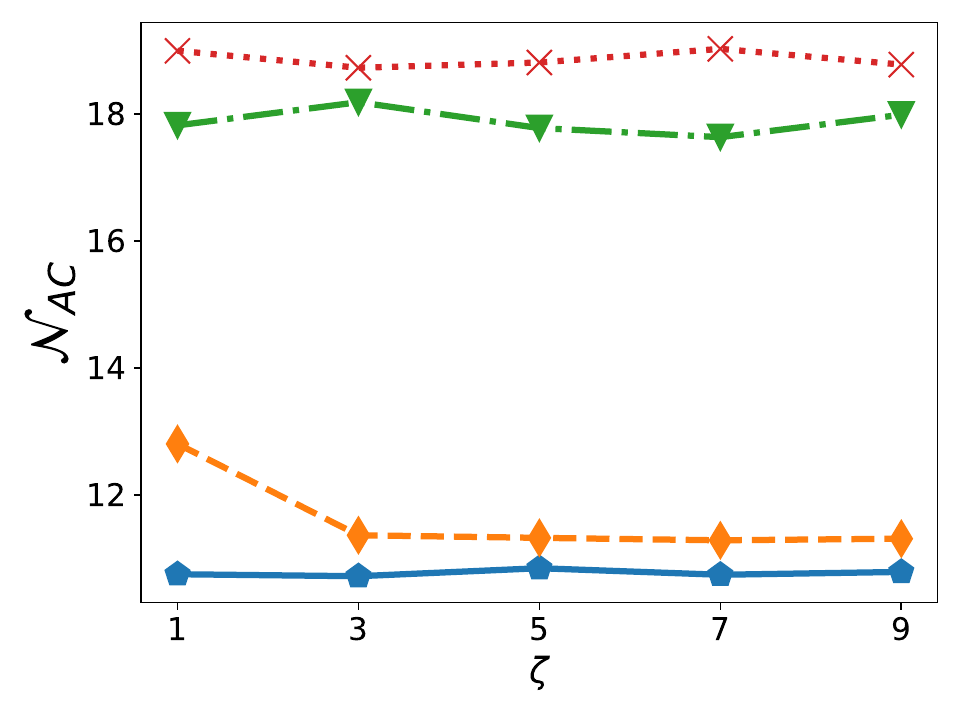}}
\hfil
\subfloat[$\mathcal{N}_{AC}$ under HT attack.]{\includegraphics[width=0.25\textwidth]{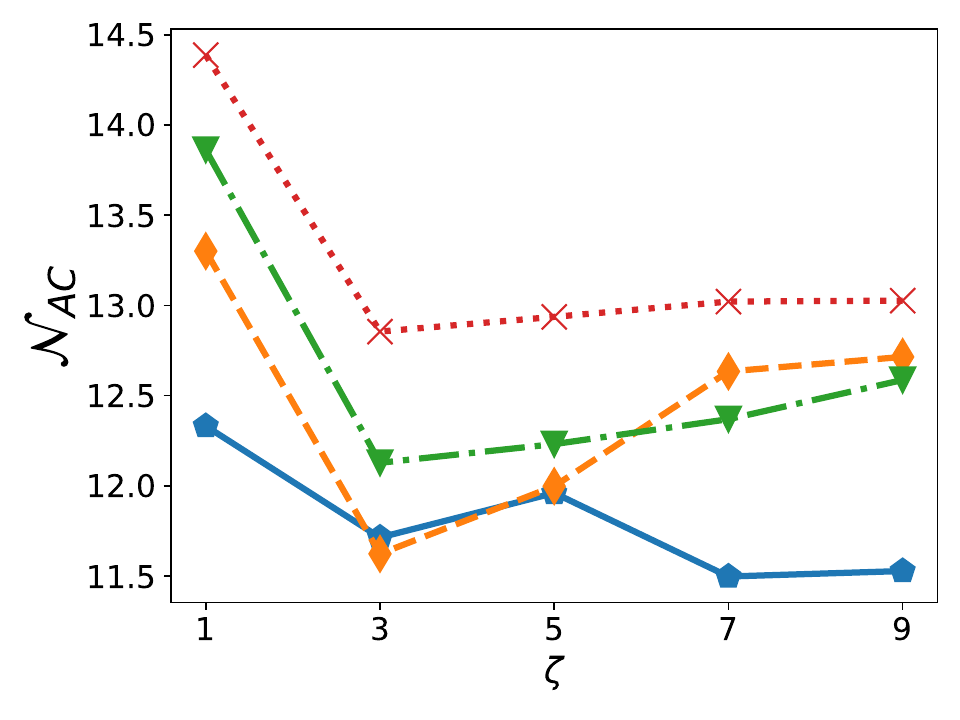}}
\hfil
\subfloat[$\mathcal{N}_{AC}$ under DRL attack.]{\includegraphics[width=0.25\textwidth]{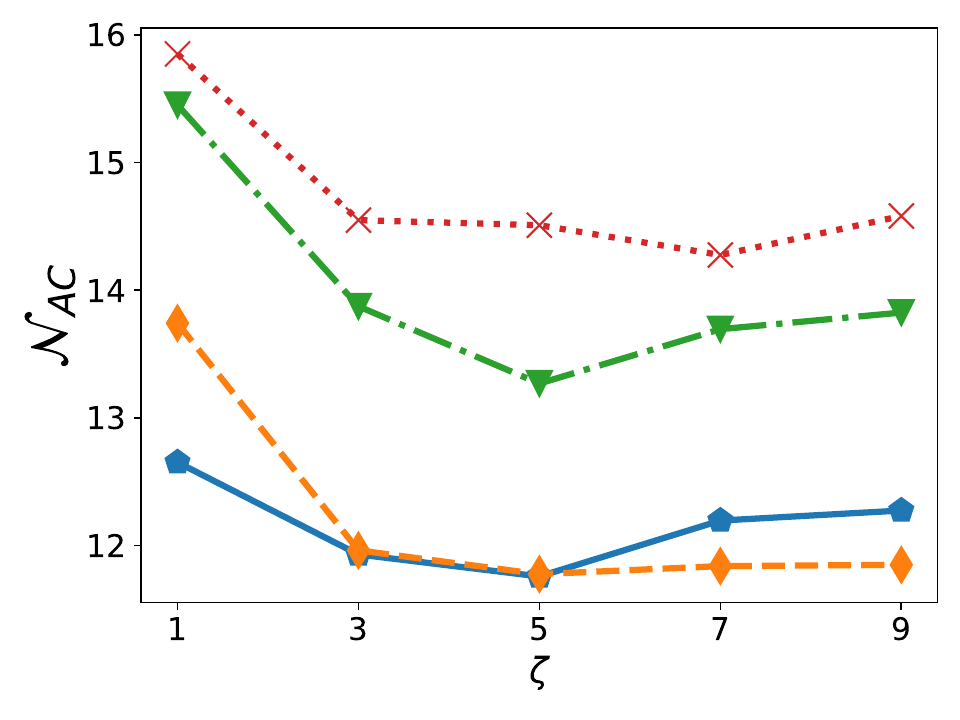}}
\hfil
\subfloat[$\mathcal{N}_{AC}$ under HT-DRL attack.]{\includegraphics[width=0.25\textwidth]{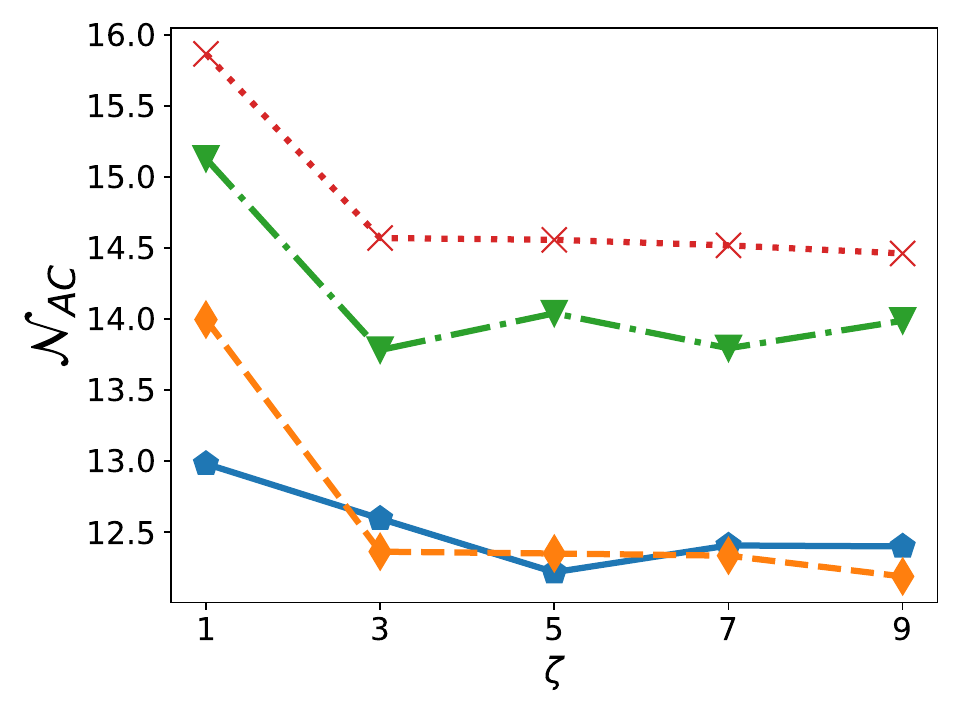}}
\hfil
\caption{Effect of varying the attack budget ($\zeta$) on the number of active, connected drones ($\mathcal{N}_{AC}$).}
\label{fig: SenseAnalysis-Attack Budget-Number of Active Connected Drone}
\end{figure*}

Fig.~\ref{fig: SenseAnalysis-Attack Budget-Number of Active Connected Drone} demonstrates how a different level of the attack budget ($\zeta$) affects the number of active, connected drones ($\mathcal{N}_{AC}$). As observed in Fig.~\ref{fig: SenseAnalysis-Attack Budget-Ratio of Completed Mission}, in Fig.~\ref{fig: SenseAnalysis-Attack Budget-Number of Active Connected Drone}, HT-DRL outperforms among all showing the highest $\mathcal{N}_{AC}$. The same discussions apply as in Appendix~\ref{Appendix: Sensitivity Analysis, Ratio of Mission Completion} with the result of the ratio of mission completion. 

\subsubsection{\bf Accumulated Reward}
\begin{figure*}[t]
\centering
\subfloat{\includegraphics[height=0.04\textwidth]{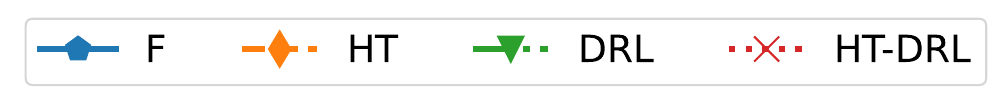}}
\hfil

\vspace{-3mm}
\setcounter{subfigure}{0}
\subfloat[$\mathcal{G}^A$ under fixed attack.]{\includegraphics[width=0.25\textwidth]{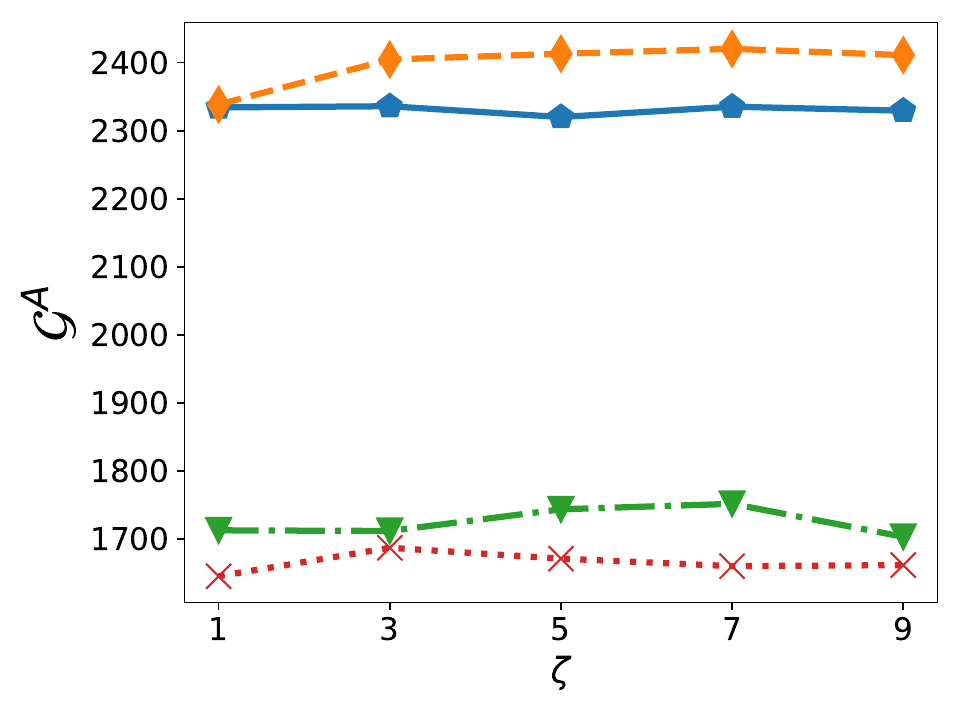}}
\hfil
\subfloat[$\mathcal{G}^A$ under HT attack.]{\includegraphics[width=0.25\textwidth]{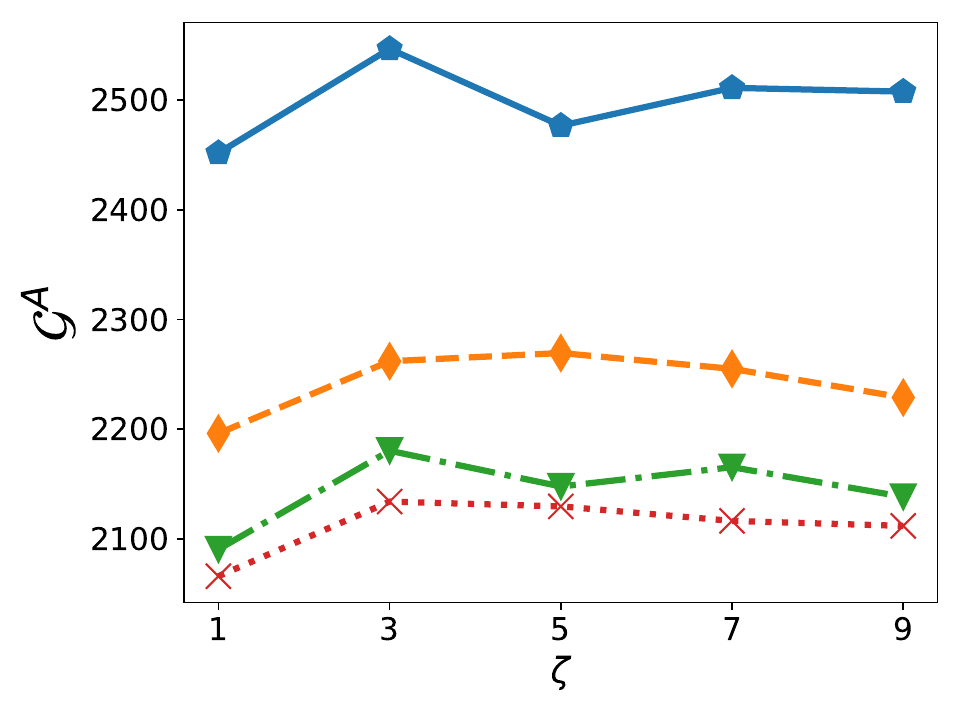}}
\hfil
\subfloat[$\mathcal{G}^A$ under DRL attack.]{\includegraphics[width=0.25\textwidth]{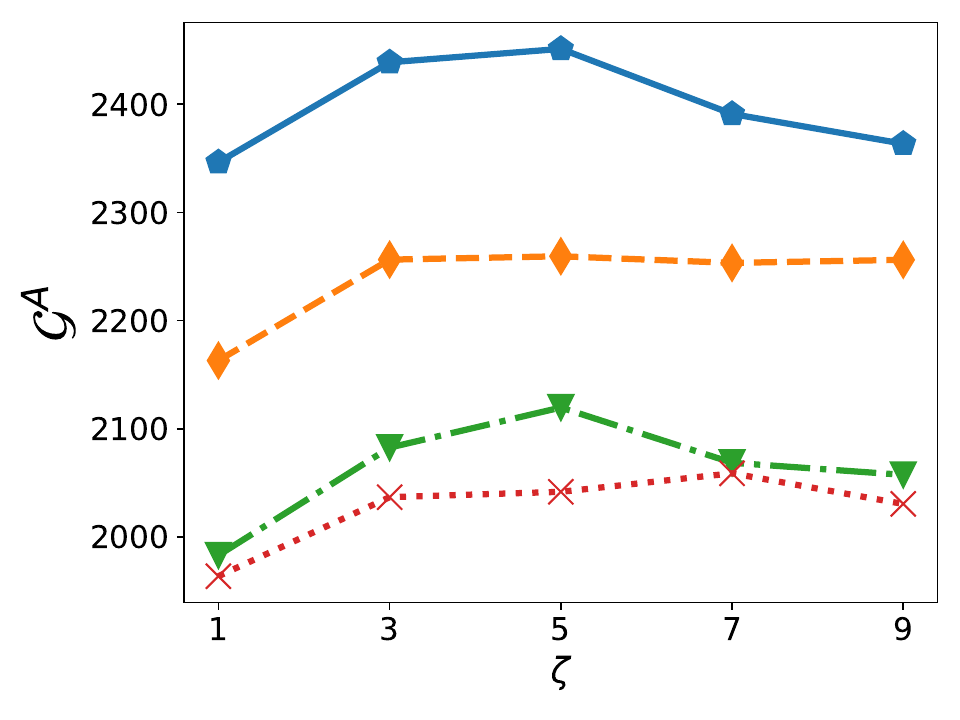}}
\hfil
\subfloat[$\mathcal{G}^A$ under HT-DRL attack.]{\includegraphics[width=0.25\textwidth]{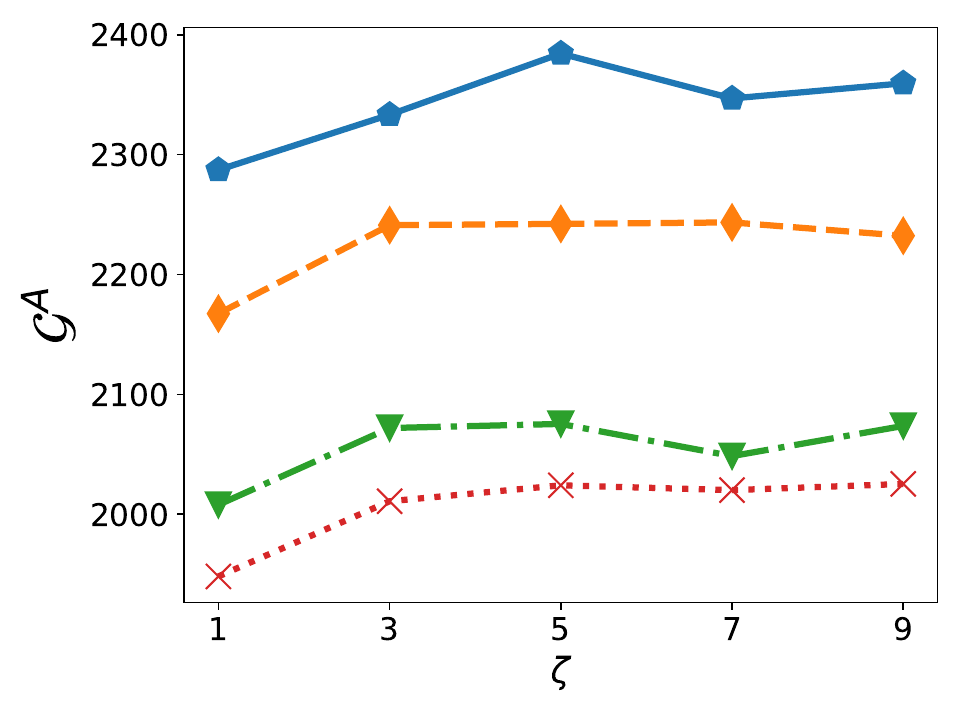}}
\hfil
\caption{Effect of varying the attack budget ($\zeta$) on the attacker's accumulated reward ($\mathcal{G}^A$).}
\label{fig: SenseAnalysis-Attack Budget-Accumulated Reward of Attacker}
\end{figure*}

\begin{figure*}[t]
\centering
\subfloat{\includegraphics[height=0.04\textwidth]{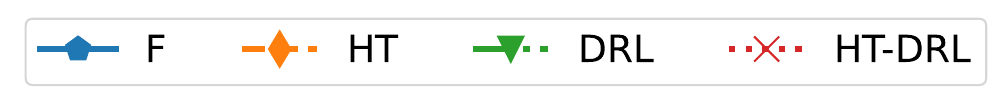}}
\hfil

\vspace{-3mm}
\setcounter{subfigure}{0}
\subfloat[$\mathcal{G}^D$ under fixed attack.]{\includegraphics[width=0.25\textwidth]{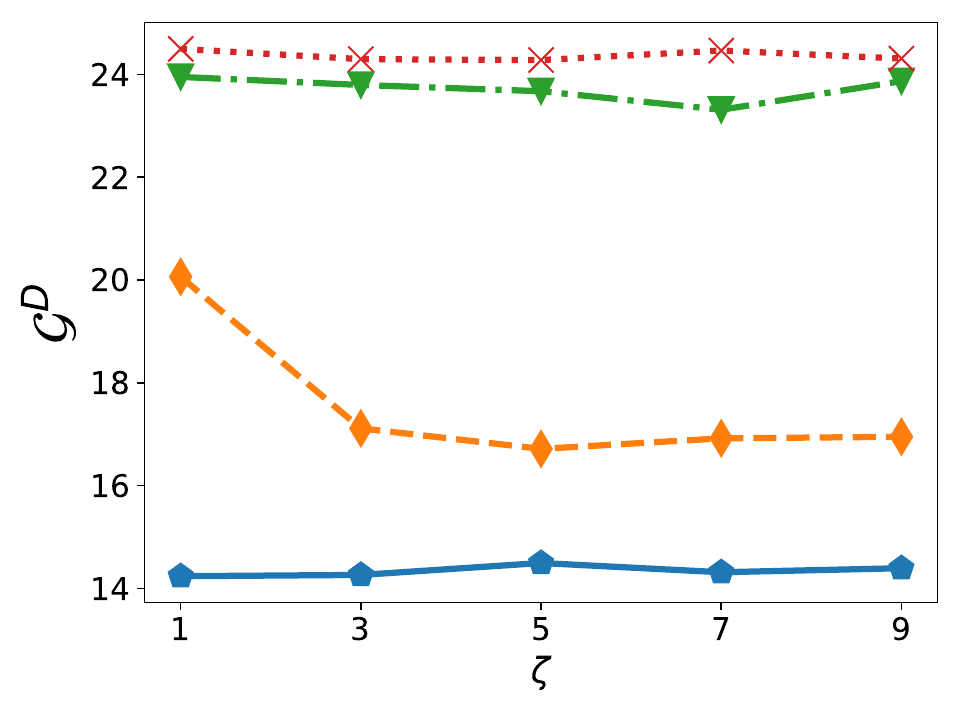}}
\hfil
\subfloat[$\mathcal{G}^D$ under HT attack.]{\includegraphics[width=0.25\textwidth]{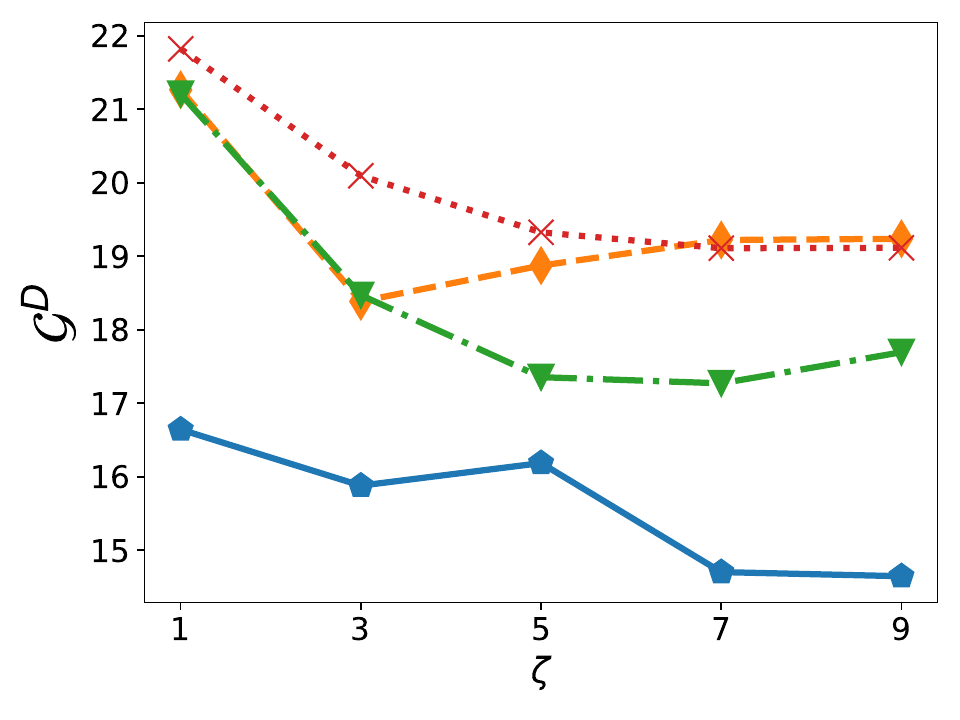}}
\hfil
\subfloat[$\mathcal{G}^D$ under DRL attack.]{\includegraphics[width=0.25\textwidth]{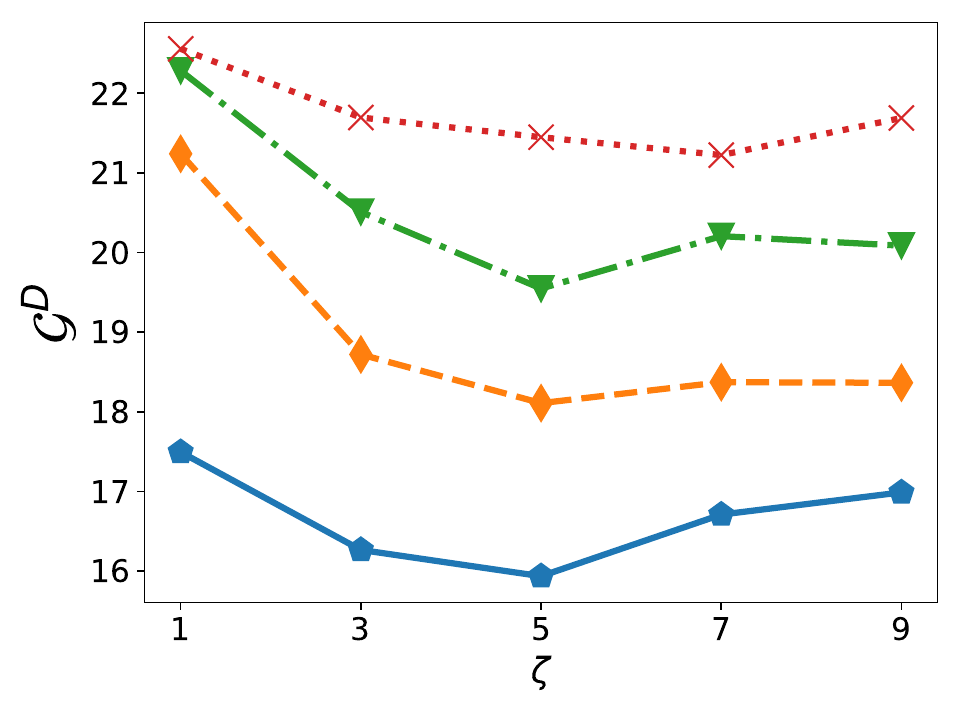}}
\hfil
\subfloat[$\mathcal{G}^D$ under HT-DRL attack.]{\includegraphics[width=0.25\textwidth]{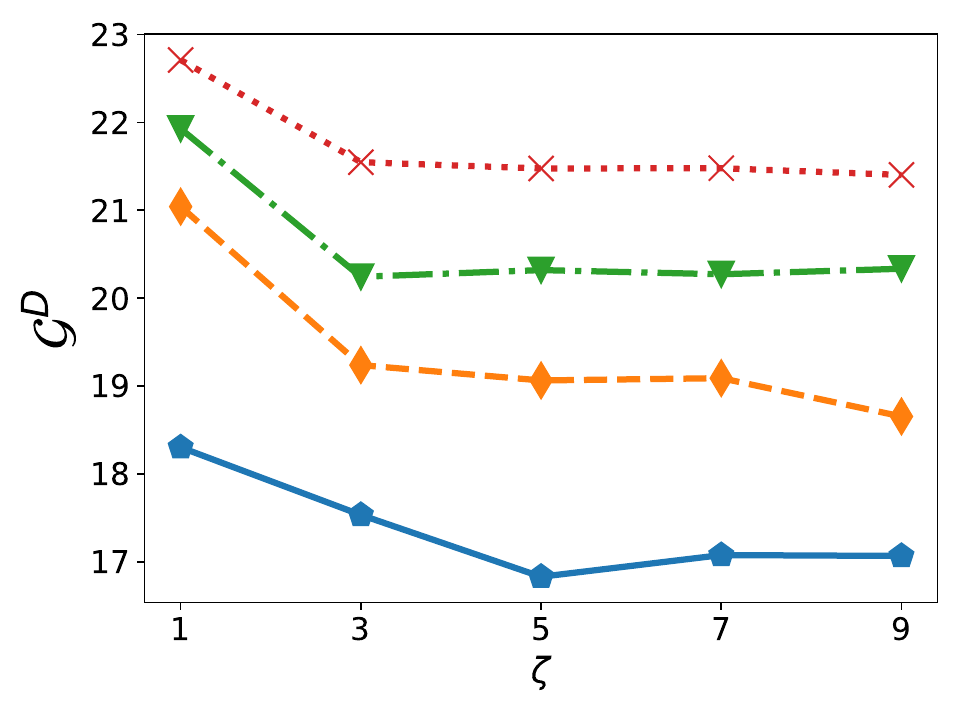}}
\hfil
\caption{Effect of varying the attack budget ($\zeta$) on the defender's accumulated reward ($\mathcal{G}^D$).}
\label{fig: SenseAnalysis-Attack Budget-Accumulated Reward of Defender}
\end{figure*}

Figs. \ref{fig: SenseAnalysis-Attack Budget-Accumulated Reward of Attacker} and \ref{fig: SenseAnalysis-Attack Budget-Accumulated Reward of Defender} show the effect of varying the attack budget ($\zeta$) on the attacker's reward ($\mathcal{G}^A$) and the defender's reward ($\mathcal{G}^D$). Interestingly, we observed a reversed ordering between $\mathcal{G}^A$ and $\mathcal{G}^D$. Specifically, Fig.~\ref{fig: SenseAnalysis-Attack Budget-Accumulated Reward of Attacker} shows the performance order of F $>$ HT $>$ DRL $>$ HT-DRL. On the other hand, Fig. \ref{fig: SenseAnalysis-Attack Budget-Accumulated Reward of Defender} demonstrates the performance order of HT-DRL $>$ DRL $>$ HT $>$ F. This is because the attacker's immediate reward is the number of mission tasks not completed in round $t$ (see more details in Section III-B3),
while the defender's immediate reward is the number of mission tasks completed in round $t$ (see the details in Section III-C3).
As a result, this reward design represents close to a zero-sum game based on the rationale that one player's gain is the opponent player's loss and vice-versa.

\section{Distributional Analysis of Each Strategy}
This section delves into the evolution of the defender's strategy distribution across different schemes.

\begin{figure*}[t]
\centering
\subfloat{\includegraphics[height=0.04\textwidth]{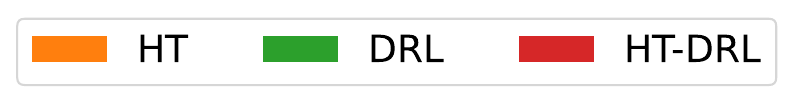}}
\hfil

\vspace{-3mm}
\setcounter{subfigure}{0}
\subfloat[Under fixed attack.]{\includegraphics[width=0.25\textwidth]{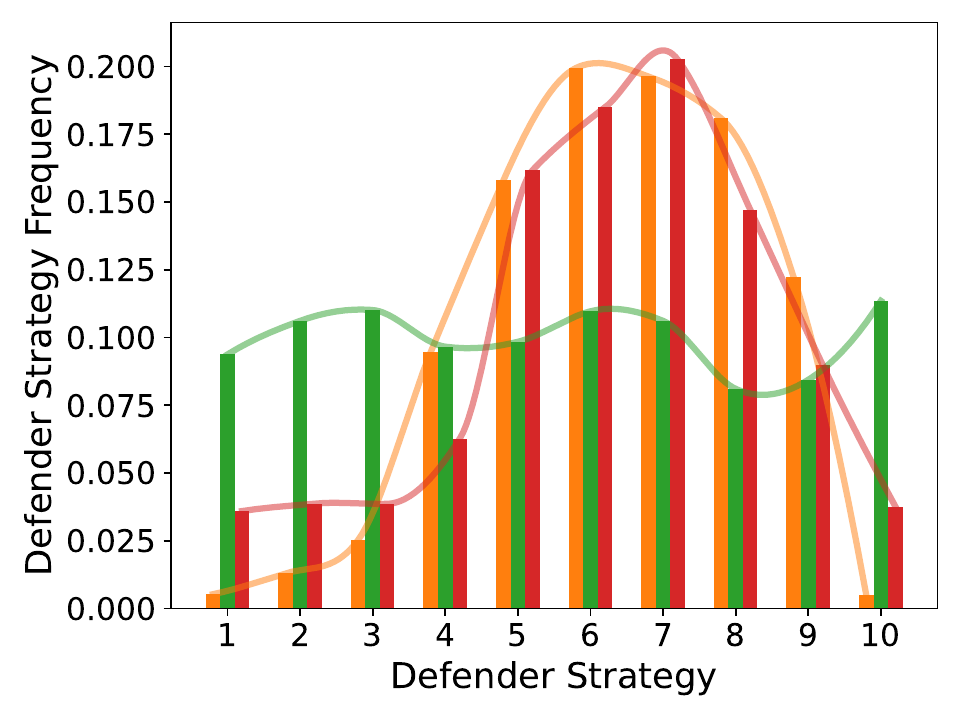}}
\hfil
\subfloat[Under HT attack.]{\includegraphics[width=0.25\textwidth]{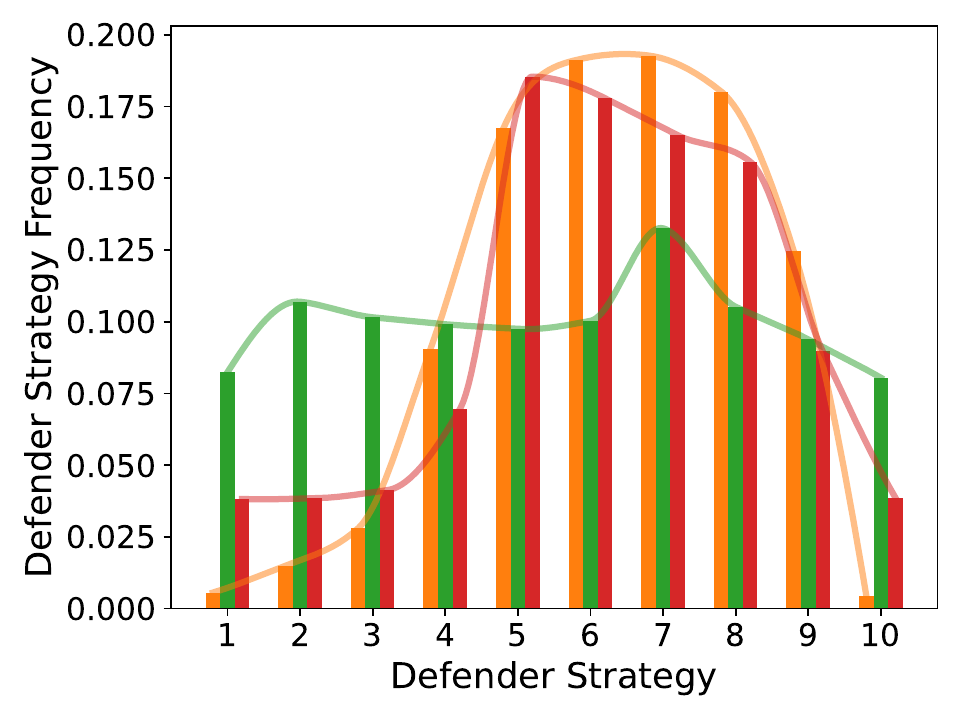}}
\hfil
\subfloat[Under DRL attack.]{\includegraphics[width=0.25\textwidth]{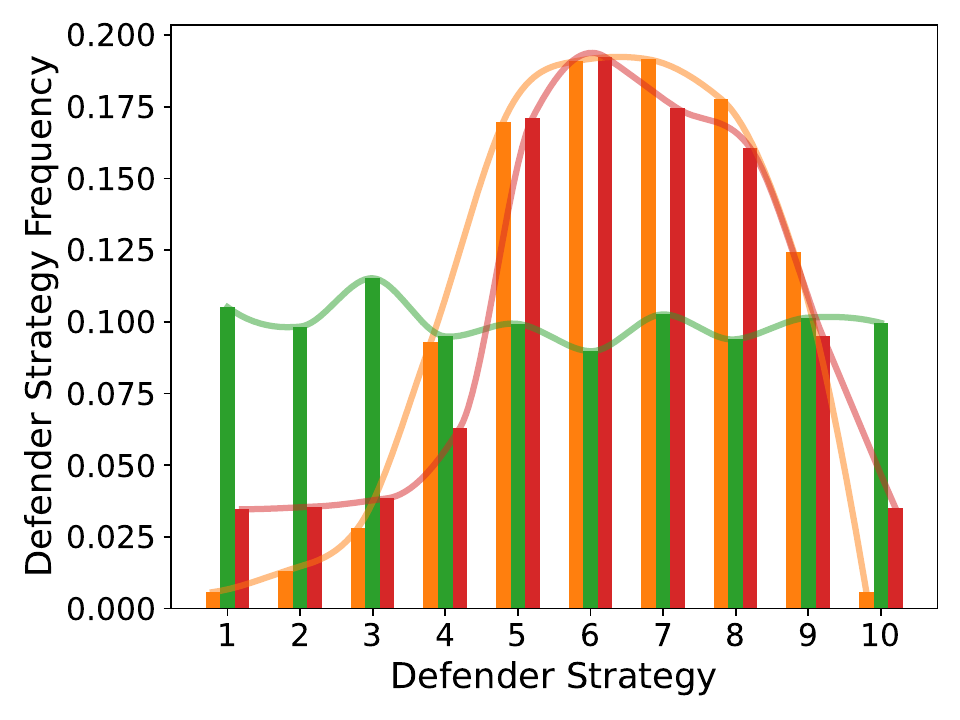}}
\hfil
\subfloat[Under HT-DRL attack.]{\includegraphics[width=0.25\textwidth]{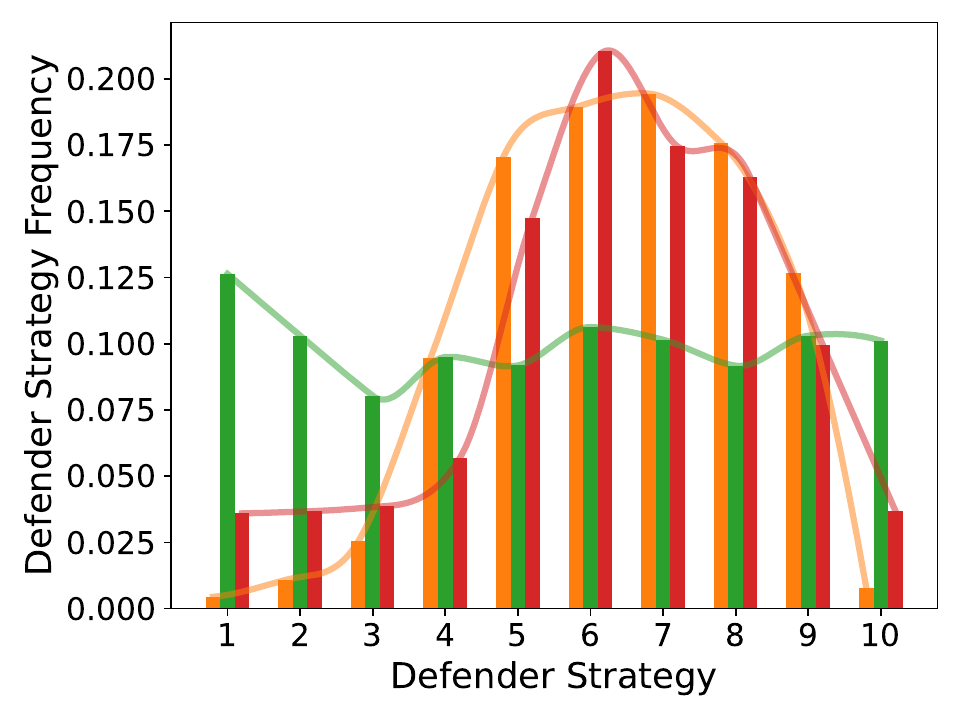}}
\hfil
\caption{Probability distributions of the strategy selection in HT, DRL, and HT-DRL after the training with the two episodes.}

\label{fig: Performance analysis: defender strategy frequency: beginning of training}
\end{figure*}

Fig.~\ref{fig: Performance analysis: defender strategy frequency: beginning of training} illustrates the probability distribution of taking defense strategies after the training with the first two episodes. These probability distributions of taking defense strategies explain: (1) DRL-based defense mainly explores based on the random selection, which shows a uniform probability distribution of the strategy selection.  This ensures a consistent output distribution of neural networks (NNs) for all schemes.  (2) HT and HT-DRL exhibit similar distribution patterns caused by the NN layer added to the original NNs before the output layer of HT-DRL.  By initializing this particular layer using the action distribution from the HT agent, HT-DRL's performance is close to that of HT.  It can start its exploration to achieve global optimal solutions from the local optima identified by HT.

\begin{figure*}[t]
\centering
\subfloat{\includegraphics[height=0.04\textwidth]{figs_v2/Perform_Analy/legendDefender_Strategy_Frequency.pdf}}
\hfil

\vspace{-3mm}
\setcounter{subfigure}{0}
\subfloat[Under fixed attack.]{\includegraphics[width=0.25\textwidth]{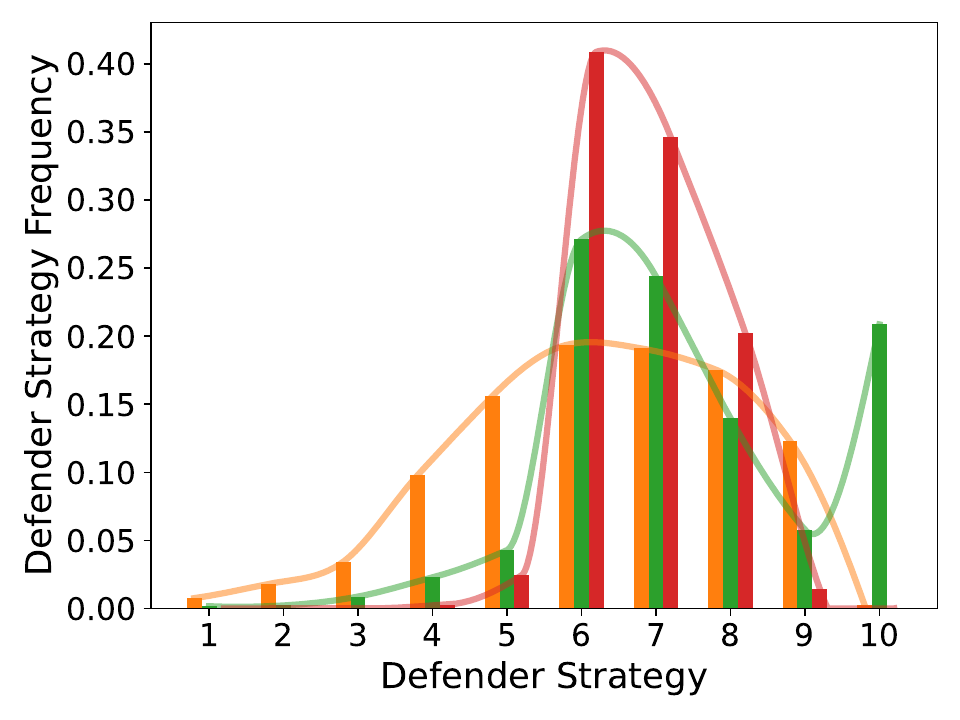}}
\hfil
\subfloat[Under HT attack.]{\includegraphics[width=0.25\textwidth]{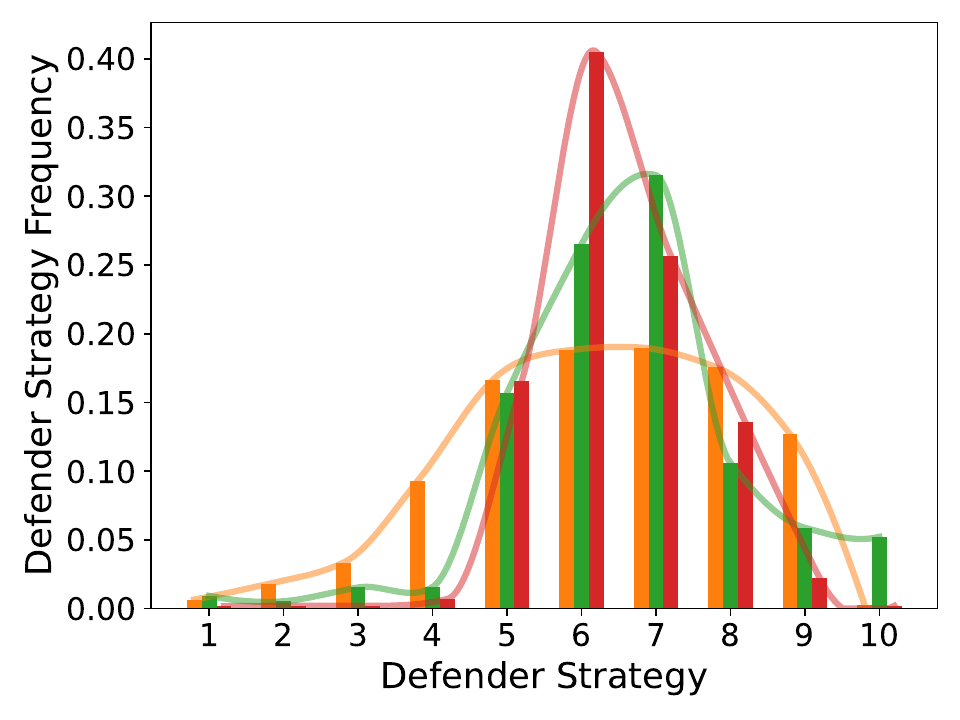}}
\hfil
\subfloat[Under DRL attack.]{\includegraphics[width=0.25\textwidth]{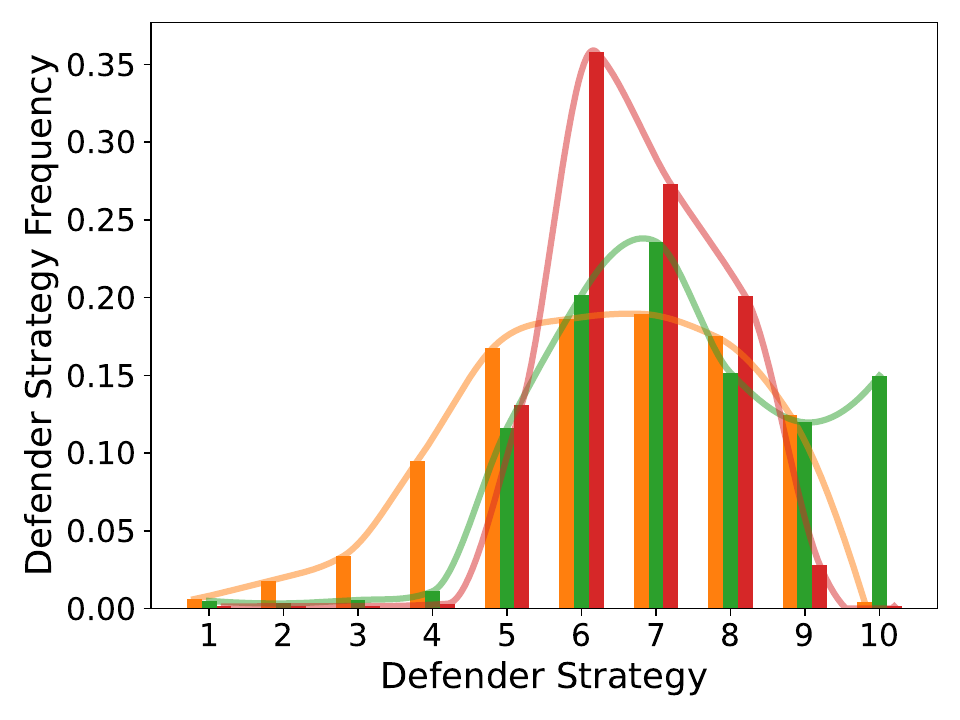}}
\hfil
\subfloat[Under HT-DRL attack.]{\includegraphics[width=0.25\textwidth]{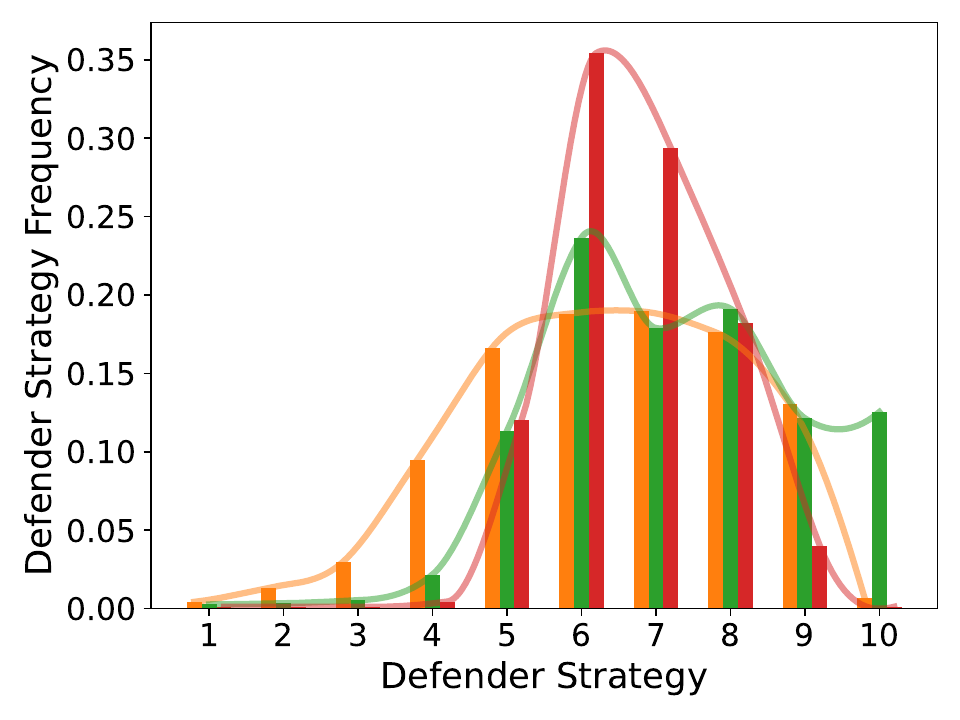}}
\hfil
\caption{Probability distributions of the strategy selection in HT, DRL, and HT-DRL after the models are fully trained.}
\label{fig: Performance analysis: defender strategy frequency: end of training}
\end{figure*}

Fig.~\ref{fig: Performance analysis: defender strategy frequency: end of training} presents the probability distributions of taking defense strategies after the models are fully trained.  The overall key findings are: (1) HT-DRL shows a skewed distribution with clearly dominant strategies compared to the distributions of the strategies taken in HT and DRL. We can expect lower entropies under HT-DRL compared to the ones under HT or DRL.  (2) A notable trend emerges that a sharper distribution often correlates with a higher $\mathcal{R}_{MC}$, as shown in Fig. 7 of the main paper. This trend suggests that those frequently employed strategies likely strike a balance between minimizing detection probability by the attacker and maximizing the connectivity between mission drones to effectively execute the given mission.

